\documentclass[11pt]{article}
\usepackage{jheppub,amsmath,amsfonts,microtype,mathtools,pdflscape}

\usepackage{tikz}
\usetikzlibrary{arrows,calc,patterns,decorations.markings,decorations.pathreplacing,plotmarks,shapes.arrows,decorations.pathmorphing,backgrounds}
\tikzset{snake it/.style={decorate, decoration=snake}}

\newcommand{\pictures}[2]{{#2}}

\usepackage{bbm}
\usepackage{parskip}
\usepackage{amssymb}
\usepackage{braket}
\usepackage[all]{xy}
\usepackage{subfig}
\usepackage[normalem]{ulem}
\usepackage{rotating}
\usepackage{rotfloat}
\usepackage{array}

\numberwithin{equation}{section}
\numberwithin{table}{section}

\newcommand{\gex}[1]{(\alpha G^1_{#1} + \beta G^2_{#1})}
\newcommand{\gbex}[1]{(\gamma\bar G^1_{#1} + \delta\bar G^2_{#1})}

\newcommand{\lex}[1]{(L^1_{#1} + L^2_{#1})}
\newcommand{\lbex}[1]{(\bar L^1_{#1} + \bar L^2_{#1})}

\newcommand{\nn}{\nonumber}
\newcommand{\One}{\mathbf{1}}

\def\SVIR2{{SVIR}$_3{}^{\otimes 2}$} 
\def\SV{\ensuremath{\mathrm{SVIR}}}
\def\sVir{\ensuremath{\mathrm{sVir}} }
\def\sTCIM{\ensuremath{\mathrm{SVIR}_3}} 

\newcommand{\tq}{{\tilde q}}
\newcommand{\DS}{\ensuremath{D_6}}
\newcommand{\ES}{\ensuremath{E_6}}

\newcommand{\wtNS}{{{\scriptstyle \widetilde{N\!S}}}}
\newcommand{\NS}{{{\scriptstyle N\!S}}}

\newcommand{\R}{{\scriptstyle R}}
\newcommand{\ch}{{\mathrm {ch}}}
\newcommand{\chTi}{{\widetilde{\mathrm {ch}}}}

\newcommand{\chthree}{\mathrm {ch}^{3}}
\newcommand{\chten}{\mathrm {ch}^{10}}

\newcommand{\hthree}{{h}^{(3)}}
\newcommand{\hten}{{h}^{(10)}}

\newcommand{\Tr}{\mathrm{Tr}}
\renewcommand{\vec}[1]{{| #1 \rangle }}
\newcommand{\kett}[1]{{| #1 \rangle\!\rangle }}
\newcommand{\braa}[1]{{\langle\!\langle {#1}|}}
\newcommand{\kkett}[1]{{\| #1 \rangle\!\rangle }}
\newcommand{\bbraa}[1]{{\langle\!\langle {#1}\|}}
\newcommand{\vac}{{\ket 0}}
\newcommand{\ketbra}[2]{{\ket{#1}\!\!\bra{#2}}}
\newcommand{\kettbraa}[2]{{\kett{#1}\!\braa{#2}}}
\newcommand{\kkettbbraa}[2]{{\kkett{#1}\!\bbraa{#2}}}
\newcommand{\chii}{\chi^{\vphantom{|}}}
\newcommand{\chiim}{\chi^{(3)}}
\newcommand{\blank}[1]{}%

\newcommand{\be}{\begin{equation}}
\newcommand{\ee}{\end{equation}}

\newcommand{\bc}{\begin{cases}}
\newcommand{\ec}{\end{cases}}

\newcommand{\eps}{{\epsilon}}
\newcommand{\veps}{{\varepsilon}}

\newcommand{\cD}{{\cal D}}
\newcommand{\cE}{{\cal E}}
\newcommand{\cH}{{\cal H}}
\newcommand{\cI}{{\cal I}}
\newcommand{\cT}{{\cal T}}
\newcommand{\cW}{{\cal W}}

\newcommand{\abgd}{{\alpha\beta\gamma\delta}}

\newcommand{\WM}{{W\!\!M}}
\renewcommand{\WM}{{SW}}
\newcommand{\ds}{\displaystyle}

\newcommand{\TiNS}{\wtNS}

\title{Defects in the Tri-critical Ising Model}

\author[a]{Isao Makabe,}
\author[a]{G\'erard M.\ T.\ Watts}

\affiliation[a]{Dept.\ of Mathematics, King's College London,\\
Strand, London WC2R\;2LS, UK}

\emailAdd{isao.makabe@kcl.ac.uk,gerard.watts@kcl.ac.uk}

\abstract{We consider two different conformal field theories with central charge
$c=7/10$.
One is the diagonal invariant minimal model in which all
fields have integer spins; the other is the local fermionic theory
with superconformal symmetry in which fields can have half-integer
spin. We construct new conformal (but not topological or factorised)
defects in the 
minimal model. We do this by first constructing defects in the
fermionic model as boundary conditions in a fermionic theory of
central charge $c=7/5$, using the folding trick as first proposed by
Gang and Yamaguchi \cite{GY}. We then acting on these with
interface defects to find the new conformal defects. As part of
the construction, we find the topological defects in the fermionic
theory and the interfaces between the fermionic theory and the
minimal model. We also consider the simpler case of defects in the
theory of a single free fermion and interface defects between
the Ising model and a single fermion as a prelude to calculations in
the tri-critical Ising model.}

\preprint{KCL-MTH-17-01}
\begin{document}
\maketitle

\section{Introduction}
\label{char calc}

A conformal defect is a local line of discontinuity in a conformal
field theory or between two different conformal field theories.
The simplest situation is that of a cylinder with the defect wrapping
the cylinder once, or equivalently of a plane with the defect placed
on the unit circle. The defect can be represented as an operator $D$
mapping from the Hilbert space of the theory inside the unit circle to
that of the theory outside the circle and it is easy to state the
condition that the defect is conformal:
\be
 ( L_m - \bar L_{-m} ) \; D
= D (L_m - \bar L_{-m} )
\;.
\label{eq:conf.def}
\ee
There are very few cases, however, in which the general solution to
this equation can be found as this is equivalent (via the folding
trick) to the equations of a conformal boundary condition in the
folded model. If the central charges of the holomorphic and anti-holomorphic Virasoro
algebras of two theories inside and
outside the defect are $(c_1,\bar c_1)$ and $(c_2,\bar c_2)$, then the
folded model has 
central charges $(c_{tot}=c_1 + \bar c_2, \: \bar c_{tot}=\bar c_1 +
c_2)$. A conformal defect between the two theories can only exist if
$c_1 - \bar c_2 = \bar c_1 - c_2$, or equivalently a conformal 
conformal boundary condition on the folded model can only exist if
$c_{tot}=\bar c_{tot}$. 
Conformal boundary conditions have been completely classified for
minimal models of the Virasoro Algebra \cite{BPPZ1,BPPZ2}
(which have $c<1$) and for free
boson theories \cite{Friedan,GRW01,GR01,Janik} 
(with $c=1$) but not for higher values of $c$. 
Since the tri-critical Ising model (TCIM) has $c=\bar c =7/10$, the
folded model has $c_{tot}=7/5$, the general conformal boundary
condition for the folded model is not known and so the 
general solution to \eqref{eq:conf.def} is not known for the TCIM.
From now on, we will only consider theories with $c=\bar c$ and will
not mention $\bar c$ again.

Some particular solutions to \eqref{eq:conf.def} are known - these are
the Topological and Factorising defects. A defect is topological if it
satisfies 
\be
 L_m \; D
= D L_m
\;,\;\;\;\;
  \bar L_{m}  \; D
= D \bar L_{m} 
\;.
\label{eq:top.def}
\ee
Such defects are classified for unitary minimal models such as the
TCIM for which there are 6 fundamental topological defects.

A defect is factorising if it satisfies
\be
 ( L_m - \bar L_{-m} ) \; D = 0
\;,\;\;
 D (L_m - \bar L_{-m} ) = 0
\;,
\label{eq:fact.def}
\ee
which is equivalent to a cut in the worldsheet separating the inner
and outer theories with a conformal boundary condition for each
theory. Conformal boundary conditions have again been classified for
unitary minimal models and so all factorising defects are also known
for the TCIM: there are 6 fundamental conformal boundary
conditions for the TCIM leading to 36 fundamental factorising
defects. 

Sums of topological and factorising defects will of course also
satisfy \eqref{eq:conf.def} but there is strong evidence that these do
not exhaust the list 
of conformal defects for the TCIM. Notably, one can consider relevant
perturbations of topological defects in the TCIM. Such perturbations
define a renormalisation group flow in the space of defects,
with the defect at the IR fixed point also being a conformal
defect. Both perturbative and numerical TCSA (truncated conformal
space approach) calculations suggest \cite{KRW} that there are
non-topological, non-factorising, 
conformal defects which can be found this way.

In \cite{GY}, Gang and Yamaguchi considered defects in TCIM
constructed as GSO projections of boundary states in a folded
supersymmetric model. This provided candidate expressions for new
non-topological non-factorising conformal defects, but their
construction also produced factorising defects which did not agree
with the known expressions. For this reason, we think that their paper
deserves re-examination: if the construction produces factorising
defects which fall outside the known classification, then it is quite
possible that the new candidate defects they proposed are also incorrect.
 
\begin{figure}[t]

\pictures{
$$  
  \begin{picture}(440,225)
  \put(0,0){\scalebox{1.}{\includegraphics{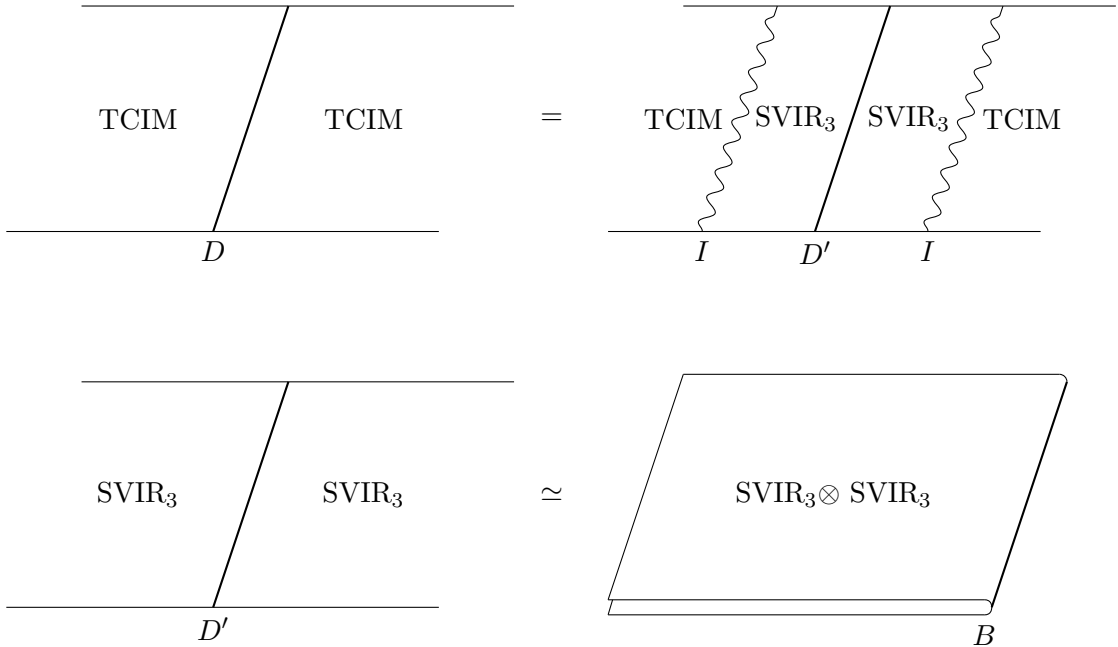}}}
  \end{picture}
$$
}{
\begin{center}
\begin{tikzpicture}
	\pgfmathsetmacro{\hshift}{4}
	\pgfmathsetmacro{\vshift}{5}

	\draw (1-\hshift,3.5) -- (6.75-\hshift,3.5);
	\draw (0-\hshift,0.5) -- (5.75-\hshift,0.5);

	\draw[thick] (2.75-\hshift,0.5) -- (3.75-\hshift,3.5);
	
	\node at (1.75-\hshift,2) {TCIM};
	\node at (4.75-\hshift,2) {TCIM};	

	\node[below] at (2.75-\hshift,0.5) {$D$}; 

	\node at (3.25,2) {$=$};
	\node at (3.25,2-\vshift) {$\simeq$};

	\draw (1+\hshift,3.5) -- (6.75+\hshift,3.5);
	\draw (0+\hshift,0.5) -- (5.75+\hshift,0.5);
	
	\draw[snake it] (1.25+\hshift,0.5) -- (2.25+\hshift,3.5);
	\draw[thick] (2.75+\hshift,0.5) -- (3.75+\hshift,3.5);
	\draw[snake it] (4.25+\hshift,0.5) -- (5.25+\hshift,3.5);
	
	\node at (1+\hshift,2) {TCIM};
	\node at (2.5+\hshift,2) {\sTCIM};
	\node at (4+\hshift,2) {\sTCIM};
	\node at (5.5+\hshift,2) {TCIM};	
	
	\node[below] at (1.25+\hshift,0.5) {$I$}; 
	\node[below] at (2.75+\hshift,0.5) {$D'$}; 
	\node[below] at (4.25+\hshift,0.5) {$I$}; 
	
	\draw (1-\hshift,3.5-\vshift) -- (6.75-\hshift,3.5-\vshift);
	\draw (0-\hshift,0.5-\vshift) -- (5.75-\hshift,0.5-\vshift);

	\draw[thick] (2.75-\hshift,0.5-\vshift) -- (3.75-\hshift,3.5-\vshift);
	
	\node at (1.75-\hshift,2-\vshift) {\sTCIM};
	\node at (4.75-\hshift,2-\vshift) {\sTCIM};	

	\node[below] at (2.75-\hshift,0.5-\vshift) {$D'$}; 
	
	\draw (1+\hshift,3.6-\vshift) -- (6+\hshift,3.6-\vshift);
	\draw (0+\hshift,0.6-\vshift) -- (5+\hshift,0.6-\vshift);
	\draw (0+\hshift,0.4-\vshift) -- (5+\hshift,0.4-\vshift);
	\draw (0+\hshift,0.4-\vshift) -- (0.06+\hshift,0.6-\vshift);
	
	\draw plot [smooth,tension=2] coordinates {(5+\hshift,0.6-\vshift) (5.1+\hshift,0.5-\vshift) (5+\hshift,0.4-\vshift)};
	\draw plot (6+\hshift,3.6-\vshift) to[out=0,in=90] (6.1+\hshift,3.5-\vshift);
	
	\draw (0+\hshift,0.6-\vshift) -- (1+\hshift,3.6-\vshift);
	\draw[thick] (5.1+\hshift,0.5-\vshift) -- (6.1+\hshift,3.5-\vshift);
	
	\node at (3+\hshift,2-\vshift) {\sTCIM $\otimes$ \sTCIM};
	\node[below] at (5+\hshift,0.4-\vshift) {$B$}; 
	
\end{tikzpicture}
\end{center}
}

\caption{The equivalence between a boundary condition $B$ in the
  folded fermionic model, a defect $D'$ in the fermionic model and a
  defect $D$ in the TCIM. $I$ is an interface defect.} 
\label{fig:fold1}
\end{figure}

In this paper we take a first step towards re-examining the results of
\cite{GY} through a related, but different, route to constructing
defects in TCIM. We adapt ideas from Gaiotto \cite{G1}
and from Gang and Yamaguchi \cite{GY} to construct defects in TCIM from
conformal boundaries in the folded version of the Neveu-Schwarz sector
of a fermionic theory through the use of interfaces between the
fermionic theory and the TCIM.
The idea is that there is a fermionic $c=7/10$ theory (\sTCIM) with
superconformal symmetry which is related to the TCIM. 
Superconformal
defects in \sTCIM\ are then equivalent to superconformal boundary
conditions in \SVIR2$ = $\sTCIM$\otimes$\sTCIM. As 
\SVIR2 has central charge $c=7/5$ less than
$3/2$, its {\it superconformal} boundary conditions can be
classified. We propose a set of 
boundary conditions $B$ for \SVIR2 (related to but not the same as
those in \cite{GY}), leading to superconformal defects $D'$ in \sTCIM.
We also propose a set of topological interfaces $I$ between
\sTCIM\ and TCIM.
One can then sandwich the superconformal defects $D'$ between the
topological interface defects $I$ 
interpolating between the TCIM and the fermionic model, leading to
conformal defects $D = I D' I^\dagger$ in the TCIM. 
One can then easily show (by considering the defect entropy) that
there are defects $D$ that are not
expressible in terms of elementary topological and factorised defects
in TCIM, and hence are new defects.
This construction is summarised in figure \ref{fig:fold1}.

As a warm-up exercise, in section \ref{sec:ising} we consider first the
simpler case of the 
$c=1/2$ Ising model and the related free-fermion. We propose defects
in the free-fermion model which preserve the fermion algebra and
interfaces between the Ising model and the free-fermion model. 

We then turn in section \ref{sec:svir} to TCIM, the tri-critical Ising
model, and \sTCIM, the fermionic model at $c=7/10$, and propose
superconformal topological defects in \sTCIM\ and
topological interfaces between \sTCIM\ and TCIM.

Next, in section \ref{sec:svir2}, we consider \SVIR2 and propose a set
of superconformal boundary conditions in this model. This is related
to, but not the same as, the construction in \cite{GY}. By considering
partition functions of \SVIR2 on the cylinder and comparing them with
partition functions of \sTCIM\ on the torus, we show how to interpret
some of these boundary conditions in terms of known defects in \sTCIM,
and how some are new defects in \sTCIM.

Finally, we consider the construction of conformal defects in TCIM
from the superconformal defects in \sTCIM\ in section \ref{sec:tcim} and
show that some of these cannot be constructed as superpositions of
known topological and factorised defects in TCIM.
We also explain that we are unable to compare our results with those
of \cite{GY} which do not seem compatible with our general approach.

We end with some comments on the new defects and on possible further work.

\section{The Ising model and the free fermion}
\label{sec:ising}

In this section we recall the basics of the Ising model, set up our
definition of the Neveu-Schwarz free fermion theory, and propose a set
of defects in the free fermion theory and a set of
interfaces between the Ising model and the free-fermion theory. We present
some results exhibiting the consistency of these proposals.
We give our conventions for the characters of the free fermion and
Ising model in appendix \ref{app:ff}.

\subsection{The Ising model}

There is a single modular invariant unitary conformal field theory
with $c=1/2$, the Ising model.
This is the first non-trivial value of $c$ in the minimal unitary series
$c(m)=1 - 6/(m(m+1))$ for the Virasoro algebra, corresponding to $m=3$.

The Virasoro algebra with $c=1/2$  has three unitary irreducible highest weight
representations with $h\in\{0, \: 1/2, \: 1/16\}$ with characters%
\footnote{We denote the character of the representation of weight $h$ in the $m$-th unitary Virasoro minimal model by $\chi^{(m)}_h$}
$\chiim_h$ . 
As the Ising model is a diagonal modular invariant theory, the
theory correspondingly has three  
primary fields $(I \equiv\phi_0, \: \epsilon \equiv \phi_{1/2}, \:
\sigma \equiv \phi_{1/16})$, three topological defects
$(D_0, \: D_{1/2}, \: D_{1/16})$ and three elementary conformal boundary
conditions $(B_0, \: B_{1/2}, \: B_{1/16})$, each labelled by the set of
highest weight representations \cite{BPPZ2}. The bulk Hilbert space is
\be
\cH_\text{Ising}
= (\cH_0\otimes\bar\cH_0)
\oplus
(\cH_{1/2}\otimes\bar\cH_{1/2})
\oplus
(\cH_{1/16}\otimes\bar\cH_{1/16})
\;,
\ee
and the partition function on the torus is 
\be
\Tr_{\cH_\text{Ising}}(q^{L_0 - c/24}\bar q^{\bar L_0 - c/24})
= | \chiim_0(q)|^2
+ | \chiim_{1/2}(q)|^2
+ | \chiim_{1/16}(q)|^2
\;,
\ee
which is modular invariant.

Modular invariance is, however, not a necessary condition to have a
well defined field theory on the plane or on the cylinder, and amongst
the possible field theories one can consider is the ``Neveu-Schwarz
free fermion'' (FF) defined below.

\subsection{The Neveu-Schwarz free fermion}

We will consider the
case of a symmetric theory with $c=\bar c=1/2$ of a holomorphic
fermion $\psi(z)$ and an
anti-holomorphic fermion $\bar\psi(\bar z)$. We shall also only consider the
Neveu-Schwarz sector, in which the fermions on the plane have mode decompositions
\be
 \psi(z) = \sum_{m\in\mathbb Z} \psi_m z^{-m-1/2}
\;,\;\;
 \bar\psi(\bar z) = \sum_{m\in\mathbb Z} \bar \psi_m \bar z^{-m-1/2}
\;,
\label{eq:psim}
\ee
and anti-commutators
\be
 \{\psi_m,\psi_n\} = \{\bar\psi_m,\bar\psi_n\} = \delta_{m+n,0}
\;,\;\;
 \{\psi_m,\bar\psi_n\}=0
\;.
\ee
The Hilbert space, $\cH_\text{FF}$,  of the Neveu-Schwarz fermion  is the Fock space generated by the action of
negative fermion modes acting on the unique vacuum state 
\def\vac{{|0\rangle}}%
$\vac$,
\be
\cH_\text{FF}
 = \cH_{\NS}\otimes \bar\cH_{\NS}
 = (\cH_0 \oplus \cH_{1/2})\otimes(\bar\cH_0 \oplus \bar\cH_{1/2})
\;.
\ee
The partition function on the cylinder is 
\be
 Z_\text{FF} = \Tr_{\cH_\text{FF}}( q^{L_0 - c/24}\bar q^{L_0 - c/24})
= | \chii_{\NS}(q) |^2
\;,
\ee
where
\be
\chii_{\NS}(q) = \chiim_0(q) + \chiim_{1/2}(q)
\;.
\ee
$Z_\text{FF}$ is invariant under the modular transformation $\tau \to
-1/\tau$, that is $q = \exp(2\pi i\tau) \to \tq = \exp(-2\pi
i/\tau)$. The function is not invariant under $\tau \to 1+\tau$ but we
shall in general consider the theory defined on right torus with $q$
real, and the theory is well defined on such a space.

\subsubsection{Defects in the free fermion theory}

We say a defect $D_{\eps,\eps'}$ in the free-fermion model conserves the fermion
algebra (up to automorphism) if 
\be
 \psi_m D_{\eps,\eps'} = \eps D_{\eps,\eps'} \psi_m
\;,\;\;
 \bar \psi_n D_{\eps,\eps'} = \eps' D_{\eps,\eps'} \bar\psi_m
\;,
\label{eq:Dsigns}
\ee
where $\eps=\pm 1$ and $\eps'=\pm 1$.
These conditions entirely determine the defect operators up to
normalisation constants $\alpha_{\eps,\eps'}$ as
\be
 D_{++} = \alpha_{++} \One\;,\;\;
 D_{+-} = \alpha_{+-} (-1)^{\bar F}
\;,\;\;
 D_{-+} = \alpha_{-+} (-1)^F
\;,\;\;
 D_{--} = \alpha_{--} (-1)^{F + \bar F}
\;,
\ee
where $\One$ is the identity operator on $\cH_\text{FF}$.

We would like to impose two conditions: firstly, the Cardy condition
that the trace over the 
cylinder with the insertion of a defect $D_A$ is an
integer combination of characters of the free fermion in the dual
channel, ie the constants $N^A_{\alpha\beta}$ defined by
\be
 \Tr_{\cH_\text{FF}}( D_A \: q^{L_0 - c/24} \: {\bar q}^{\bar L_0 - c/24})
= \sum_{\alpha,\beta} \: N^A_{\alpha\beta} \:
  \chii_\alpha(\tq)\:
  \chii_\beta(\bar{\tq})
\;,
\label{eq:sum}
\ee
should be non-negative integers.
This reflects the requirement that the torus partition function can be
interpreted as a trace over a space which carries a representation of
the free-fermion algebra. 
Note that we will allow $\alpha$ and $\beta$ in the sum in
\eqref{eq:sum} to run over $\NS$ and $\R$ which is necessary as the
defect can change the periodicities of the 
field $\psi$ and $\bar \psi$ - see the appendix for details.


The Hilbert space of the Neveu-Schwarz fermions is 
$\cH_\text{FF}$ and so we have
\newcommand{\bea}{\begin{eqnarray}}
\newcommand{\eea}{\end{eqnarray}}
\be
\begin{array}{rclcl}
    \Tr_{\cH}( D_{++} \: q^{L_0 -c/24} \bar q^{\bar L_0 - c/24} )
&=& \alpha_{++}|\chii_{\NS}(q)|^2
&=& \alpha_{++} | \chii_{\NS}(\tq)|^2
\\
    \Tr_{\cH}( D_{+-} \: q^{L_0 -c/24} \bar q^{\bar L_0 - c/24} )
&=&   \alpha_{+-} \chii_{\NS}(q)\chii_{\widetilde{\NS}}(\bar q)
&=&  \sqrt 2\alpha_{+-} \chii_{\NS}(\tq) \chii_{\R}(\bar{\tq})
\\
    \Tr_{\cH}( D_{-+} \: q^{L_0 -c/24} \bar q^{\bar L_0 - c/24} )
&=&   \alpha_{-+} \chii_{\widetilde{\NS}}(q)\chii_{\NS}(\bar q)
&=& \sqrt 2\alpha_{-+} \chi_{\R}(\tq) \chii_{\NS}(\bar{\tq})
\\
    \Tr_{\cH}( D_{--} \: q^{L_0 -c/24} \bar q^{\bar L_0 - c/24} )
&=&   \alpha_{--} \chii_{\widetilde{\NS}}(q)\chii_{\widetilde{\NS}}(\bar q)
&=& 2 \alpha_{--} |\chii_{\R}(\tq)|^2
\end{array}
\label{eq:dtraces}
\ee
Secondly, we would like the structure constants $M_{AB}^C$ in the
algebra of defect operators, 
\be
D_A \: D_B \; = \; \sum_C\; M_{AB}^C D_C
\;,
\ee
to be
non-negative integers. We clearly have
\be
 D_{\eps,\eps'} \: D_{\eta,\eta'} = \frac{ \alpha_{\eps,\eps'} \: \alpha_{\eta,\eta'} }{\alpha_{\eps\eta,\eps'\eta'}} D_{\eps\eta,\eps'\eta'}
\;.
\label{eq:dprods}
\ee
The simplest solution that makes the right-hand side of equations
\eqref{eq:dtraces}  and the coefficients in equation \eqref{eq:dprods}
integers is
\be
 \alpha_{++} = \alpha_{--} = 1
\;,\;\;
 \alpha_{+-} = \alpha_{-+} = \sqrt 2
\;.
\label{eq:alphasol}
\ee

\begin{figure}[htb]

\pictures{
$$  
  \begin{picture}(440,325)
  \put(-20,0){\scalebox{.95}{\includegraphics{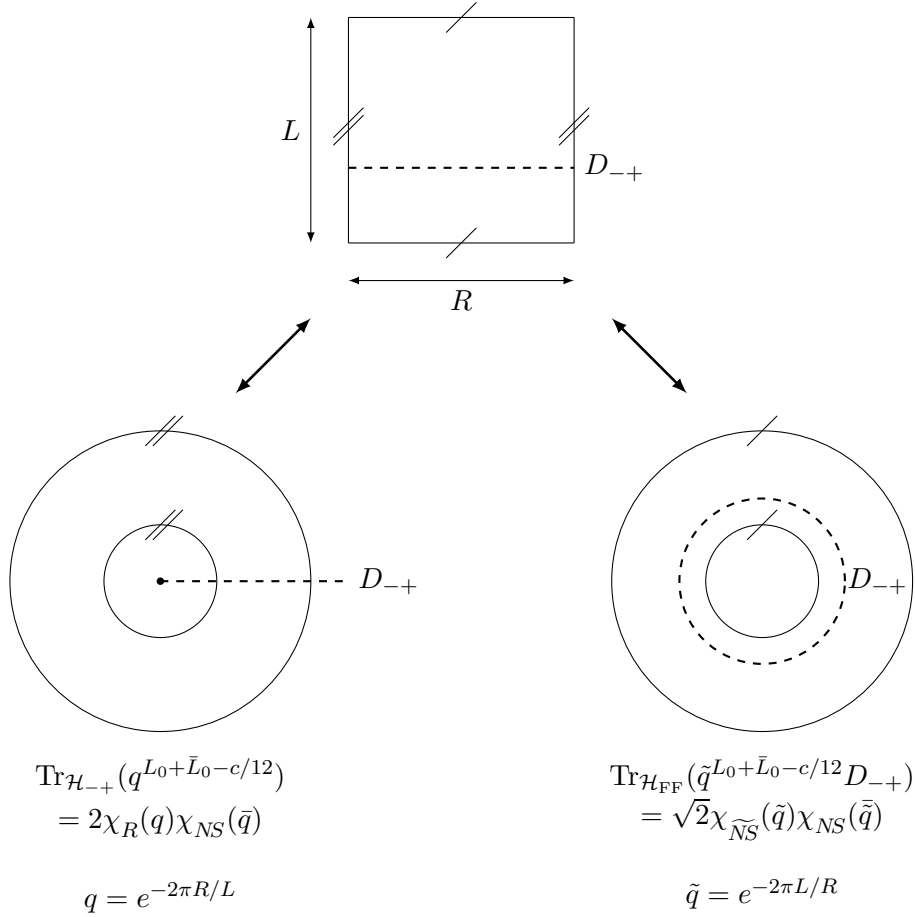}}}
  \end{picture}
$$
}{
\begin{center}
\begin{tikzpicture}
	\pgfmathsetmacro{\hshift}{4}
	\pgfmathsetmacro{\vshift}{6}

	\draw (-1.5,-1.5) -- (1.5,-1.5);
	\draw (1.5,-1.5) -- (1.5,1.5);
	\draw (1.5,1.5) -- (-1.5,1.5);
	\draw (-1.5,1.5) -- (-1.5,-1.5);
	
	\draw (-0.2,-1.7) -- (0.2,-1.3);
	\draw (-0.2,1.3) -- (0.2,1.7);
	\draw (-1.7,-0.1) -- (-1.3,0.3);
	\draw (-1.7,-0.2) -- (-1.3,0.2);
	\draw (1.3,-0.1) -- (1.7,0.3);
	\draw (1.3,-0.2) -- (1.7,0.2);
	
	\draw[latex-latex] (-2,-1.5) -- (-2,1.5);
	\draw[latex-latex] (-1.5,-2) -- (1.5,-2);
	\node[left] at (-2,0) {$L$};
	\node[below] at (0,-2) {$R$};
	
	\draw[dashed,thick] (-1.5,-0.5) -- (1.5,-0.5);
	\node[right] at (1.5,-0.5) {$D_{-+}$};
	
	\draw (0-\hshift,0-\vshift) circle (2);
	\draw (0-\hshift,0-\vshift) circle (0.75);
	\node[circle,fill,inner sep=1pt] at (0-\hshift,0-\vshift) {};
	
	\draw[dashed,thick] (0-\hshift,0-\vshift) -- (2.5-\hshift,0-\vshift);
	\node[right] at (2.5-\hshift,0-\vshift) {$D_{-+}$};
	
	\draw (-0.2-\hshift,0.55-\vshift) -- (0.2-\hshift,0.95-\vshift);
	\draw (-0.1-\hshift,0.55-\vshift) -- (0.3-\hshift,0.95-\vshift);
	\draw (-0.2-\hshift,1.8-\vshift) -- (0.2-\hshift,2.2-\vshift);
	\draw (-0.1-\hshift,1.8-\vshift) -- (0.3-\hshift,2.2-\vshift);
	
	\node[below,align=center] at (0-\hshift,-2.2-\vshift) {$\Tr_{\cH_{-+}}( q^{L_0 + \bar L_0 - c/12} )$ \\ $=2 \chii_\R(q) \chii_{\NS}(\bar q)$ \\ \\ $q=e^{-2\pi R/L}$};
	
	\draw (0+\hshift,0-\vshift) circle (2);
	\draw (0+\hshift,0-\vshift) circle (0.75);
	
	\draw[dashed,thick] (0+\hshift,0-\vshift) circle (1.1);
	\node[right] at (1+\hshift,0-\vshift) {$D_{-+}$};
	
	\draw (-0.2+\hshift,0.55-\vshift) -- (0.2+\hshift,0.95-\vshift);
	\draw (-0.2+\hshift,1.8-\vshift) -- (0.2+\hshift,2.2-\vshift);
	
	\node[below,align=center] at (0+\hshift,-2.2-\vshift) {$\Tr_{\cH_\text{FF}}( \tq^{L_0 + \bar L_0 - c/12} D_{-+} )$ \\ $=\sqrt 2 \chii_{\widetilde{\NS}}(\tq)\chii_{\NS}(\bar{\tilde{q}})$ \\ \\ $\tq=e^{-2\pi L/R}$};
		
	\draw [latex-latex,line width=1pt] (-2,-2.5) -- (-3,-3.5);
	\draw [latex-latex,line width=1pt] (2,-2.5) -- (3,-3.5);
	
\end{tikzpicture}
\end{center}
}

\caption{
The trace over the space of fields on which the $D_{-+}$ defect can end is
related by a modular transformation to the trace with the defect
inserted}

\label{fig:defectop}
\end{figure}

It is, at first sight, surprising that this means that the operator
$(-1)^F$ is not represented by a defect, 
the defect instead being
$D_{-+} = \sqrt 2(-1)^F$. However, this seems necessary for there to
be an integer number of operators that create the $D_{-+}$ defect. The
primary operators that create the defect, i.e. the operators on which the defect
can end, are counted by the partition function on the torus with a
single defect inserted, as shown in figure \ref{fig:defectop}, and the
proposal here ensures that this space is two-dimensional, which is the
smallest dimension possible given that this space has to carry a
representation of the Ramond algebra.

The space of fields on which the $D_{-+}$ defect can end is
$\cH_{-+}$. This is related by a modular S transformation to the
trace over the free fermion space with the insertion of $D_{-+}$ and
so (taking $q$ to be real) 
$
 \Tr_{\cH_{-+}}( q^{L_0 + \bar L_0 - c/12})
= \Tr_{\cH_\text{FF}}( \tq^{L_0 + \bar L_0 - c/12} D_{-+} )
= \sqrt 2 \chii_{\widetilde{\NS}}(\tq)\chii_{\NS}(\tq)
= 2 \chii_\R(q) \chii_{\NS}(q)
\;,
$
i.e. a two dimensional space of primary operators on which the defect
can end.

\subsubsection{Boundary states for the free fermion theory}

We would like to define boundary conditions which preserve the free
fermion algebra, up to automorphism. 
This implies that a boundary
state $\kkett{B}$ must satisfy the condition
\be
  (\psi_m - i \eps \bar\psi_{-m}) \kkett{B}
\;,
\ee
for $\eps = \pm 1$.
The space of Ishibashi states
for a fixed choice of $\epsilon$ is
one-dimensional, that is there are only two Ishibashi states of
interest,
\be
 \kett{N\!S,\eps}
= \prod_{m=0}^\infty e^{i \eps \psi_{-m{-}1/2} \bar\psi_{-m{-}1/2}} \vac
\;.
\ee
The overlaps of these Ishibashi states are
\be
\begin{array}{l}
 \braa{N\!S,\pm } q^{\tfrac{1}{2}(L_0 + \bar L_0 - c/12)}
 \kett{N\!S,\pm} = \chii_{\NS}(q) = \chii_{\NS}(\tq)
\;,\;\;\\
 \braa{N\!S,\pm } q^{\tfrac{1}{2}(L_0 + \bar L_0 - c/12)}
 \kett{N\!S,\mp} = \chii_{\widetilde{\NS}}(q) = \sqrt 2\chii_{\R}(\tq)
\;.
\end{array}
\label{eq:overlaps}
\ee
Note that we use `double kets with single vertical bar' $\kett{i}$ to denote Ishibashi states and `double kets with double vertical bars' $\kkett{B}$ to denote (elementary) boundary states. Since we would like the cylinder partition function of two physical
boundary states to be expressible in the crossed channel as the trace
of the Hamiltonian along a strip, it should be a
non-negative integer combination of the characters
$\chii_{\alpha}(\tq)$. 
It would be nice to be able to choose the boundary states to be  
$\kett{N\!S,+}$ and $\kett{N\!S,-}$, but the mutual overlap is not an
integer multiple of $\chii_\R(\tq)$ and so we have to make a choice
for the boundary states. We can make either of the two choices
\be
1.:\;\;\{ \kkett A = \kett{N\!S,+}
\;,\;\;
 \kkett B = \sqrt 2 \kett{N\!S,-} \}
\;,
\ee\be
2.:\;\;\{ \kkett A = \kett{N\!S,-}
\;,\;\;
 \kkett B = \sqrt 2 \kett{N\!S,+} \}
\;.
\ee
These then have cylinder partition functions
\be
\begin{array}{l}
Z_{AA} =  \bbraa A q^{\tfrac{1}{2}(L_0 + \bar L_0 - c/12)}
 \kkett{A} = \chii_{\NS}(\tq)
\;,\;\;\\
Z_{AB} =  \bbraa A q^{\tfrac{1}{2}(L_0 + \bar L_0 - c/12)}
 \kkett{B} = 2\chii_{\R}(\tq)
\;,\;\;\\
Z_{BB} =  \bbraa B q^{\tfrac{1}{2}(L_0 + \bar L_0 - c/12)}
 \kkett{B} =
 2\chii_{\NS}(\tq)
\;.
\end{array}
\label{eq:overlaps2}
\ee 
Since $\chii_{\NS} = \chii_0 + \chii_{1/2}$, the identity Virasoro
representation appears twice in $Z_{BB}$ and so this is not actually
an elementary defect with respect to the Virasoro algebra. 
The usual conclusion \cite{BBR,N1,A1} would be that we need to
introduce the Ramond sector with Ishibashi states $
\kett{R,\pm}$. This will allow us to write the boundary state $\kkett
B$ as the superposition of two states $\kkett{B_\pm}$, which take the
form 
\bea
1.:\;\;&& \kkett{B_\pm} =
\frac{1}{\sqrt 2} \kett{N\!S,-} \pm \frac{1}{2^{1/4}} \kett{R,-} \;,\;\;
\\
2.:\;\;&& \kkett{B_\pm} = \frac{1}{\sqrt 2} \kett{N\!S,+} \pm
\frac{1}{2^{1/4}} \kett{R,+} \;.\;\;
\eea
These, however, do not have
cylinder partition functions that can be interpreted as the trace over
a representation of the fermion algebra. 
They can be
considered as sums over different spin structures for the fermion, or
as partition functions for the Ising model obtained from the free
fermion by projecting onto even fermion number states.
The Ising model has three fundamental boundary conditions,
usually denoted $(-)$, $(f)$ and $(+)$ for Ising spin fixed down,
free, or fixed up and their boundary states can be identified as
\be
\kkett{f} = \kkett{A}
\;,\;\;
\kkett{\pm} = \kkett{B_\pm}
\;,
\ee
and the cylinder partition functions are traces over representations
of the Virasoro algebra,
\be
\begin{array}{l}
   \bbraa A q^{\tfrac{1}{2}(L_0 + \bar L_0 - c/12)}
 \kkett{B_\pm} = \chii_{\R}(\tq) = \chiim_{1/16}(\tq)\;,
\\
   \bbraa{ B_\pm } q^{\tfrac{1}{2}(L_0 + \bar L_0 - c/12)}
 \kkett{B_\pm} = \frac{1}{2} \big( \chii_{\NS}(\tq) +
 \chii_{\widetilde{\NS}}(\tq) \big)
= \chiim_0(\tq)
\\
   \bbraa { B_\pm } q^{\tfrac{1}{2}(L_0 + \bar L_0 - c/12)}
 \kkett{B_\mp} = \frac{1}{2} \big( \chii_{\NS}(\tq) -
 \chii_{\widetilde{\NS}}(\tq) \big)
= \chiim_{1/2}(\tq)
\end{array}
\ee
There is, however, no requirement for us to extend the space of
boundary conditions in this way: it is perfectly consistent to ask
that the fermion boundary condition be defined for a single choice of
spin structure and we will work with the boundary
conditions $\kkett A$ and $\kkett B$.

This agrees with the analysis in 
\cite{GZ,ZC1}
in which the free fermion is stated to have two boundary conditions,
``free'' and ``fixed''\footnote{Note that this is not the same as the usual naming
conventions for the Ising model, for which ``free'' and ``fixed''
would have one and two ground states 
respectively, but since these are interchanged by duality
\cite{ortiz}, this should not be of concern.
}, distinguished by having two and one ground
states respectively, with the ``fixed'' boundary condition having a
two-dimensional space of weight zero fields, one bosonic and one
fermionic. 
This fermionic weight zero field also arises in the Ising model, 
where the ``$\epsilon$'' defect can end at the junction of a $(+)$ and
and a $(-)$ boundary condition in a field of weight zero.
Such a space with one bosonic and one fermionic degree of freedom,
$\mathbb{C}^{1|1}$, also arises when constructing spin fields in the Ising
model in \cite{NR}, and we think it is probable that a treatment of boundary
conditions in the manner of \cite{NR} will give our result in a
rigorous manner.

\subsection{Interfaces between the Ising model and the free fermion}

We would now like to consider the case of topological interfaces
between the Ising model and the Neveu-Schwarz free fermion.
Consider an operator $I$ from the Hilbert space of the free fermion to
the Hilbert space of the Ising model; the operator $I^\dagger$ will
map from the Ising model to the free fermion model. The topological
conditions,
\be
 L_m \: I = I \: L_m
\;,\;\;
 \bar L_m \: I = I \: \bar L_m
\;,\;\;
\label{eq:topI}
\ee
mean that $I$ must be a sum of projectors on representations of the
Virasoro algebra. Since the Hilbert spaces of the two theories are
\be
\begin{array}{rcl}
  \cH_\text{Ising} 
&=&      (\cH_0\otimes \bar\cH_0)\;
\oplus \;(\cH_{1/2} \otimes \bar\cH_{1/2} )\;
\oplus \;(\cH_{1/16}\otimes \bar\cH_{1/16})
\;,\;\;
\\  \cH_\text{FF} 
&=&      (\cH_0\oplus \cH_{1/2})\otimes 
       (\bar\cH_0 \oplus \bar\cH_{1/2})
\;,\;\;
\end{array}
\ee
we see that the operator $I$ is determined up to two constants, $a$
and $b$, as
\be
 I_{a,b} = a P_0\bar P_0 + b P_{1/2}\bar P_{1/2}
\;,
\ee
where $P_0\bar P_0$ is the projector onto $\cH_0\otimes\bar\cH_0$ and 
$P_{1/2}\bar P_{1/2}$ is the projector onto $\cH_{1/2}\otimes\bar\cH_{1/2}$.

As before, we would like the coefficients $M_{A\beta}^\gamma$ and
$M_{\beta A}^\gamma$ in the algebra of defect operators 
\be
 D_A\: I_\beta = \sum_\gamma\, M_{A\beta}^\gamma \: I_\gamma
\;,\;\;
 I_\beta \: D_A  = \sum_\gamma\, M_{\beta A}^\gamma \: I_\gamma
\;,
\ee
to be
non-negative integer coefficients. The known Ising defects
\cite{OA,PZ} and free 
fermion defects can be expressed in terms of the projectors $P_h
\bar{P}_{\bar h}$ as 
\be
\begin{array}{rll}
\hbox{Ising}:&
 D_0 &= P_0\bar P_0 + P_{1/2}\bar P_{1/2} + P_{1/16}\bar P_{1/16}
\;,
\\
& D_{1/2} &= P_0\bar P_0 + P_{1/2}\bar P_{1/2} - P_{1/16}\bar P_{1/16}
\;,
\\
& D_{1/16} &= \sqrt{2} ( P_0\bar P_0 - P_{1/2}\bar P_{1/2} )
\;,
\\[3mm]
\hbox{Free fermion}:&
 D_{++} &= (P_0 + P_{1/2})(\bar P_0 + \bar P_{1/2})
\;,
\\
& D_{+-} &= \sqrt 2 (P_0 + P_{1/2})(\bar P_0 - \bar P_{1/2})
\;,
\\
& D_{-+} &= \sqrt 2(P_0 - P_{1/2})(\bar P_0 + \bar P_{1/2})
\;,
\\
& D_{++} &= (P_0 - P_{1/2})(\bar P_0 - \bar P_{1/2})
\;.
\\
\end{array}
\ee 
This gives the algebra
\be
D_0 \: I_{a,b} = I_{a,b}
\;,\;\;
D_{1/2} \: I_{a,b} = I_{a,b}
\;,\;\;
D_{1/16} \: I_{a,b} = \sqrt 2 I_{a,-b}
\;,\;\;
\ee
\be
I_{a,b} \: D_{++} = I_{a,b}
\;,\;\;
I_{a,b} \: D_{+-} = \sqrt 2I_{a,-b}
\;,\;\;
I_{a,b} \: D_{-+} = \sqrt 2I_{a,-b}
\;,\;\;
I_{a,b} \: D_{--} = I_{a,b}
\;.\;\;
\ee
We would also like the action of the interfaces on boundary states to
give integer combinations of boundary states in the other model, that
is the coefficients $M_{\alpha A}^B$ and $\tilde M_{\alpha A}^B$ in the
algebra 
\be
  I_\alpha \: \kkett{B^\text{FF}_A} 
= \sum_B M_{\alpha A}^B \: \kkett{B^\text{Ising}_B}
\;,\;\;
 I^\dagger _\alpha \: \kkett{B^\text{Ising}_A} 
= \sum_B \tilde M_{\alpha A}^B \, \kkett{B^\text{FF}_B}
\;,
\ee
should be non-negative integers.
If $I_{a,b}$ is an interface that acts from the free-fermion space
to the Ising model and $I^\dagger_{a,b}$ acts in the opposite
direction, we have
\be
\begin{array}{l}
I_{a,b} \: \kkett A 
= \tfrac{a+b}{2\sqrt 2} \big( \kkett{B_0} + \kkett{B_{1/2}} \big)
+ \tfrac{a-b}{2} \kkett{B_{1/16}}
\;,
\\
I_{a,b} \: \kkett B 
= \tfrac{a-b}{2} \big( \kkett{B_0} + \kkett{B_{1/2}} \big)
+ \tfrac{a+b}{\sqrt 2} \kkett{B_{1/16}}
\;,
\\
I_{a,b}^\dagger \: \kkett{B_0}
= \tfrac{a+b}{2\sqrt 2} \kkett{A}
+ \tfrac{a-b}{4} \kkett{B}
\;,
\\
I_{a,b}^\dagger \: \kkett{B_{1/2}}
= \tfrac{a+b}{2\sqrt 2} \kkett{A}
+ \tfrac{a-b}{4} \kkett{B}
\;,
\\
I_{a,b}^\dagger \: \kkett{B_{1/16}}
= \tfrac{a-b}{2} \kkett{A}
+ \tfrac{a+b}{2\sqrt 2} \kkett{B}
\;.
\end{array}
\ee
The general solution to the integrality conditions is $a = \sqrt 2 m + 2 n$, 
$b =\sqrt 2 m - 2 n$, for integers $m$ and $n$.
This suggests that there are two elementary interfaces, $I = I_{\sqrt
  2,\sqrt 2}$ and $I' = I_{2,-2}$ satisfying
\be
\begin{array}{ll}
  I \: \kkett A = \kkett{B_0} + \kkett{B_{1/2}} \;,
&  I^\dagger \: \kkett{B_0} = I^\dagger \: \kkett{B_{1/2}} = \kkett{A} \;,\\
  I \: \kkett B = 2 \kkett{B_{1/16}} \;,
&  I^\dagger \: \kkett{B_{1/16}} = \kkett{B} \;,\\
  I' \: \kkett A = 2 \kkett{B_{1/16}} \;,
&  I'^\dagger \: \kkett{B_0} = I'^\dagger \: \kkett{B_{1/2}} = \kkett{B} \;,\\
  I' \: \kkett B = 2 \big( \kkett{B_0} + \kkett{B_{1/2}} \big) \;,
&  I'^\dagger \: \kkett{B_{1/16}} = 2\kkett{A} \;,\\
\end{array}
\ee
and that every other interface is
formed by a linear combination of these two.

Finally, we would like the product of interfaces to be expressible as
a non-negative integer combination of topological defects, that is 
the constants $M_{\alpha\beta}^A$ and ${\tilde M}_{\alpha\beta}^A$ in 
\be
 I_\alpha \: I^\dagger_\beta = \sum_A M_{\alpha\beta}^A \: D^\text{Ising}_A
\;,\;\;
 I^\dagger_\alpha \: I_\beta = \sum_A {\tilde M}_{\alpha\beta}^A \: D^\text{FF}_A
\;,
\ee
should also be non-negative integers.
We have
\be
 I_{a,b} \: I^{\dagger}_{c,d} = 
\tfrac{ac + bd}{4} \big( D_0 + D_{1/2} \big) + \tfrac{ac - bd}{2\sqrt 2}D_{1/16}
\;,\;\;
 I^{\dagger}_{c,d} \: I_{a,b} = 
\tfrac{ac + bd}{4} \big( D_{++} + D_{--} \big) + \tfrac{ac - bd}{4\sqrt 2}
\big( D_{+-} + D_{-+} \big)
\;,\;\;
\ee
We see that two elementary interfaces $I$ and $I'$
do lead to an algebra with integer coefficients:
\be
\begin{array}{rclrcl}
&&I \: I^\dagger = D_0 + D_{1/2}
\;,\;\;
I \: I'^\dagger = 2 D_{1/16}
\;,\;\;
I^\dagger \: I = D_{++} + D_{--}
\;,\;\;
I'^\dagger \: I = D_{+-} + D_{-+}
\;,\;\;
\\
&&I \: D_{++} = I \: D_{--} = I
\;,\;\;
I \: D_{+-} = I \: D_{-+} = I'
\;,\;\;
D_0 \: I = D_{1/2} \: I = I
\;,\;\;
D_{1/16} \: I = I'
\;.
\end{array}
\ee


\subsection{Consistency tests}

\begin{figure}[htb]

\pictures{
$$  
  \begin{picture}(440,290)
  \put(-20,0){\scalebox{.9}{\includegraphics{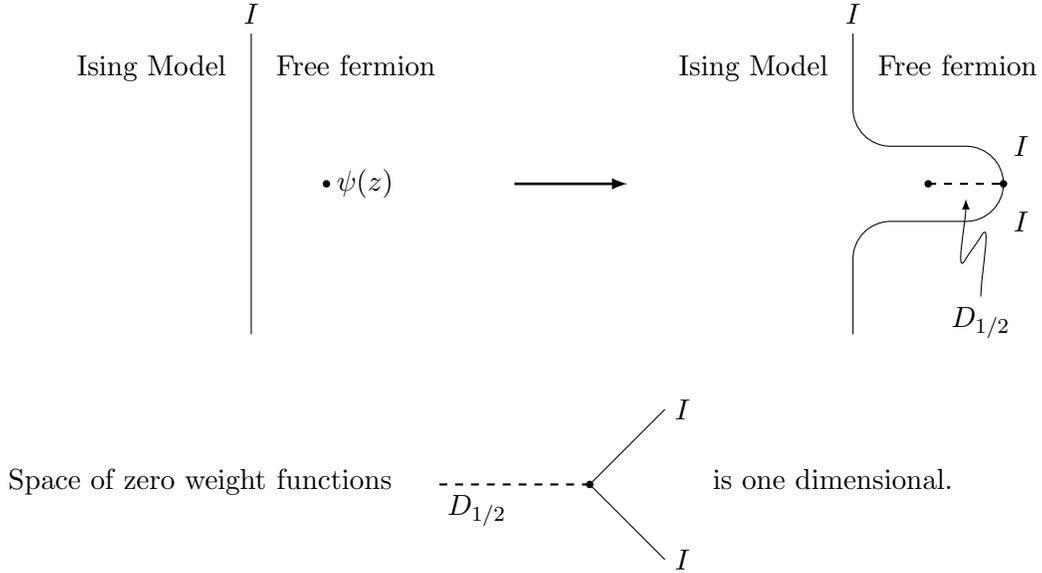}}}
  \end{picture}
$$
}{
\begin{center}
\begin{tikzpicture}
	\pgfmathsetmacro{\hshift}{4}
	\pgfmathsetmacro{\vshift}{2}

	\draw (2-\hshift,0) -- (2-\hshift,4);
	\node[left,text height=1ex,text depth=.25ex] at (1.8-\hshift,3.5) {Ising Model};
	\node[right,text height=1ex,text depth=.25ex] at (2.2-\hshift,3.5) {Free fermion};
	\node[above] at (2-\hshift,4) {$I$};
	
	\node[circle,fill,inner sep=1pt] at (3-\hshift,2) {};
	\node[right] at (3-\hshift,2) {$\psi (z)$};
	
	\draw (2+\hshift,0) -- (2+\hshift,1) to[out=90,in=180] (2.5+\hshift,1.5) -- (3.5+\hshift,1.5) to[out=0,in=-90] (4+\hshift,2) to[out=90,in=0] (3.5+\hshift,2.5) -- (2.5+\hshift,2.5) to[out=180,in=-90] (2+\hshift,3) -- (2+\hshift,4);
	\draw[dashed,thick] (3+\hshift,2) -- (4+\hshift,2);
	\node[right] at (4+\hshift,2.5) {$I$};
	\node[right] at (4+\hshift,1.5) {$I$};
	
	\node[circle,fill,inner sep=1pt] at (4+\hshift,2) {};
	
	\draw[latex-] (3.5+\hshift,1.8) to[out=-90,in=-120] (3.5+\hshift,1) to[out=60,in=-120] (3.7+\hshift,1.3) to[out=60,in=90] (3.7+\hshift,0.5);
	\node[below] at (3.7+\hshift,0.5) {$D_{1/2}$};
	
	\node[left,text height=1ex,text depth=.25ex] at (1.8+\hshift,3.5) {Ising Model};
	\node[right,text height=1ex,text depth=.25ex] at (2.2+\hshift,3.5) {Free fermion};
	\node[above] at (2+\hshift,4) {$I$};
	
	\node[circle,fill,inner sep=1pt] at (3+\hshift,2) {};
	
	\draw [-latex,line width=1pt] (5.5-\hshift,2) -- (-1+\hshift,2);
	
	\node[left,text height=1ex,text depth=.25ex] at (0,0-\vshift) {Space of zero weight functions};
	
	\draw[dashed,thick] (0.5,0-\vshift) -- (2.5,0-\vshift);
	\draw (2.5,0-\vshift) -- (3.5,1-\vshift);
	\draw (2.5,0-\vshift) -- (3.5,-1-\vshift);
	\node[circle,fill,inner sep=1pt] at (2.5,0-\vshift) {};
	
	\node[below] at (1,0-\vshift) {$D_{1/2}$};
	\node[right] at (3.5,1-\vshift) {$I$};
	\node[right] at (3.5,-1-\vshift) {$I$};
	
	\node[right,text height=1ex,text depth=.25ex] at (4,0-\vshift) {is one dimensional.};
\end{tikzpicture}
\end{center}
}

\caption{
The $\psi$ field in the free-fermion model is realised as a defect-creation operator in the Ising model. The associated space of three-defect junctions is one-dimensional}

\label{fig:fermionpull}
\end{figure}

The main object of this section was to construct topological interface operators
between the Ising model and the free-fermion model (as a warm-up for
the tri-critical Ising case). Since these
interfaces are topological, one should be able to move the interfaces
past field insertions without changing their conformal properties. As an
example, we can ask whether we can pull the interface past the free
fermion field, so that the free fermion field's insertion point is now
on the Ising model
side of the interface. Since the free fermion $\psi(z)$ is not a local field in the
Ising model, it must arise as a defect creation operator, that is as
the termination point of a defect, which is the defect $D_{1/2}$. This
is shown in figure \ref{fig:fermionpull}. This is only possible if there is
a one-dimensional space of zero-weight interface-interface-defect
junctions. These are counted by the partition function with insertions
of $I$, $D_{1/2}$ and $I^\dagger$. 
We can calculate this using $ I \: I^\dagger \: D_{1/2} = (D_0
+ D_{1/2}) \: D_{1/2} = (D_0 + D_{1/2})$, 
\bea
 \Tr_{\cH_{IID_{1/2}}}(q^{L_0 -c/24}\bar q^{\bar L_0 - c/24})
&=&
  \Tr_{\cH_\text{FF}}(I^\dagger D_{1/2} I \;q^{L_0 -c/24}\bar q^{\bar L_0 - c/24})
\nonumber \\
&=&
  \Tr_{\cH_\text{Ising}}(I I^\dagger D_{1/2} \; q^{L_0 -c/24}\bar q^{\bar L_0 - c/24})
\nonumber \\
&=& 
\Tr_{\cH_\text{Ising}}(D_0\, q^{L_0 -c/24}\bar q^{\bar L_0 - c/24})
+ 
\Tr_{\cH_\text{Ising}}(D_{1/2}\, q^{L_0 -c/24}\bar q^{\bar L_0 - c/24})
\nonumber \\
&=& 
\left( |\chiim_0(q)|^2 + |\chiim_{1/2}(q)|^2 + |\chiim_{1/16}(q)|^2\right) \nonumber \\
&& \hspace{3cm} +
\left( |\chiim_0(q)|^2 + |\chiim_{1/2}(q)|^2 - |\chiim_{1/16}(q)|^2\right)
\nonumber \\
&=& 2 |\chiim_0(q)|^2 + 2|\chiim_{1/2}( q)|^2
\nonumber \\
&=& |\chiim_0(\tq) + \chiim_{1/2}(\tq)|^2 + 2
|\chiim_{1/16}(\tq)|^2
\;.
\eea
The coefficient of $|\chiim_0(\tq)|^2$ counts the dimension of
weight zero junction fields so this is indeed
one-dimensional.

As a second example, consider a defect $D_{-+}$ terminating in the FF
model on fields of conformal weights $(1/16,0)$. Since the partition
function with $D_{-+}$ is given from \eqref{eq:dtraces} and \eqref{eq:alphasol} as
\be
 \Tr_{\cH_\text{FF}}( D_{-+} q^{L_0 - c/24}\bar q^{\bar L_0 - c/24})
= 2 \chii_\R(\tq) \chii_{\NS}(\bar{\tq})
\;,
\ee
the space of such fields is two dimensional. What happens when the
interface is pulled past this defect-terminating field?
The only way a field of weights $(1/16,0)$ can arise in the Ising
model is as a field on the end of a $D_{1/16}$ defect, and the 
space of such fields is only one dimensional. The resolution
is that the space of zero-weight fields at the point where the
$D_{-+}$ defect crosses the interface $I$ to become the $D_{1/16}$
defect is two-dimensional, as calculated by
\bea
 \Tr_{\cH_{I D_{-+} I D_{1/16}}}
(q^{L_0 -c/24}\bar q^{\bar L_0 - c/24})
&=&
 \Tr_{\cH_\text{Ising}}( I D_{-+} I^\dagger D_{1/16}\,
q^{L_0 -c/24}\bar q^{\bar L_0 - c/24})
\nonumber \\
&=&
 \Tr_{\cH_\text{Ising}} \big( (2 D_0 + 2 D_{1/2})
q^{L_0 -c/24}\bar q^{\bar L_0 - c/24} \big)
\nonumber \\
&=& 2 | \chii_0(\tq) | ^2 + \ldots
\;.
\eea
In every case we have considered, the result of pulling local fields
through the interface result, in a similar fashion, in consistent
interpretations in terms of the space of interface-defect junctions.
 
We now turn to the main objects of interest, the tri-critical Ising
model and the related supersymmetric theory.  
 

\section{The tri-critical Ising model  and the fermionic model at $c=7/10$}
\label{sec:svir}

The tri-critical Ising model (TCIM for short) is 
the unitary
minimal model of the Virasoro algebra with $c=7/10$ with diagonal
modular invariant and all the local fields have 
integer spin.
The Virasoro algebra with $c=7/10$ has six unitary highest weight
representations, and correspondingly the TCIM has 
six primary fields, six
elementary topological defects and six elementary conformal boundary
conditions, each labelled by these representations.

The TCIM is the unique local field theory with $c=7/10$ with a modular
invariant partition function but we can define a related model with local
fields with integer and half-integer spins which is well-defined on
the cylinder, in the same way the FF model is related to the Ising
model. This fermionic model has superconformal symmetry and we will
denote it by \sTCIM. 
There are four irreducible unitary highest-weight representations of
the super Virasoro algebra with $c=7/10$ of which two are
Neveu-Schwarz and two are Ramond. The local field theory \sTCIM\ is
the diagonal theory formed from the Neveu-Schwarz representations.

Our aim is to find interfaces between the TCIM and \sTCIM\ theories
and use these to construct defects in TCIM from defects in
\sTCIM. Before we do that we will first have to construct defects in
\sTCIM\ and we will do this by identifying them with boundary
conditions on the doubled model, \SVIR2. This is the main technical
challenge of the paper.
Before we come to this point, we first introduce some notation and
recall some facts about the TCIM and \sTCIM\ models. 
(We list
our conventions for the representations, their labels, characters,
fusion rules, etc, in appendix \ref{app:tcim}.)

\subsection{TCIM}

The tri-critical Ising model is the unitary, diagonal modular
invariant theory at $c=7/10$ and is the $m=4$ member of the unitary
minimal series.
The Virasoro algebra has six unitary highest weight representations at
$c=7/10$ with characters $\chi^{(4)}_h$; associated to each of these
there is a topological defect 
$D_{\hat\eps}\equiv 
D_{1/10} \equiv D_{(1,2)}$, etc and a 
conformal boundary condition $\kett{\hat\eps}$ etc.
We list these together with their defect and and boundary entropies in
table \ref{tab:tcim}. 

\begin{table}[htb]
\[
\begin{array}{r|cccccc}
\hline
\hbox{name} &
I & \eps & \sigma & \hat I & \hat \eps & \hat \sigma
\\
\hline
\hbox{Kac labels $(r,s)$} &
(1,1) & (3,1) & (2,1) & (1,3) & (1,2) & (2,2) \\
\hline
\hbox{weight } h &
0 & 3/2 & 7/16 & 3/5 & 1/10 & 3/80 
\\ \hline
\hbox{Entropy of defect $D_h$} &
1 & 1 & \sqrt 2 & \frac{1 + \sqrt 5}{2} & \frac{1 + \sqrt 5}{2} & \frac{1 + \sqrt 5}{\sqrt 2} \\ \hline
\hbox{Entropy of boundary $B_h$} &
\left(\frac{5 - \sqrt 5}{40}\right)^{\frac 14} & 
\left(\frac{5 - \sqrt 5}{40}\right)^{\frac 14} & 
\left(\frac{5 - \sqrt 5}{10}\right)^{\frac 14} & 
\left(\frac{5 + 2 \sqrt 5}{20}\right)^{\frac 14} & 
\left(\frac{5 + 2 \sqrt 5}{20}\right)^{\frac 14} & 
\left(\frac{5 + 2 \sqrt 5}{5}\right)^{\frac 14} 
 \\ \hline
\end{array}
\]
\caption{TCIM data}
\label{tab:tcim}
\end{table}

The shorthand names are chosen to make clear the fusion rules, which
are (Lee-Yang)$\times$(Ising). The fields $I$, $\eps$ and $\sigma$
have Ising fusion rules, and $I$ and $\hat I$ have Lee-Yang rules,
that is
\be
 \hat I \star \hat I = I + \hat I
\;,
\ee
and in general
\be
 X \star \hat Y = \widehat{( X \star Y)}
\;,\;\;
 \hat X \star \hat Y = (X \star Y) +  \widehat{(X \star Y)}
\;.
\ee

\subsection{\sTCIM}

The \sTCIM\ model has local fields $G(z)$ and $\bar G(\bar z)$ of
conformal weights $(3/2,0)$ and $(0,3/2)$ respectively, which are
fermionic and generate two copies of the $c=7/10$ superconformal
algebra. This is the $m=3$ member of the superconformal unitary
minimal series for which $c(m) = 3/2(1 - 8/(m(m+2)))$.
There are two Neveu-Schwarz representations and two Ramond
representations, which we label as in table \ref{tab:stcim}.

\begin{table}[htb]
\[
\begin{array}{r|cccc}
\hline
\hbox{name} &
I_{\NS} & \varphi_\R & \varphi_{\NS} & I_\R \\
\hline
\hbox{Kac labels $(r.s)$} &
(1,1)\equiv(2,4) & (1,2)\equiv(2,3) & (1,3)\equiv(2,2) & (1,4)\equiv(2,1) \\ 
\hbox{weight $h$} &
0 & 3/80 & 1/10 & 7/16 \\
\hline
\end{array}
\]
\caption{\sTCIM\ data}
\label{tab:stcim}
\end{table}

The fusion rules are (Lee-Yang)$\times$(Free-fermion).
The local field theory \sTCIM\ consists of the Neveu-Schwarz fields and,
in terms of representations of the Virasoro algebra,
has Hilbert space
\be
\cH_{\sTCIM} = 
(\cH_0 \oplus \cH_{3/2})\otimes(\bar\cH_0 \oplus \bar\cH_{3/2})
\;\;\oplus\;\;
(\cH_{1/10} \oplus \cH_{3/5})\otimes(\bar\cH_{1/10} \oplus
\bar\cH_{3/5})
\;.
\ee
The partition function of the Neveu-Schwarz sector of \sTCIM\ model 
can be expressed in terms of the characters $\chthree_h$ of
the superconformal algebra and $\chi^{(4)}_h$ of the Virasoro algebra,
\bea
\Tr_{\cH_{\sTCIM}}( q^{L_0 - c/24}\bar q^{\bar L_0 - c/24})
&=& | \chthree_0(q)|^2 +   | \chthree_{1/10}(q) |^2
  \nn\\
  &=& | \chi^{(4)}_0(q) + \chi^{(4)}_{3/2}(q) |^2
  + | \chi^{(4)}_{1/10}(q) + \chi^{(4)}_{3/5}(q) |^2
\;.
\eea
It is not fully
modular invariant, but it is invariant under $\tau \to -1/\tau$ and
$\tau \to \tau +2$.

The operators $(-1)^F$ and $(-1)^{\bar F}$ each have eigenvalue $+1$
on the state $\ket 0$ and value $-1$ on the state $\ket{1/10}$. The
field $\phi_{1/10}$ is still a bosonic field, though, as $(-1)^{F+\bar F}$ 
has eigenvalue $+1$.

The Ramond representations of the $N=1$ superconformal algebra do not
correspond to local fields in the \sTCIM\ theory - as fields, they can
only arise as defect-creation fields (that is fields on which defect
lines end) or 
fields at which two or more defects join.

\subsubsection{Superconformal defects in \sTCIM}

We will consider superconformal topological defects in \sTCIM, that is
defects which preserve the $N=1$ algebra up to automorphism and
so satisfy
\be
  L_m \: D = D \: L_m
\;,\;\;
  G_m \: D = \eps D \: G_m
\;,\;\;
  \bar L_m \: D = D \: \bar L_m
\;,\;\;
  \bar G_m \: D = \eps' D \: \bar G_m
\;,
\label{eq:sd}
\ee
where $\eps$ and $\eps'$ are $\pm 1$.
A defect satisfying \eqref{eq:sd} is determined up to two constants as
\be
 D_{\eps\eps'}
= a (P_0 + \eps P_{3/2})(\bar P_0 + \eps' \bar P_{3/2})
+ b (P_{1/10} + \eps P_{3/5})(\bar P_{1/10} + \eps' \bar P_{3/5})
\;,
\ee
where 
$P_i\bar P_j$ is the projector onto the Virasoro representations
$\cH_i\otimes\bar\cH_j$. 

We have found a complete set of eight elementary defects satisfying these
conditions, namely
\be
\{
  D_I , \:
  D_\varphi, \:
 \sqrt 2(-1)^F D_I, \:
 \sqrt 2(-1)^F D_\varphi, \:
 \sqrt 2(-1)^{\bar F} D_I, \:
 \sqrt 2(-1)^{\bar F} D_\varphi, \:
 (-1)^{F+\bar F} D_I, \:
 (-1)^{F+\bar F} D_\varphi
\}
\;,
\ee
where the two fundamental defects with $\eps=\eps'=1$ are $D_I =
\One$ (the identity operator on $\cH_{\sTCIM}$) and
$D_\varphi$, given by
\be
\begin{array}{rcl}
 D_I 
&=&  (P_0 + P_{3/2})(\bar P_0 + \bar P_{3/2})
   + (P_{1/10} + P_{3/5})(\bar P_{1/10} + \bar P_{3/5})
\;,\;\;
\\
 D_\varphi 
&=& \tfrac{1+\sqrt 5}{2} (P_0 + P_{3/2})(\bar P_0 + \bar P_{3/2})
   +  \tfrac{1-\sqrt 5}{2} (P_{1/10} + P_{3/5})(\bar P_{1/10} + \bar P_{3/5})
\;.
\end{array}
\ee
These defects have a product algebra with non-negative integer
structure constants, 
the defect product algebra being determined by
\be
 D_\varphi \: D_\varphi \;=\; D_I + D_\varphi
\;.
\ee
Further, the trace over the Hilbert space with any
number of defects
inserted can be expressed as an integer combination of characters of
the superconformal algebra in the dual channel.
All such traces are determined by the elementary traces
\bea
 \Tr( q^H \bar q^{\bar H}) &=& |\chthree_{1,1}(\tq)|^2 + |\chthree_{1,3}(\tq)|^2 
\;,
\\
 \Tr( \sqrt 2 (-1)^F q^H \bar q^{\bar H} ) &=& 2 \left[
  \chthree_{1,2}(\tq)\chthree_{1,3}(\bar \tq) +
  \chthree_{1,4}(\tq)\chthree_{1,1}(\bar \tq)
                                 \right]
\;,
\\
 \Tr( (-1)^{F + \bar F} q^H \bar q^{\bar H} ) &=& 2     \left[ 
|\chthree_{1,2}(\tq)|^2 + |\chthree_{1,4}(\tq)|^2  \right]
\;,
\\
 \Tr( D_\varphi (-1)^F q^H \bar q^{\bar H} ) &=&
 |\chthree_{1,3}(\tq)|^2 
+ \chthree_{1,3}(\tq) \chthree_{1,1}(\bar \tq)
+ \chthree_{1,1}(\tq) \chthree_{1,3}(\bar \tq)
\;,
\\
 \Tr( \sqrt 2 (-1)^F D_\varphi \, q^H \bar q^{\bar H} ) &=&
  2 \left[
  \chthree_{1,2}(\tq) (\chthree_{1,1}(\bar \tq) + \chthree_{1,3}(\bar \tq))
+
  \chthree_{1,4}(\tq)\chthree_{1,3}(\bar \tq) \right] \; ,
\\
 \Tr( (-1)^{F+\bar F} D_\varphi \, q^H \bar q^{\bar H} ) &=& 
 2 \left[ |\chthree_{1,2}(\tq)|^2 
+ \chthree_{1,2}(\tq) \chthree_{1,4}(\bar \tq)
+ \chthree_{1,4}(\tq) \chthree_{1,2}(\bar \tq) \right]
\;,
\label{eq:dtraces2}
\eea
where $H= L_0 - \frac 7{240}\;,\;\; \bar H= \bar L_0 -  \frac 7{240}$.

\subsubsection{Boundary states for \sTCIM}

As with the free fermion, we shall consider boundary states with
components in the Neveu-Schwarz sector only. Some of these will not be
elementary with respect to the Virasoro algebra, with the
introduction of the Ramond sector allowing this degeneracy to be
lifted, but we do not need this sector to construct non-conformal
defects in TCIM.

There are two gluing conditions for the superconformal algebra, $\pm$,
which we take to be 
\be
 (G_m + i \eps \bar{G}_{-m}) \kett{h,\eps} = 0
\;.
\ee
Given that there are two Neveu-Schwarz representations, there are then
four Ishibashi states,
\be
\begin{array}{rcl}
 \ket{0,\pm} 
&=& \ket 0 \mp \frac{i}{2c/3}G_{-3/2}\bar G_{-3/2}\ket 0 +
 \frac{1}{c/2} L_{-2} \bar L_{-2}\ket 0 + \ldots
\\
 \ket{\tfrac 1{10},\pm} 
&=& \ket{\tfrac 1{10}} \mp \frac{i}{1/5}G_{-1/2}\bar
G_{-1/2}\ket{\tfrac 1{10}} +
 \frac{1}{1/5} L_{-1} \bar L_{-1}\ket {\tfrac{1}{10}} + \ldots
\end{array}
\label{eq:expansions}
\ee
Their normalisation is given by
\bea
 \braa{h,\pm} q^{\tfrac{1}{2}(L_0 + \bar L_0 - c/12)} \kett{h',\pm} &=& \delta_{h,h'} \: \chthree_h (q) \;, \\
 \braa{h,\pm} (-1)^F q^{\tfrac{1}{2}(L_0 + \bar L_0 - c/12)} \kett{h',\pm} &=& \delta_{h,h'} \: \widetilde{\ch}^{3}_h (q) \;.
\eea
In addition, the fermion parity operators act on the Ishibashi states as
\be
 (-1)^F \kett{h,\pm} = \veps(h)\kett{h,\mp} \;, \qquad
 (-1)^{\bar F} \kett{h,\pm} = \veps(h) \kett{h,\mp} \;, \qquad
  (-1)^{F+\bar F} \kett{h,\pm} = \kett{h,\pm} \;,
\ee
where $\veps(h) = \pm 1$ is the fermion parity of the highest weight
state $\vec{h}$. For \sTCIM\ we take $\veps(0) = 1$ and $\veps(1/10) =
-1$. 

We find that there are four consistent fundamental boundary
conditions, which we call 
$\kkett{I_{\NS}}$, $\kkett{\varphi_{\NS}}$, $\kkett{I_\R}$, and
$\kkett{\varphi_\R}$.  
There are, again, two choices for the way to construct these boundary
states, corresponding to the two gluing conditions. This means that
one consistent choice is
\be
\begin{array}{rcl}
\kkett{I_{\NS}} &=& 
\left( \frac{5 - \sqrt 5}{10} \right)^{\frac 14} \kett{0,+} +
\left( \frac{5 + \sqrt 5}{10} \right)^{\frac 14} \kett{\tfrac{1}{10},+} 
\;,
\\
 \kkett{\varphi_{\NS}} = D_\varphi \kkett{I_{NS}}
&=& 
\left( \frac{\sqrt 5 + 2}{\sqrt 5} \right)^{\frac 14} \kett{0,+} -
\left( \frac{\sqrt 5 - 2}{\sqrt 5} \right)^{\frac 14} \kett{\tfrac{1}{10},+} 
\;,
\\
 \kkett{I_{\R}} = \sqrt{2}(-1)^F \kkett{ I_{NS}}
&=& 
\left( \frac{2(5 - \sqrt 5)}{5} \right)^{\frac 14} \kett{0,-} -
\left( \frac{2(5 + \sqrt 5)}{5} \right)^{\frac 14} \kett{\tfrac{1}{10},-} 
\;,
\\
 \kkett{\varphi_{\R}} = \sqrt{2}(-1)^F \kkett{ \varphi_{NS}}
&=&
\left( \frac{4(\sqrt 5 + 2)}{\sqrt 5} \right)^{\frac 14} \kett{0,-} +
\left( \frac{4(\sqrt 5 - 2)}{\sqrt 5} \right)^{\frac 14} \kett{\tfrac{1}{10},-} 
\;.
\end{array}
\label{eq:ketdef}
\ee
and the other is given by using the Ishibashi states of the opposite
gluing condition.

These have overlaps/cylinder partition functions as follows:
\be
\begin{array}{r|llll}
 & \kett{I_{NS}} & \multicolumn{1}{c}{\kett{\varphi_{NS}}} &
 \kett{I_R} & 
 \multicolumn{1}{c}{\kett{\varphi_R}}
\\
\hline
& \\[\dimexpr-\normalbaselineskip+3pt]
\langle\!\langle I_{NS}| 
 & \chthree_{1,1}(\tq) & \chthree_{1,3}(\tq) 
 & 2\, \chthree_{1,4}(\tq)
 & 2\, \chthree_{1,2}(\tq)\\
\langle\!\langle \varphi_{NS}| 
 &&\chthree_{1,1}(\tq){+}\chthree_{1,3}(\tq)
 & 2\, \chthree_{1,2}(\tq)
 & 2\,\chthree_{1,2}(\tq){+}2\,\chthree_{1,4}(\tq)
\\
\langle\!\langle I_{R}|
 &&&2\, \chthree_{1,1}(\tq)&2\, \chthree_{1,3}(\tq)\\
\langle \varphi_{R}|
 &&&&2\, \chthree_{1,1}(\tq){+}2\, \chthree_{1,3}(\tq)\\
\end{array}
\ee
The boundary states $\kkett{I_\R}$ and $\kkett{\varphi_R}$ are not
fundamental as conformal 
boundary conditions, since the identity representation appears twice
in their cylinder partition functions.
If we allow the introduction of the Ramond
sector then they can each be written as a superposition of two
boundary states, but the cylinder partition functions will not then be  
expressible as a sum of characters of the super Virasoro algebra.

\subsection{Interface operators}
\label{sec:tciminterface}

The common sectors of $\cH_\text{TCIM}$ and $\cH_{\sTCIM}$ are, 
in terms of representations of the Virasoro algebra,
\be
 (\cH_0\otimes \bar\cH_0)
\;\oplus\;
 (\cH_{3/2}\otimes \bar\cH_{3/2})
\;\oplus\;
 (\cH_{1/10}\otimes \bar\cH_{1/10})
\;\oplus\;
 (\cH_{3/5}\otimes \bar\cH_{3/5})
\;.
\ee
This means that a topological interface operator $I$ satisfying
\eqref{eq:topI} and acting from the space of TCIM to \sTCIM\ is
a constant on each of these four sectors and so is determined by four
constants 
\be
I(a,b,c,d) = 
  a \: P_0\bar P_0
+ b \: P_{3/2}\bar P_{3/2}
+ c \: P_{1/10}\bar P_{1/10}
+ d \: P_{3/5}\bar P_{3/5}
\;,
\ee
as well as a map identifying the Virasoro highest weights states of
weights $(3/2,3/2)$ and $(3/5,3/5)$. We take this to be
\bea
 \ket{(1,3)}_{\text{TCIM}} 
&=& i \,\xi_{1,3} G_{-3/2}\bar G_{-3/2}\ket  0_{\sTCIM}
\;,
\nn\\
 \ket{(3,1)}_{\text{TCIM}} 
&=& i \,\xi_{3,1} G_{-1/2}\bar G_{-1/2}\ket {\tfrac 1{10}}_{\sTCIM}
\;,
\label{eq:idents}
\eea
where $\xi_{1,3}$ and $\xi_{3,1}$ are signs.

Requiring integer coefficients in the expansions
\be
  I_\alpha \: \kkett{B^{\sTCIM}_A} 
= \sum_B M^B_{\alpha A} \kkett{B^\text{TCIM}_B}
\;,\;\;
  I^\dagger_\alpha \: \kkett{B^\text{TCIM}_A} 
= \sum_B \tilde{M}^B_{\alpha A} \kkett{B^{\sTCIM}_B}
\;,
\ee
allows us to solve for $(a,b,c,d)$. In particular, from the
definitions in \eqref{eq:ketdef}, the expansions
\eqref{eq:expansions} and the identifications \eqref{eq:idents}, 
we get
\be
I(a,b,c,d)^\dagger \: \kkett{B_0}
= 
m_1 \kkett{I_{\NS}}
+m_2 \kkett{\varphi_{\NS}}
+m_3 \kkett{I_{\R}}
+m_4 \kkett{\varphi_{\R}}
\;,
\ee
with
\be
\begin{array}{rcl}
a &=& \sqrt 2 m_1 + \frac{1 + \sqrt 5}{\sqrt 2}m_2 + 2 m_3 + (1+\sqrt 5) m_4 
\;,
\\
-\xi_{1,3}\,b 
&=& \sqrt 2 m_1 + \frac{1 + \sqrt 5}{\sqrt 2}m_2 - 2 m_3 -(1+ \sqrt 5) m_4 
\;,
\\
c &=& \sqrt 2 m_1 + \frac{1 - \sqrt 5}{\sqrt 2}m_2 - 2 m_3 - (1-\sqrt 5) m_4 
\;,
\\
-\xi_{3,1} d 
&=& \sqrt 2 m_1 
+ \frac{1 - \sqrt 5}{\sqrt 2}m_2 + 2 m_3 + (1-\sqrt 5) m_4 
\;.
\end{array}
\ee
We will choose $\xi_{1,3} = \xi_{3,1} = -1$.
This means that any interface can be expressed as a combination
\be
 I(a,b,c,d) 
= m_1 \: I  + m_2 \: I_2 + m_3 \: I_3 + m_4 \: I_4
\;,
\label{eq:idefabcd}
\ee
where the interfaces are given by
\be
\begin{array}{rcl}
I   &=& I(\sqrt 2,\sqrt 2,\sqrt 2,\sqrt 2)\\
I_2 &=& I(\frac{1 + \sqrt 5}{\sqrt 2},\frac{1 + \sqrt 5}{\sqrt 2},\frac{1 - \sqrt 5}{\sqrt 2},\frac{1 - \sqrt 5}{\sqrt 2})\\
I_3 &=& I( 2,- 2,- 2, 2)\\
I_4 &=& I(1{+}\sqrt 5,-1{-}\sqrt 5,-1{+}\sqrt 5,1{-}\sqrt 5)
\end{array}
\label{eq:idef}
\ee
These act on the boundary states as follows
\be
\begin{array}{r@{\,}c@{\,}lr@{\,}c@{\,}l}
 I\, \kkett{I_{\NS}} &=& \kkett{B_{(1,1)}} + \kkett{B_{(3,1)}}        \;,
&I^\dagger\, \kkett{B_{(1,1)}} &=& I^\dagger\, \kkett{B_{(3,1)}} = \kkett{I_{\NS}}  \;,
\\
 I\, \kkett{\varphi_{\NS}} &=& \kkett{B_{(1,2)}} + \kkett{B_{(1,3)}}  \;,
&I^\dagger\, \kkett{B_{(1,2)}} &=& I^\dagger\, \kkett{B_{(1,3)}} = \kkett{\varphi_{\NS}}  \;,
\\
 I\, \kkett{I_{\R}} &=& 2 \,\kkett{B_{(2,1)}}       \;,
&  I^\dagger\, \kkett{B_{(2,1)}} &=& \kkett{I_{\R}}       \;,
\\
 I\, \kkett{\varphi_{\R}} &=& 2\, \kkett{B_{(2,2)}} \;,
&I^\dagger\, \kkett{B_{(2,2)}} &=& \kkett{\varphi_{\R}} \;,
\end{array}
\label{eq:iactions}
\ee
and satisfy the relations
\be
I_2 = D_{1/10} \: I = I \: D_{\varphi}
\;,\;\;
I_3 = D_{7/16} \: I = I \cdot \sqrt{2} (-1)^F 
\;,\;\
I_4 = D_{1/10} \: I = I \cdot \sqrt{2} (-1)^F D_{\varphi}
\;.\;\;
\ee

Requiring $ I \: I^\dagger$ be expressible as a sum of topological
defects in TCIM and $I^\dagger \: I$ be expressible as a sum of
topological defects in \sTCIM\ provides a strong constraint, just as it
did in the Ising/FF case. Taking $a,b,c,d$ to be real, we have, for example,
\bea
 I(a,b,c,d) \: I(a,b,c,d)^\dagger &=& 
\tfrac{ (5 - \sqrt 5)(a^2 + b^2) + (5 + \sqrt 5)(c^2 + d^2)}{40}
( D_0 + D_{3/2})
\nn\\
&&+\tfrac{ a^2 + b^2 - c^2 - d^2}{4 \sqrt 5}(D_{1/10} + D_{3/5})
\nn\\
&&+
\tfrac{ (5 - \sqrt 5)(a^2 - b^2) - (5 + \sqrt 5)(c^2 - d^2)}{20\sqrt 2}
D_{7/16}
\nn\\
&&+ \tfrac{a^2 - b^2 + c^2 - d^2}{2 \sqrt{10}} D_{3/80}
\;.
\eea
We find the interfaces given by \eqref{eq:idefabcd} indeed give integer
coefficients, and the fundamental interface $I$ given by \eqref{eq:idef} satisfies
\be
I \: I^\dagger = D_0 + D_{3/2}
\;,\;\;
I^\dagger \: I = D_I + (-1)^{F + \bar F} D_I
\;,
\ee
\be
D_{3/2} \: I = I \cdot (-1)^{F + \bar F} = I
\;.
\ee

To summarise, we have found a set of four elementary boundary conditions for
\sTCIM, a set of eight elementary topological superconformal defects in
\sTCIM\ and a set of four fundamental interfaces between TCIM and \sTCIM. 
All we need to do now is to find superconformal
defects in \sTCIM\ and use the interface operators to generate defects
in TCIM. 

\section{The doubled model, \SVIR2}
\label{sec:svir2}

The basic idea behind finding superconformal non-topological defects
in \sTCIM\ is that every superconformal defect in \sTCIM\ is equivalent to a
superconformal boundary condition on \SVIR2. Since the central
charge
of \SVIR2 is $7/5$, which is less than 3/2, there are only a finite set of fundamental
superconformal boundary conditions, and that it is quite possible that
not all of these are topological. 

We first have to identify the local fields in \SVIR2. Since we are
only considering Neveu-Schwarz sectors, the partition function of
\SVIR2 is exactly the square of the partition function of \sTCIM. 
This has been identified in \cite{GY} as the Neveu-Schwarz sector of
the \DS--\ES\ modular invariant of the $N=1$ super Virasoro
algebra, and so we can consider \SVIR2 as a model with $\sVir_{10}$
symmetry. 
There
are numerous character identities relating characters of the
superconformal algebras $\sVir_3$ at $c=7/10$ and $\sVir_{10}$ at
$c=7/5$; these allow one to 
write the partition function in terms of characters of $\sVir_{10}$ as
\be
\begin{array}{rl}
 Z_\text{\SVIR2} &= ( Z_{\sTCIM} )^2
= \left( | \chthree_{1,1} |^2 + | \chthree_{1,3} |^2 \right)^2
\\
&= 
| \chten_{1,1} + \chten_{1,5} + \chten_{1,7} + \chten_{1,11} |^2
+ 
| \chten_{3,1} + \chten_{3,5} + \chten_{3,7} + \chten_{3,11} |^2 \\
& \hspace{9.5cm} + 2
| \chten_{(5,1)} + \chten_{(5,5)} |^2
\;.
\end{array}
\label{eq:zd6e6}
\ee

We will be constructing boundary states which respect the $\sVir_{10}$
symmetry, and the building blocks are Ishibashi states for the
$\sVir_{10}$ algebra. There are two Ishibashi states (one for each
gluing condition) for each diagonal
term (those with $h=\bar h$) in the partition function
\eqref{eq:zd6e6}. Hence there are 24 Ishibashi states in total.

The diagonal terms in the partition function \eqref{eq:zd6e6} can be
identified by Kac labels $(r,s)$ 
with $r$ and $s$ taking values in the exponents of \DS\ and a subset
of the exponents of \ES\
respectively. These exponents are $\{1,3,5,5',7,9\}$ and
$\{1,4,5,7,8,11\}$ (the label $5$ appears twice, as do fields with
labels $(5,s)$ in the Neveu-Schwarz sector \eqref{eq:zd6e6}). This
over-count the set of fields by a factor of two, as it does not take
into account the symmetry of the Kac labels $(r,s)\simeq(10-r,12-s)$.
Furthermore, it also includes the Ramond sector (those with $s=4$ or
$s=8$). the result is that we can label the Neveu-Schwarz diagonal
fields in \eqref{eq:zd6e6} by $(r,s)$
with $r\in\{1,3,5,5',7,9\}$ and $s\in\{1,5,7,11\}$ modulo
$(r,s)\simeq(10-r,12-s)$. 

The full chiral algebra of \SVIR2 is, of course, larger than 
$\sVir_{10}$ and so can be described as an extension of the
$\sVir_{10}$ by super-primary fields, i.e. as a
super W-algebra. From the character decomposition of the full chiral
algebra $\sVir_3 \otimes \sVir_3$,
\be
(\chthree_{1,1})^2
= 
\chten_{1,1} + \chten_{1,5} + \chten_{1,7} + \chten_{1,11} 
\;,
\ee
we see that the chiral algebra contains three super-primary fields 
$\cW^{(3/2)}$, 
$\cW^{(7/2)}$, and
$\cW^{(10)}$ 
of weights
$\hten_{1,5}=3/2$, 
$\hten_{1,7}=7/2$, and
$\hten_{1,11}=10$ respectively.
Expressions for these fields are given in appendix \ref{app:chiralalgebra}.
Extending $\sVir_{10}$ by the field of weight 3/2 just recovers the
full $\sVir_{3}\times\sVir_3$ algebra.
Extending $\sVir_{10}$ by the field of weight 7/2 gives an algebra
SW(7/2) which has been considered before in \cite{FS,H1,BEHH}.
Extending $\sVir_{10}$ by the field of weight 10 gives a new algebra
SW(10). We will return to these algebras when we consider the boundary
states in section \ref{sec:newboundaries}.

Finally, we note here that 
it is also possible to          
express a single copy of the \sTCIM\
partition function in terms of characters of the
$c=7/5$ algebra:
\be
\begin{array}{rl}
 Z_{\sTCIM} 
&=
 ( \chthree_{1,1} )^2 + ( \chthree_{1,3} )^2 
\\
&= 
 \chten_{1,1} + \chten_{1,5} + \chten_{1,7} + \chten_{1,11} 
+ 
 \chten_{3,1} + \chten_{3,5} + \chten_{3,7} + \chten_{3,11} 
\;,
\end{array}
\label{eq:zd6e62}
\ee
where we note again that $q$ is real.
The reason is that one can embed the $c=7/5$ superconformal algebra
into the two (holomorphic and anti-holomorphic) copies of the $c=7/10$
algebra in \sTCIM.

\subsection{Relating boundary conditions on \SVIR2 to defects in SVIR${}_3$}

Central to the idea of the folding procedure is that a boundary
condition in \SVIR2 is equivalent to a defect in \sTCIM. 
The folding condition is simple if the boundary is along the real axis
with the folded model living entirely in the upper half plane: 
for each field $\phi(z,\bar z)$ in \sTCIM, there are two copies $\phi^{(a)}$ in
the folded theory, with the $a=1$ copy being the original field,
$\phi^{(1)}(z,\bar z)=\phi(z,\bar z)$ and  the $a=2$ copy being the
folded field,
$\phi^{(2)}(z,\bar z)=\phi(\bar z, z)$.
When we consider boundary states and defect operators, it is more usual
for the boundary/defect to be on the unit circle, and we can define a
map  $\rho$ from states in \SVIR2 to
defects in \sTCIM\ as follows. 

If the boundary lies on the unit circle with \SVIR2
defined on the exterior of the unit circle, copy $a=1$ corresponds to
the superconformal algebra on the outside of 
the unit circle and copy $a=2$, the inside.
If $\rho(\kett B) = \hat D$, then we can define
\be
\begin{array}{l}
\rho( G^1_m \kkett B ) =  G_m \,\hat D
\;,\\
\rho( \bar G^1_m \kkett B ) =  \bar G_m \,\hat D
\;,\\
   \rho(  G^2_m \kkett B ) 
=  -i (-1)^{F + \bar F} \hat D (-1)^{F + \bar F} \bar G_{-m}
\;,\\
  \rho( \bar G^2_m \kkett B ) 
=  i (-1)^{F + \bar F} \hat D (-1)^{F + \bar F} G_{-m}
\;.
\end{array}
\label{eq:rhodef}
\ee
The signs in \eqref{eq:rhodef} are determined by finding a family of
M\"obius maps which interpolate the identity map preserving the real
axis and a map which send the real axis to the unit circle, as explained
in appendix \ref{app:glue}.

To complete the definition, we need the image of 
the highest weight states in \SVIR2 which are tensor products,
$\ket{h_1}\otimes\ket{h_2}$. We can take the simplest choice, i.e.
$\rho( \ket{h_1}\otimes\ket{h_2} )= \ketbra{h_1}{h_2}$, 
but as we will see later, it will be helpful to define in addition the map
$\rho'( \ket{h_1}\otimes\ket{h_2} )
= \ketbra{h_1}{h_2}(-1)^F$. These maps are summarised in table \ref{tab:rhorho'}.

\begin{table}[htb]
\[
\renewcommand{\arraystretch}{1.2}
\begin{array}{c|rrrr}
(r,s) &
(1,1) & (3,5)~ & (5,5)~ & (5',5)
\\ \hline
\rho(\ket{r,s}) &
\ketbra 00 & 
\ketbra{\tfrac{1}{10}}{\tfrac{1}{10}} & 
\ketbra{0}{\tfrac{1}{10}} & 
\ketbra{\tfrac{1}{10}}{0} 
\\ 
\rho'(\ket{r,s}) &
\ketbra 00 & 
-\ketbra{\tfrac{1}{10}}{\tfrac{1}{10}} & 
-\ketbra{0}{\tfrac{1}{10}} & 
\ketbra{\tfrac{1}{10}}{0} 
\end{array}
\]
\caption{Images of the highest weight states}
\label{tab:rhorho'}
\end{table}

We can now identify the 24 known conformal defects in $\SV_3$ (8
topological and 16 factorised) with boundary conditions on \SVIR2, and
in particular the gluing conditions satisfied by the boundary
conditions.  

For a topological defect, we have two free signs given in equation
\eqref{eq:sd}. These imply that the corresponding boundary state
satisfies
\be
\begin{rcases*}  
  G_m D = \eta D G_m
\\
 \bar G_m D = \eta' D \bar G_m
\end{rcases*}
\Rightarrow
\begin{cases}
 (G^1_m + i \eta \bar G^2_{-m}) \kkett{B} = 0
\\
 (G^2_m + i \eta' \bar G^1_{-m}) \kkett{B} = 0
\end{cases}
\label{eq:sd2}
\ee
For a factorising defect, we have again two free signs coming from the
two gluing conditions of the two boundary states, which imply for the
corresponding defect
\be
\begin{rcases*}  
 ( G_m + i \eta \,\bar G_{-m} ) \kkettbbraa{A,\eta}{B,\eta'} = 0
\\
 \kkettbbraa{A,\eta}{B,\eta'}
 ( G_m - i \eta'\,\bar G_{-m} ) = 0
\end{rcases*}
\Rightarrow
\begin{cases}
 (G^1_m + i \eta \bar G^1_{-m}) \kkett{B} = 0
\\
 (G^2_m - i \eta' \bar G^2_{-m}) \kkett{B} = 0
\end{cases}
\label{eq:sd3}
\ee
It is clear from equations \eqref{eq:sd2} and \eqref{eq:sd3} that we
will not be able to find a way to express all the gluing conditions that arise
as gluing conditions on a single set of combinations $G^1_m\pm G^2_m$
and $\bar G^1_m \pm \bar G^2_m$, and so, in the next section, we
consider exactly how we can organise the boundary states of \SVIR2
corresponding to the known defects 
into boundary states of the algebra $\sVir_{10}$.

\subsection{Relating the boundary conditions on \SVIR2 to boundary
  conditions for  sVir$\mathbf{_{10}}$ }

To view \SVIR2 as a model of the $c=7/5$
superconformal algebra $\sVir_{10}$ we need to define an embedding of
$\sVir_{10}$  into the two copies of the $c=7/10$ algebra $\sVir_3$ in
\sTCIM. 
An embedding of $\sVir_{10}$ into $\sVir_3 \times \sVir_3$ is defined
by four signs $\{\alpha,\beta,\gamma,\delta\}$:
\be
  \iota_{\alpha\beta\gamma\delta}( G_m ) 
= \alpha G^1_m + \beta G^2_m
\;,\;\;
  \iota_{\alpha\beta\gamma\delta}( \bar G_m ) 
= \gamma \bar G^1_m + \delta \bar G^2_m
\;.
\label{eq:embed}
\ee
We will denote the combined map $(\rho \circ \iota_{\alpha\beta\gamma\delta})$ by 
$\rho_{\alpha\beta\gamma\delta}$.

We also need to define a map from the Ishibashi states with
respect to $\sVir_{10}$ to the states in \SVIR2.
The Ishibashi states are determined by a highest weight states
and a gluing condition, and the highest weight condition depends on the
choice of embedding, so we need to take some care over this.

For the moment we restrict attention to the gluing condition.
Suppose we have an Ishibashi state $\kett{h,\eps}$ of $\sVir_{10}$ satisfying 
\be
 ( G_m + i \epsilon \bar G_{-m} ) \kett{h,\eps} = 0
\;.
\ee
With the embedding $\iota_\abgd$, this state satisfies
\be
 (\alpha G^1_m + \beta G^2_m + i \epsilon \gamma \bar G^1_{-m} + i
 \epsilon \delta \bar G^2_{-m} ) \kett{h,\epsilon} = 0
\;.
\ee
Hence a purely transmitting defect with gluing conditions
$\{\eta,\eta'\}$ corresponds to $\eta = \alpha\delta\eps$ and $\eta' =
\beta\gamma\eps$, that is $\alpha\beta\gamma\delta = \eta\eta'$
and a pure reflecting defects 
with gluing conditions
$\{\eta,\eta'\}$ corresponds to $\eta = \alpha\gamma\eps$ and $\eta' =
- \beta\delta\eps$, that is $\abgd = -\eta\eta'$.
Hence we find that there are two equivalence classes of embeddings,
given by $\abgd=\pm 1$.
An embedding with $\abgd=1$ will correspond to
transmitting defects with $\eta\eta'=1$ and reflecting defects with
$\eta\eta' = -1$, 
whereas an embedding with $\abgd=-1$ will correspond to
transmitting defects with $\eta\eta'=-1$ and reflecting defects with
$\eta\eta' = 1$.

The result is that we can expect two sets of boundary states with
respect to two different choices of embeddings $\sVir_{10}$, with one
set giving half the defects of $\sTCIM$ and the other set giving the
other half.
The precise expressions for these boundary states in terms of
Ishibashi states will of course depend on the definitions of the
highest weight states. 
There are eight highest weight states with respect to $\sVir_{10}$
whose definitions depend on the embedding, namely those with Kac labels
$(1,5), \: (1,7), \: (1,11), \: (3,1), \: (3,7), \: (3,11), \: (5,1)$,
and $(5',1)$. 
For example, the state $\ket{1,5}$ of weight $(3/2,3/2)$ is given by
\be
 \iota_\abgd ( \vec{1,5} )
= \frac{i \eta_{1,5}   }{4c/3} 
( \alpha G^1_{-3/2} - \beta G^2_{-3/2} )
( \gamma \bar G^1_{-3/2} - \delta \bar G^2_{-3/2} ) \vec 0
\;,\;\;
\ee
where $c=7/10$ and $\eta_{1,5}$ is a free sign,
and the state $\ket{3,1}$ of weight $(7/10,7/10)$ is given by 
\be
 \iota_\abgd (\ket{3,1} )= \frac{i\eta_{3,1}}{2h}(\alpha G^1_{-1/2} - \beta
 G^2_{-1/2})(\gamma \bar G^1_{-1/2} - \delta \bar G^2_{-1/2})\ket{3,5}
\;,
\ee
where $h=1/10$ and $\eta_{3,1}$ is a free sign.  We have a free sign
$\eta_{r,s}$ for each of these eight such states.
Expanded expressions for these states and others can be found in
appendix \ref{app:states}.

Using these facts, it is possible to construct the boundary states 
corresponding to all the known topological and factorising
defects in \sTCIM. 
We find that they can all be written
in terms of the states $\kkett{(a,b)_\NS}$ and
$\kkett{(a,b)_\wtNS}$ defined in \cite{GY} in (at least) two
ways.
We illustrate this in the next two sections with the case of the identity
defect in \sTCIM, and the factorising defect
$\kkettbbraa{I_{\NS}}{I_{NS}}$, before summarising the results in 
section \ref{sec:summary}.

\subsubsection{The identity defect in \sTCIM}

The identity defect in \sTCIM\ takes the very simple form of a sum over
an orthonormal basis of $\cH_{\sTCIM}$:
\be
 \One = \sum_\psi \ketbra{\psi}{\psi}
\;.
\ee
Expanding this out, 
\bea
\One &=& \ketbra 00 + \frac{1}{2c/3}\left(
G_{-3/2}\ketbra 00 G_{3/2} +
\bar G_{-3/2}\ketbra 00 \bar G_{3/2} \right)
+ \ldots
\nn\\
&+&
\ketbra {\tfrac1{10}}{\tfrac1{10}} + \frac{1}{1/5}\left(
G_{-1/2} \ketbra{\tfrac1{10}}{\tfrac1{10}} G_{1/2} +
\bar G_{-1/2}\ketbra{\tfrac1{10}}{\tfrac1{10}}\bar G_{1/2} \right)
+ \ldots
\label{eq:iddef}
\eea
This must arise from a combination of the Ishibashi states 
$\kett{(1,1),\eps}$, $\kett{(1,5),\eps}$, 
$\kett{(1,7),\eps}$, $\kett{(1,11),\eps}$, 
$\kett{(3,1),\eps}$, $\kett{(3,5),\eps}$, 
$\kett{(3,7),\eps}$, and $\kett{(3,11)\eps}$. Since the Identity
defect satisfies \eqref{eq:sd} with $\eta=\eta'=1$, the gluing
condition $\eps$ and embedding $\iota_\abgd$ satisfy
$\eps=\alpha\delta=\beta\gamma$ and $\abgd=1$.

The simplest choice is $\alpha=\beta=\gamma=\delta=\eps=1$.
This still leaves the signs $\eta_{r,s}$ free.
Given the freedom to choose these signs, the boundary state
$\kkett{D_I}$ can be expressed as sum over all the Ishibashi states in
the NS sector with the $+$ gluing condition: 
\be
\begin{array}{rl}
\One =& \rho_{++++}(\kkett{D_I})\;,\\
 \kkett{D_I}
=& 
 \phantom{+\,}\kett{(1,1),+} + 
 \kett{(1,5),+} + 
 \kett{(1,7),+} + 
 \kett{(1,11),+} \\
&+\, 
 \kett{(3,1),+} + 
 \kett{(3,5),+} + 
 \kett{(3,7),+} + 
 \kett{(3,11),+} 
\;,
\label{eq:d1}
\end{array}
\ee
where $\rho$ acts as in table \ref{tab:rhorho'}.
Looking at the explicit expressions in appendix \ref{app:states},  this
fixes in particular $\eta_{1,5} = 1$, $\eta_{3,1} = 1$ and $\eta_{3,7} = -1$.

We could, by choosing the signs of $\eta$ in a different fashion,
instead have the equally symmetric expression
\be
\begin{array}{rl}
\One =& \rho_{++++}'(\kkett{D_I})\;,\\
 \kkett{D_I}
=& \phantom{+\,}
 \kett{(1,1),+} - 
 \kett{(1,5),+} + 
 \kett{(1,7),+} - 
 \kett{(1,11),+} \\
&+\, 
 \kett{(3,1),+} -
 \kett{(3,5),+} + 
 \kett{(3,7),+} - 
 \kett{(3,11),+} 
\;,
\end{array}
\label{eq:d1'}
\ee
where $\rho'$ acts as in table \ref{tab:rhorho'}.
Looking again at the explicit expressions in appendix \ref{app:states},
this implies the opposite choices,
\be
\eta_{1,5} = -1
\;,\;\;
\eta_{3,1} = -1
\;,\;\;
\eta_{3,7} = 1
\;.
\label{eq:signfix}
\ee

Note that in equations \eqref{eq:d1} and \eqref{eq:d1'} we have
suppressed all the information regarding the choices of signs for the
descendent states $\ket{1,5}$ etc, and have only kept the information
regarding the choice of the signs for the maps of the highest weight
states from table \eqref{tab:rhorho'}.

\subsubsection{The factorising defect $\kkettbbraa{I_{\NS}}{I_{\NS}}$
  in \sTCIM}

We will take the boundary state $\kkett{I_{\NS}}$ 
to be given in terms of the
Ishibashi states in \sTCIM\ as
\be
\kkett{I_{\NS}} =
\left( \tfrac{5 - \sqrt 5}{10} \right)^{\frac 14} \kett{0{,}+} +
\left( \tfrac{5 + \sqrt 5}{10} \right)^{\frac 14} \kett{\tfrac{1}{10}{,}+} 
\;,
\ee
which is one of the two possibilities shown in \eqref{eq:ketdef}.
This means the factorising defect $\kkettbbraa{I_{\NS}}{I_{\NS}}$ is given
as
\be
\begin{array}{rl}
&\kkettbbraa{I_{\NS}}{I_{\NS}}
\\
=&\left( \frac{5 - \sqrt 5}{10} \right)^{\frac 12} \kettbraa{0{,}+}{0{,}+} 
+
\left( \frac{1}{5} \right)^{\frac 14} \kettbraa{0{,}+}{\tfrac{1}{10}{,}+} 
+
\left( \frac{1}{5} \right)^{\frac 14} \kettbraa{\tfrac{1}{10}{,}+}{0{,}+} 
\left( \frac{5 + \sqrt 5}{10} \right)^{\frac 12} \kettbraa{\tfrac{1}{10}{,}+}{\tfrac{1}{10}{,}+} 
\;.
\end{array}
\ee
Since this factorising
defect satisfies \eqref{eq:sd3} with $\eta=\eta'=1$, the gluing
condition $\eps$ and embedding $\iota_\abgd$ satisfy
$\eps=\alpha\gamma=-\beta\delta$ and $\abgd=-1$.

The simplest choice, which we take from now on, is
$\alpha=\beta=\gamma=\eta=1$, $\delta=-1$.
Given the freedom to choose the signs $\eta_{r,s}$, we can express 
this factorising defect in many ways, but the one that we will use later
is this:
\begin{align}
\kkettbbraa{I_{\NS}}{I_{\NS}} =& \rho_{+++-}(\kkett{I_{\NS}I_{\NS}})\;, \nn\\
 \kkett{I_{\NS}I_{\NS}}
=& 
\left(\frac{5 - \sqrt 5}{10}\right)^{\frac 12}
  \big( \kett{(1,1),+} + \kett{(1,5),+} +  \kett{(1,7),+}
    +  \kett{(1,11),+} \big) 
\nn\\
+& \left( \frac{1}{5} \right)^{\frac 14} 
  \big( \kett{(5,1),+} + \kett{(5,5),+} +  \kett{(5',1),+} 
    +  \kett{(5',5),+} \big)
\nn\\
+& \left(\frac{5 + \sqrt 5}{10}\right)^{\frac 12}
  \big( \kett{(3,1),+} + \kett{(3,5),+} +  \kett{(3,7),+} 
    +  \kett{(3,11),+} \big) 
\;,
\label{eq:insins}
\end{align}
where $\rho$ acts as in table \ref{tab:rhorho'} and the signs
$\eta_{1,5}$,$\eta_{3,1}$ and $\eta_{3,7}$ are again as in equation
\eqref{eq:signfix}. We cannot say whether the remaining signs are also
fixed as for the identity defect as we have not calculated them.
Note that in equation \eqref{eq:insins} 
we have again suppressed all the information regarding the choices of
signs for the 
descendent states $\ket{1,5}$ etc, and have only kept the information
regarding the choice of the signs for the maps of the highest weight
states from table \eqref{tab:rhorho'}.

\subsection{Boundary conditions in \SVIR2}
\label{sec:bcinsvir2}

The boundary states are linear combinations of the Ishibashi states
corresponding to diagonal ($h=\bar h$) Neveu-Schwarz fields. As noted
above, there are 24 of these, $\kett{(r,s),\pm}$, labelled by $(r,s)$
odd exponents of \DS\ and \ES\ modulo the Kac symmetry, and a gluing
condition.

The set of boundary states for the GSO projection of \SVIR2 were proposed in
\cite{GY} although there are some difficulties with these, as
explained in \ref{sec:conclusions}. The states in \cite{GY} have
both Neveu-Schwarz and Ramond contributions; here we only need 
components in the Neveu-Schwarz sector. 

The boundary conditions themselves are labelled by pairs of modes on the
Dynkin diagrams of \DS\ and \ES, together with a choice of gluing
condition. However, this again over-counts the number of Neveu-Schwarz
boundary states in two ways. Firstly, the nodes on the \ES\ diagram
that are related by the diagram symmetry, $r: 1 \mapsto 5$ and
$r: 2\mapsto 4$ lead to the same Neveu-Schwarz
contribution. (We can think of this as replacing the \ES\ Dynkin
diagram by the $F_4$ diagram, with the nodes related by the $Z_2$
symmetry corresponding to the short simple roots of $F_4$).
This gives 24 pairs of nodes.
Secondly, we can bi-colour the Dynkin diagrams and split these 24
pairs into those with nodes of the same colour and those with nodes of
opposite colour, giving two sets of 12 pairs nodes, combined with the
gluing condition.

\begin{figure}[htb]

\pictures{
$$  
  \begin{picture}(440,170)
  \put(0,-20){\scalebox{.45}{\includegraphics{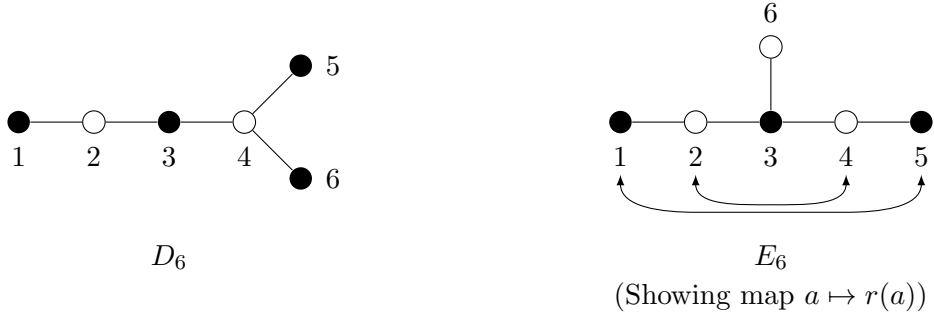}}}
  \end{picture}
$$
}{
\begin{center}
\begin{tikzpicture}
	\pgfmathsetmacro{\hshift}{4}
	
	\node[circle,fill,inner sep=3pt] (d1) at (-2-\hshift,0) {};
	\node[circle,draw,inner sep=3pt] (d2) at (-1-\hshift,0) {};
	\node[circle,fill,inner sep=3pt] (d3) at (0-\hshift,0) {};
	\node[circle,draw,inner sep=3pt] (d4) at (1-\hshift,0) {};
	\node[circle,fill,inner sep=3pt] (d5) at (1.75-\hshift,0.75) {};
	\node[circle,fill,inner sep=3pt] (d6) at (1.75-\hshift,-0.75) {};
	
	\draw (d1) -- (d2) -- (d3) -- (d4);
	\draw (d4) -- (d5);
	\draw (d4) -- (d6);
	
	\node[below] at (-2-\hshift,-0.2) {$1$};
	\node[below] at (-1-\hshift,-0.2) {$2$};
	\node[below] at (0-\hshift,-0.2) {$3$};
	\node[below] at (1-\hshift,-0.2) {$4$};
	\node[right] at (1.95-\hshift,0.75) {$5$};
	\node[right] at (1.95-\hshift,-0.75) {$6$};
	
	\node[below] at (0-\hshift,-1.5) {$D_6$};
	
	\node[circle,fill,inner sep=3pt] (e1) at (-2+\hshift,0) {};
	\node[circle,draw,inner sep=3pt] (e2) at (-1+\hshift,0) {};
	\node[circle,fill,inner sep=3pt] (e3) at (0+\hshift,0) {};
	\node[circle,draw,inner sep=3pt] (e4) at (1+\hshift,0) {};
	\node[circle,fill,inner sep=3pt] (e5) at (2+\hshift,0) {};
	\node[circle,draw,inner sep=3pt] (e6) at (0+\hshift,1) {};
	
	\draw (e1) -- (e2) -- (e3) -- (e4) -- (e5);
	\draw (e3) -- (e6);
	
	\node[below] at (-2+\hshift,-0.2) {$1$};
	\node[below] at (-1+\hshift,-0.2) {$2$};
	\node[below] at (0+\hshift,-0.2) {$3$};
	\node[below] at (1+\hshift,-0.2) {$4$};
	\node[below] at (2+\hshift,-0.2) {$5$};
	\node[above] at (0+\hshift,1.2) {$6$};
	
	\draw[latex-latex] (-2+\hshift,-0.7) to[out=-90,in=180] (-1+\hshift,-1.2) -- (1+\hshift,-1.2) to[out=0,in=-90] (2+\hshift,-0.7);
	\draw[latex-latex] (-1+\hshift,-0.7) to[out=-90,in=180] (0+\hshift,-1.1) to[out=0,in=-90] (1+\hshift,-0.7);
	
	\node[below,align=center] at (0+\hshift,-1.5) {$E_6$ \\ (Showing map $a \mapsto r(a)$)};
	
\end{tikzpicture}
\end{center}
}

\caption{
The Dynkin diagrams of $D_6$ and $E_6$ showing the bi-colouration and the map $r$}

\label{fig:dynkin}
\end{figure}

With the bi-colouration as in figure \ref{fig:dynkin}, and making
a choice for the representatives of the nodes related by the $Z_2$
symmetry of the \ES\ diagram $a \mapsto r(a)$, we take the nodes
with the same colouration to be 
\be
{\cal N}_e = \{ (1,1),(3,1),(5,1),(6,1),
   (2,2),(4,2),
   (1,3),(3,3),(5,3),(6,3),
   (2,6),(4,6)
\}
\;,
\label{eq:singleset}
\ee
and the nodes with opposite colouration to be 
\be
{\cal N}_o = 
\{ (2,1),(4,1),
   (1,2),(3,2),(5,2),(6,2)
   (2,3),(4,3),
   (1,6),(3,6),(5,6),(6,6)
\}
\;.
\label{eq:otherset}
\ee

The key ingredient in the boundary states proposed in \cite{GY} are
matrices
$\Psi^{(a,b)}_{(r,s)}$,
\be
  \Psi^{(a,b)}_{(r,s)} 
= \frac{ \psi^r_a(D_6)\,\psi^s_b(E_6) }
       {\sqrt{S_{1r}^{(8)} S_{1s}^{(10)} }}
\;,
\label{eq:Psis}
\ee
formed from eigenvectors $\psi^r_a(G)$ of the adjacency matrices of
the Dynkin diagram of $G$ and modular S-matrices $S_{rs}^{(k)}$ for affine
$su(2)$ characters at level $k$. We give the vectors $\psi^r_a(G)$ and
a table of numerical values of $\Psi^{(a,b)}_{(r,s)}$ in appendix
\ref{app:bcs} for convenience.
These matrices have the property that under
the Kac-symmetry
\be
 \Psi^{(a,b)}_{(r,s)} = \begin{cases}
 \Psi^{(a,b)}_{(10-r,12-s)} & (a, b) \in {\cal N}_e\;,\hbox{ same colouration,}
\\
 - \Psi^{(a,b)}_{(10-r,12-s)} & (a, b) \in {\cal N}_o\;, \hbox{ opposite colouration.}
\end{cases}
\ee

Following \cite{GY}, we can define boundary states $\kkett{(a,b)_\NS}$
using these matrices, but we take a slightly different choice to
\cite{GY},
\be
\kkett{(a,b)_\NS}
= 
\sum_{\substack{r\in\{1,3,5,5',7,9\}\\ s\in\{1,7\}}} \Psi^{(a,b)}_{(r,s)} \: \kett{(r,s), +}
\;,
\ee
where in \cite{GY} the sum over $s$ is $s\in\{1,5\}$. The sums are
over exactly the same representations, but the choice of different
representatives results in expressions which differ by a sign for
$s=7$ when the nodes are of opposite colour (eg the labels $(1,7)$ and $(9,5)$ denote the same representation, but $\Psi^{(1,2)}_{(1,7)} = - \Psi^{(1,2)}_{(9,5}$).

Our choice of representatives was motivated by the fact the \ES\  $\hat{su}(2)_{10}$ WZW model has an extended symmetry algebra consisting of the representations $(1)\oplus(7)$ and so our choice seems natural when considering fusion of the $\sVir_{10}$ model. We think it results in more natural expression for the final boundary states.

One consequence is that (unlike the situation in  \cite{GY}) our choice of representatives results in sets of states which only differ by factors
of $\sqrt 2$,
\be
 \kkett{(a,6)_\NS} = \sqrt 2 \: \kkett{(a,1)_\NS}
\;,\;\;
 \kkett{(a,3)_\NS} = \sqrt 2 \: \kkett{(a,2)_\NS}
\;.
\label{eq:nsdef}
\ee
These two different ways of expressing the same boundary state may
seem redundant, but it helps a great deal when it 
comes to providing consistent descriptions of all the possible
boundary states for \SVIR2.

We also define states $\kkett{(a,b)_\wtNS}$ in a slightly
different way to \cite{GY}, as
\be
  \kkett{(a,b)_\wtNS} 
= (-1)^F \kkett{(a,b)_\NS}
\;.
\label{eq:nstdef}
\ee
These differ from the states defined in \cite{GY} by an extra sign for
each of the Ishibashi states which corresponds to a fermionic highest
weight state, that is for the states
$\ket{r,s}$ with
$(r,s)\in\{(1,5),(1,7),(3,1),$ $(3,11)$$,(5,5),(5',5)\}$
which again simplifies the identification of the known boundary states.

\subsection{Identifying known defects}
\label{sec:summary}

With the definitions \eqref{eq:nsdef} and \eqref{eq:nstdef} we can
identify all the known defects in $\sTCIM$.
These split into two sets, those with $\abgd=1$ for
which we need the highest-weight state map $\rho'$, supplemented by suitable
choices of signs for the 
descendent states,
and those with
$\abgd=-1$ for which we need the map $\rho$, shown in 
in table
\ref{tab:knownd}.

\begin{table}[htb]
\[
\renewcommand{\arraystretch}{1.2}\small
\begin{array}{c|c@{\,,\;}r}
\multicolumn{3}{c}{\abgd=-1,\, \alpha{=}\beta{=}\gamma{=}1,\,\delta{=}{-}1, \hbox{ map = $\rho_{+++-}$}}\\
\hline
\hbox{Defect} & \multicolumn{2}{c}{\hbox{Boundary states}}\\
\hline
\kkettbbraa{I_{\NS}}{I_{\NS}} & \sqrt 2 \: \kkett{(1,1)_\NS} & \kkett{(1,6)_\NS} \\
\kkettbbraa{I_{\NS}}{\varphi_{\NS}} & \sqrt 2 \: \kkett{(5,1)_\NS} & \kkett{(6,6)_\NS} \\
\kkettbbraa{\varphi_{\NS}}{I_{\NS}} & \sqrt 2 \: \kkett{(6,1)_\NS} & \kkett{(6,6)_\NS} \\
\kkettbbraa{\varphi_{\NS}}{\varphi_{\NS}} & \sqrt 2 \: \kkett{(3,1)_\NS} & \kkett{(3,6)_\NS} \\
\sqrt 2 (-1)^F & \sqrt 2 \: \kkett{(2,6)_\NS} & 2 \: \kkett{(2,1)_\NS} \\
\sqrt 2 (-1)^{\bar F} & \sqrt 2 \: \kkett{(2,6)_\wtNS} & 2 \: \kkett{(2,1)_\wtNS} \\
\sqrt 2 (-1)^F D_\varphi & \sqrt 2\: \kkett{(4,6)_\NS} & 2 \: \kkett{(4,1)_\NS} \\
\sqrt 2 (-1)^{\bar F} D_\varphi & \sqrt 2 \: \kkett{(4,6)_\wtNS} & 2 \: \kkett{(4,1)_\wtNS} \\
(-1)^F\kkettbbraa{I_{\NS}}{I_{\NS}}(-1)^F & \sqrt 2 \: \kkett{(1,1)_\wtNS} & \kkett{(1,6)_\wtNS} \\
(-1)^F\kkettbbraa{I_{\NS}}{\varphi_{\NS}}(-1)^F & \sqrt 2\:\kkett{(5,1)_\wtNS} & \kkett{(6,6)_\wtNS} \\
(-1)^F\kkettbbraa{\varphi_{\NS}}{I_{\NS}}(-1)^F & \sqrt 2\:\kkett{(6,1)_\wtNS} & \kkett{(6,6)_\wtNS} \\
(-1)^F\kkettbbraa{\varphi_{\NS}}{\varphi_{\NS}}(-1)^F & \sqrt
2 \: \kkett{(3,1)_\wtNS} & \kkett{(3,6)_\wtNS} \\
\end{array}
\;\;
\renewcommand{\arraystretch}{1.2}
\begin{array}{c|c@{\,,\;}r}
\multicolumn{3}{c}{\abgd=1,\,  \alpha{=}\beta{=}\gamma{=}\delta{=}1, \hbox{ map = $\rho'_{++++}$}}\\
\hline
\hbox{Defect} & \multicolumn{2}{c}{\hbox{Boundary states}}\\
\hline
\kkettbbraa{I_{\NS}}{I_{\R}} & \sqrt 2\:\kkett{(1,6)_\NS} & 2\:\kkett{(1,1)_\NS} \\
\kkettbbraa{I_{\NS}}{\varphi_{\R}} & \sqrt 2\:\kkett{(5,6)_\NS} & 2\:\kkett{(6,1)_\NS} \\
\kkettbbraa{\varphi_{\NS}}{I_{\R}} & \sqrt 2\:\kkett{(6,6)_\NS} & 2\:\kkett{(6,1)_\NS} \\
\kkettbbraa{\varphi_{\NS}}{\varphi_{\R}} & \sqrt 2\:\kkett{(3,6)_\NS} & 2\:\kkett{(3,1)_\NS} \\
\kkettbbraa{I_{\R}}{I_{\NS}} & \sqrt 2\:\kkett{(1,6)_\wtNS} & 2\:\kkett{(1,1)_\wtNS} \\
\kkettbbraa{I_{\R}}{\varphi_{\NS}} & \sqrt 2\:\kkett{(5,6)_\wtNS} & 2\:\kkett{(6,1)_\wtNS} \\
\kkettbbraa{\varphi_{\R}}{I_{\NS}} & \sqrt 2\:\kkett{(6,6)_\wtNS} & 2\:\kkett{(6,1)_\wtNS} \\
\kkettbbraa{\varphi_{\R}}{\varphi_{\NS}} & \sqrt 2\:\kkett{(3,6)_\wtNS} & 2\:\kkett{(3,1)_\wtNS} \\
\One & \sqrt 2 \: \kkett{(2,1)_\NS} &  \kkett{(2,6)_\NS} \\
(-1)^{F + \bar F} & \sqrt 2 \: \kkett{(2,1)_\wtNS} &  \kkett{(2,6)_\wtNS} \\
D_\varphi & \sqrt 2 \: \kkett{(4,1)_\NS} &\kkett{(4,6)_\NS} \\
(-1)^{F + \bar F} D_\varphi & \sqrt 2 \: \kkett{(4,1)_\wtNS} &  \kkett{(4,6)_\wtNS} \\
\end{array}
\]
\caption{
Identifications of the boundary states corresponding to the known defects 
} 
\label{tab:knownd}
\end{table}

Note that these two sets are not defined at the same time as they use
different embeddings; we cannot describe the defects $\One$ and
$\kkettbbraa{I_{\NS}}{I_{\NS}}$ as supersymmetric boundary conditions
for \SVIR2\ at the same time.

\begin{figure}[htb]

\pictures{
$$  
  \begin{picture}(440,355)
  \put(-20,0){\scalebox{.8}{\includegraphics{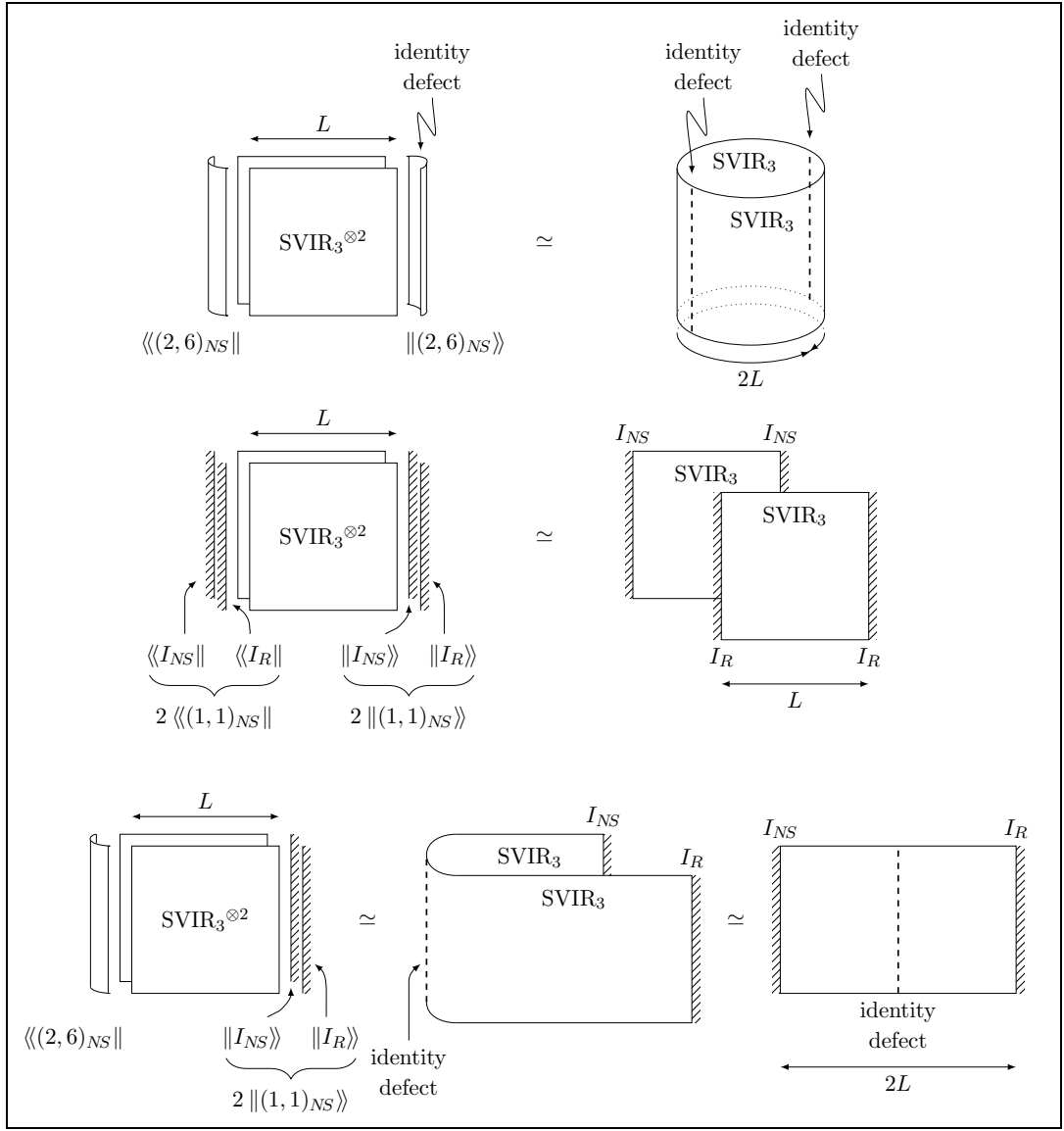}}}
  \end{picture}
$$
}{
\begin{center}
\framebox{%
\scalebox{0.8}{%
\begin{tikzpicture}
	\pgfmathsetmacro{\hshift}{3}
	\pgfmathsetmacro{\vshift}{5}
	\pgfmathsetmacro{\hshiftB}{5}
	\pgfmathsetmacro{\hshiftC}{0}
	\pgfmathsetmacro{\vshiftB}{8.5+\hshift}

	\draw (-2.5-\hshift,0) -- (0-\hshift,0) -- (0-\hshift,2.5) -- (-2.5-\hshift,2.5) -- (-2.5-\hshift,0);
	\draw (-2.5-\hshift,0.2) -- (-2.7-\hshift,0.2) -- (-2.7-\hshift,2.7) -- (-0.2-\hshift,2.7) -- (-0.2-\hshift,2.5);
	
	\draw (-3.1-\hshift,2.7) to[out=180,in=120] (-3.2-\hshift,2.6) to[out=-60,in=0] (-2.9-\hshift,2.5);
	\draw (-3.2-\hshift,0.1) to[out=-60,in=0] (-2.9-\hshift,0);
	\draw (-2.9-\hshift,2.5) -- (-2.9-\hshift,0);
	\draw (-3.2-\hshift,2.6) -- (-3.2-\hshift,0.1);
	\draw (-3.1-\hshift,2.7) -- (-3.1-\hshift,2.55);
	
	\draw (0.2-\hshift,2.7) to[out=180,in=130] (0.5-\hshift,2.65) to[out=-60,in=0] (0.4-\hshift,2.5);
	\draw (0.5-\hshift,0.1) to[out=-60,in=0] (0.4-\hshift,0);
	\draw (0.2-\hshift,0.2) to[out=180,in=120] (0.4-\hshift,0.1);
	\draw (0.5-\hshift,0.1) -- (0.5-\hshift,2.6);
	\draw (0.4-\hshift,0) -- (0.4-\hshift,2.5);
	\draw (0.2-\hshift,0.2) -- (0.2-\hshift,2.7);
	
	\draw[latex-latex] (-2.5-\hshift,3) -- (0-\hshift,3);
	\node[above] at (-1.25-\hshift,3) {$L$};
	
	\node[left,text height=1ex,text depth=.25ex] at (-2.5-\hshift,-0.5) {$\bbraa{(2,6)_\NS}$};
	\node[right,text height=1ex,text depth=.25ex] at (0-\hshift,-0.5) {$\kkett{(2,6)_\NS}$};
	
	\node at (-1.25-\hshift,1.25) {\SVIR2};
	
	\draw[latex-] (0.4-\hshift,2.8) to[out=90,in=120] (0.4-\hshift,3.4) -- (0.6-\hshift,3.1) to[out=-60,in=-90] (0.6-\hshift,3.7);
	\node[above,align=center] at (0.6-\hshift,3.7) {identity \\ defect};
	
	\draw (0+\hshift,2.5) ellipse (1.25 and 0.5);
	\draw (-1.25+\hshift,2.5) -- (-1.25+\hshift,0);
	\draw (-1.25+\hshift,0) arc (180:360:1.25 and 0.5);
	\draw [dotted] (-1.25+\hshift,0) arc (180:360:1.25 and -0.5);
	\draw (1.25+\hshift,0) -- (1.25+\hshift,2.5); 
	
	\draw[dotted] (-1.25+\hshift,-0.3) arc (180:360:1.25 and -0.5);
	\draw[] (-1.25+\hshift,-0.3) arc (180:360:1.25 and 0.5);
	\draw[-latex,,draw opacity=0] (1.4+\hshift,-0.3) -- (1+\hshift,-0.6);
	\draw[-latex,,draw opacity=0] (0+\hshift,-0.9) -- (1+\hshift,-0.62);
	\node[below] at (0+\hshift,-0.8) {$2L$};
	
	\node at (-0.1+\hshift,2.6) {\sTCIM};
	\node at (0.2+\hshift,1.6) {\sTCIM};
	
	\draw[dashed,thick] (-1+\hshift,-0.27) -- (-1+\hshift,2.23);
	\draw[dashed,thick] (1+\hshift,0.27) -- (1+\hshift,2.77);
	
	\draw[latex-] (-1+\hshift,2.4) to[out=90,in=120] (-1+\hshift,3.4) -- (-0.8+\hshift,3.1) to[out=-60,in=-90] (-0.8+\hshift,3.7);
	\node[above,align=center] at (-0.8+\hshift,3.7) {identity \\ defect};
	\draw[latex-] (1+\hshift,3) to[out=90,in=120] (1+\hshift,3.8) -- (1.2+\hshift,3.5) to[out=-60,in=-90] (1.2+\hshift,4.1);
	\node[above,align=center] at (1.2+\hshift,4.1) {identity \\ defect};
	
	\draw (-2.5-\hshift,0-\vshift) -- (0-\hshift,0-\vshift) -- (0-\hshift,2.5-\vshift) -- (-2.5-\hshift,2.5-\vshift) -- (-2.5-\hshift,0-\vshift);
	\draw (-2.5-\hshift,0.2-\vshift) -- (-2.7-\hshift,0.2-\vshift) -- (-2.7-\hshift,2.7-\vshift) -- (-0.2-\hshift,2.7-\vshift) -- (-0.2-\hshift,2.5-\vshift);
	
	\draw (-2.9-\hshift,2.5-\vshift) -- (-2.9-\hshift,0-\vshift);
	\node[fill,pattern=north east lines,scale=0.5,above,minimum width=142,rotate=90] at (-2.9-\hshift, 1.25-\vshift) {};
	\draw (-3.1-\hshift,2.7-\vshift) -- (-3.1-\hshift,0.2-\vshift);
	\node[fill,pattern=north east lines,scale=0.5,above,minimum width=142,rotate=90] at (-3.1-\hshift, 1.45-\vshift) {};
	
	\draw (0.4-\hshift,0-\vshift) -- (0.4-\hshift,2.5-\vshift);
	\node[fill,pattern=north east lines,scale=0.5,below,minimum width=142,rotate=90] at (0.4-\hshift, 1.25-\vshift) {};
	\draw (0.2-\hshift,0.2-\vshift) -- (0.2-\hshift,2.7-\vshift);
	\node[fill,pattern=north east lines,scale=0.5,below,minimum width=142,rotate=90] at (0.2-\hshift, 1.45-\vshift) {};
	
	\draw[latex-latex] (-2.5-\hshift,3-\vshift) -- (0-\hshift,3-\vshift);
	\node[above] at (-1.25-\hshift,3-\vshift) {$L$};
	
	\node[left,text height=1ex,text depth=.25ex] at (-3.1-\hshift,-0.8-\vshift) {$\bbraa{I_{\NS}}$};
	\node[right,text height=1ex,text depth=.25ex] at (-2.9-\hshift,-0.8-\vshift) {$\bbraa{I_{\R}}$};
	\draw[latex-] (-3.4-\hshift,0.5-\vshift) to[out=-140,in=90] (-3.6-\hshift,0.2-\vshift) -- (-3.6-\hshift,-0.4-\vshift);
	\draw[latex-] (-2.8-\hshift,0.1-\vshift) to[out=-30,in=90] (-2.5-\hshift,-0.3-\vshift) to[out=-90,in=60] (-2.5-\hshift,-0.4-\vshift);
	\draw [decorate,decoration={brace,mirror,amplitude=10pt}] (-4.2-\hshift,-1.1-\vshift) -- (-2-\hshift,-1.1-\vshift);
	\node[below] at (-3.1-\hshift,-1.5-\vshift) {$2\:\bbraa{(1,1)_\NS}$};
	
	\node[left,text height=1ex,text depth=.25ex] at (0.2-\hshift,-0.8-\vshift) {$\kkett{I_{\NS}}$};
	\node[right,text height=1ex,text depth=.25ex] at (0.4-\hshift,-0.8-\vshift) {$\kkett{I_{\R}}$};
	\draw[latex-] (0.2-\hshift,0.1-\vshift) to[out=-90,in=40] (0-\hshift,-0.3-\vshift) -- (-0.2-\hshift,-0.4-\vshift);
	\draw[latex-] (0.6-\hshift,0.5-\vshift) to[out=-40,in=90] (0.8-\hshift,-0.2-\vshift) -- (0.8-\hshift,-0.4-\vshift);
	\draw [decorate,decoration={brace,mirror,amplitude=10pt}] (-0.9-\hshift,-1.1-\vshift) -- (1.3-\hshift,-1.1-\vshift);
	\node[below] at (0.2-\hshift,-1.5-\vshift) {$2\:\kkett{(1,1)_\NS}$};
	
	\node at (-1.25-\hshift,1.25-\vshift) {\SVIR2};
	
	\draw (-0.5+\hshift,-0.5-\vshift) -- (2+\hshift,-0.5-\vshift) -- (2+\hshift,2-\vshift) -- (-0.5+\hshift,2-\vshift) -- (-0.5+\hshift,-0.5-\vshift);
	\draw (-0.5+\hshift,0.2-\vshift) -- (-2+\hshift,0.2-\vshift) -- (-2+\hshift,2.7-\vshift) -- (0.5+\hshift,2.7-\vshift) -- (0.5+\hshift,2-\vshift);
	
	\node[fill,pattern=north east lines,scale=0.5,above,minimum width=142,rotate=90] at (-0.5+\hshift, 0.75-\vshift) {};
	\node[fill,pattern=north east lines,scale=0.5,above,minimum width=142,rotate=90] at (-2.+\hshift, 1.45-\vshift) {};
	\node[fill,pattern=north east lines,scale=0.5,below,minimum width=142,rotate=90] at (2.+\hshift, 0.75-\vshift) {};
	\node[fill,pattern=north east lines,scale=0.5,below,minimum width=40,rotate=90] at (0.5+\hshift, 2.35-\vshift) {};
	
	\draw[latex-latex] (-0.5+\hshift,-1.25-\vshift) -- (2+\hshift,-1.25-\vshift);
	\node[below] at (0.75+\hshift,-1.25-\vshift) {$L$};
	
	\node at (-0.75+\hshift,2.3-\vshift) {\sTCIM};
	\node at (0.75+\hshift,1.6-\vshift) {\sTCIM};
	
	\node[above] at (-2+\hshift,2.7-\vshift) {$I_{\NS}$};
	\node[above] at (0.5+\hshift,2.7-\vshift) {$I_{\NS}$};
	\node[below] at (-0.5+\hshift,-0.5-\vshift) {$I_{\R}$};
	\node[below] at (2+\hshift,-0.5-\vshift) {$I_{\R}$};
	
	\draw (-2.5-\hshiftB,0-\vshiftB) -- (0-\hshiftB,0-\vshiftB) -- (0-\hshiftB,2.5-\vshiftB) -- (-2.5-\hshiftB,2.5-\vshiftB) -- (-2.5-\hshiftB,0-\vshiftB);
	\draw (-2.5-\hshiftB,0.2-\vshiftB) -- (-2.7-\hshiftB,0.2-\vshiftB) -- (-2.7-\hshiftB,2.7-\vshiftB) -- (-0.2-\hshiftB,2.7-\vshiftB) -- (-0.2-\hshiftB,2.5-\vshiftB);
	
	\draw (-3.1-\hshiftB,2.7-\vshiftB) to[out=180,in=120] (-3.2-\hshiftB,2.6-\vshiftB) to[out=-60,in=0] (-2.9-\hshiftB,2.5-\vshiftB);
	\draw (-3.2-\hshiftB,0.1-\vshiftB) to[out=-60,in=0] (-2.9-\hshiftB,0-\vshiftB);
	\draw (-2.9-\hshiftB,2.5-\vshiftB) -- (-2.9-\hshiftB,0-\vshiftB);
	\draw (-3.2-\hshiftB,2.6-\vshiftB) -- (-3.2-\hshiftB,0.1-\vshiftB);
	\draw (-3.1-\hshiftB,2.7-\vshiftB) -- (-3.1-\hshiftB,2.55-\vshiftB);
	
	\draw (0.4-\hshiftB,0-\vshiftB) -- (0.4-\hshiftB,2.5-\vshiftB);
	\node[fill,pattern=north east lines,scale=0.5,below,minimum width=142,rotate=90] at (0.4-\hshiftB, 1.25-\vshiftB) {};
	\draw (0.2-\hshiftB,0.2-\vshiftB) -- (0.2-\hshiftB,2.7-\vshiftB);
	\node[fill,pattern=north east lines,scale=0.5,below,minimum width=142,rotate=90] at (0.2-\hshiftB, 1.45-\vshiftB) {};
	
	\draw[latex-latex] (-2.5-\hshiftB,3-\vshiftB) -- (0-\hshiftB,3-\vshiftB);
	\node[above] at (-1.25-\hshiftB,3-\vshiftB) {$L$};
	
	\node[left,text height=1ex,text depth=.25ex] at (-2.5-\hshiftB,-0.8-\vshiftB) {$\bbraa{(2,6)_\NS}$};
	
	\node[left,text height=1ex,text depth=.25ex] at (0.2-\hshiftB,-0.8-\vshiftB) {$\kkett{I_{\NS}}$};
	\node[right,text height=1ex,text depth=.25ex] at (0.4-\hshiftB,-0.8-\vshiftB) {$\kkett{I_{\R}}$};
	\draw[latex-] (0.2-\hshiftB,0.1-\vshiftB) to[out=-90,in=40] (0-\hshiftB,-0.3-\vshiftB) -- (-0.2-\hshiftB,-0.4-\vshiftB);
	\draw[latex-] (0.6-\hshiftB,0.5-\vshiftB) to[out=-40,in=90] (0.8-\hshiftB,-0.2-\vshiftB) -- (0.8-\hshiftB,-0.4-\vshiftB);
	\draw [decorate,decoration={brace,mirror,amplitude=10pt}] (-0.9-\hshiftB,-1.1-\vshiftB) -- (1.3-\hshiftB,-1.1-\vshiftB);
	\node[below] at (0.2-\hshiftB,-1.5-\vshiftB) {$2\:\kkett{(1,1)_\NS}$};
	
	\node at (-1.25-\hshiftB,1.25-\vshiftB) {\SVIR2};
	
	\draw (-2+\hshiftC,-0.5-\vshiftB) -- (2+\hshiftC,-0.5-\vshiftB) -- (2+\hshiftC,2-\vshiftB) -- (-2+\hshiftC,2-\vshiftB);
	\draw (-2+\hshiftC,2.7-\vshiftB) -- (0.5+\hshiftC,2.7-\vshiftB) -- (0.5+\hshiftC,2-\vshiftB);
	\draw (-2+\hshiftC,2-\vshiftB) to[out=180,in=-90] (-2.5+\hshiftC,2.35-\vshiftB) to[out=90,in=180] (-2+\hshiftC,2.7-\vshiftB);
	\draw (-2+\hshiftC,-0.5-\vshiftB) to[out=180,in=-90] (-2.5+\hshiftC,-0.15-\vshiftB);
	
	\node[fill,pattern=north east lines,scale=0.5,below,minimum width=142,rotate=90] at (2+\hshiftC, 0.75-\vshiftB) {};
	\node[fill,pattern=north east lines,scale=0.5,below,minimum width=40,rotate=90] at (0.5+\hshiftC, 2.35-\vshiftB) {};
	
	\draw[dashed,thick] (-2.5+\hshiftC,-0.15-\vshiftB) -- (-2.5+\hshiftC,2.35-\vshiftB);
	
	\node at (-0.75+\hshiftC,2.3-\vshiftB) {\sTCIM};
	\node at (0+\hshiftC,1.6-\vshiftB) {\sTCIM};
	
	\draw[latex-] (-2.6+\hshiftC,0.5-\vshiftB) to[out=-140,in=90] (-2.8+\hshiftC,0.2-\vshiftB) -- (-2.8+\hshiftC,-0.8-\vshiftB);
	\node[below,align=center] at (-2.8+\hshiftC,-0.8-\vshiftB) {identity \\ defect};
	\node[above] at (0.5+\hshiftC,2.7-\vshiftB) {$I_{\NS}$};
	\node[above] at (2+\hshiftC,2-\vshiftB) {$I_{\R}$};
	
	\draw (-1.5+\hshiftB,0-\vshiftB) -- (2.5+\hshiftB,0-\vshiftB) -- (2.5+\hshiftB,2.5-\vshiftB) -- (-1.5+\hshiftB,2.5-\vshiftB) -- (-1.5+\hshiftB,0-\vshiftB);
	\node[fill,pattern=north east lines,scale=0.5,below,minimum width=142,rotate=90] at (2.5+\hshiftB, 1.25-\vshiftB) {};
	\node[fill,pattern=north east lines,scale=0.5,above,minimum width=142,rotate=90] at (-1.5+\hshiftB, 1.25-\vshiftB) {};
	
	\draw[dashed,thick] (0.5+\hshiftB,0-\vshiftB) -- (0.5+\hshiftB,2.5-\vshiftB);
	
	\node[above] at (-1.5+\hshiftB,2.5-\vshiftB) {$I_{\NS}$};
	\node[above] at (2.5+\hshiftB,2.5-\vshiftB) {$I_{\R}$};
	\node[below,align=center] at (0.5+\hshiftB,0-\vshiftB) {identity \\ defect};
	
	\draw[latex-latex] (-1.5+\hshiftB,-1.25-\vshiftB) -- (2.5+\hshiftB,-1.25-\vshiftB);
	\node[below] at (0.5+\hshiftB,-1.25-\vshiftB) {$2L$};
	
	\node at (-0.5,1.25) {$\simeq$};
	\node at (-0.5,1.25-\vshift) {$\simeq$};
	\node at (-3.5,1.25-\vshiftB) {$\simeq$};
	\node at (2.75,1.25-\vshiftB) {$\simeq$};
	
\end{tikzpicture}
}
}
\end{center}
}

\caption{
Different boundary conditions on \SVIR2\ result in different geometrical set-ups
for $\sTCIM$.
}

\label{fig:overlaps}
\end{figure}

As an example, we show here the overlap of the boundary states
$\sqrt 2\:\kkett{(2,1)_\NS}=\kkett{(2,6)_\NS}$ (representing the identity
defect) with
$\sqrt 2 \: \kkett{(2,1)_\wtNS}=\kkett{(2,6)_\wtNS}$ (representing the 
defect $(-1)^{F + \bar F}$), 
$2 \: \kkett{(1,1)_\NS} =
\sqrt2 \: \kkett{(1,6)_\NS}$ (representing $\kkettbbraa{I_{\NS}}{I_{\R}}$), 
and 
$2 \: \kkett{(1,1)_\wtNS} =
\sqrt2 \: \kkett{(1,6)_\wtNS}$ (representing $\kkettbbraa{I_{\R}}{I_{\NS}}$), 
all of which have $\abgd=1$. 
We have exactly the expected results:
\bea
\bbraa{(2,6)_\NS} q^H \kkett{(2,6)_\NS}
&{=}& 
\big(\chten_{1,1} + 
\chten_{1,5} + 
\chten_{1,7} + 
\chten_{1,11}  \nn \\
&& \hspace{3cm}{} + \chten_{3,1} + 
\chten_{3,5} + 
\chten_{3,7} + 
\chten_{3,11}
\big)(\tq)
\nn\\
&{=}& 
\big( \chthree_{1,1}(\tq) \big)^2 
+ 
\big( \chthree_{1,3}(\tq) \big)^2 
\;,
\\
\bbraa{(2,6)_\NS} q^H \kkett{(2,6)_\wtNS}
&{=}& 
2 
\big(\chten_{1,4} + 
\chten_{1,8} + 
\chten_{3,4} + 
\chten_{3,8}
\big)(\tq)
\nn\\
&{=}& 
2\big( \chthree_{1,2}(\tq) \big)^2 
+ 
2\big( \chthree_{1,4}(\tq) \big)^2 
\;,
\\
2 \: \bbraa{(2,6)_\NS} q^H \kkett{(1,1)_\NS}
&{=}& 
2 \big(\chten_{2,4} + \chten_{2,8}\big)(\tq)
\nn\\
&{=}& 
2\, \chthree_{1,4}(\sqrt{\tq}) 
\;,
\\
2 \: \bbraa{(2,6)_\NS} q^H \kkett{(1,1)_\wtNS}
&{=}& 
2 \big(
\chten_{2,1} + 
\chten_{2,5} + 
\chten_{2,7} + 
\chten_{2,11}
\big)(\tq)
\nn\\
&{=}& 
2\,\chthree_{1,4}(\sqrt{\tq}) \;,
\eea
where $H = (L_0 + \bar L_0 - 7/60)$, $q=\exp(-4\pi L)$ and $\tq =
\exp(-\pi/L)$. 
Note that the overlaps of $\kkett{(2,6)_\NS}$ 
with $2\:\kkett{(1,1)_\NS}$ and $2\:\kkett{(1,1)_\wtNS}$ are the same, 
$2\,\chthree_{1,4}(\sqrt{\tq})$, thanks to two different identities
relating the characters of $\sVir_{10}$ and $\sVir_3$. 
This is a function of $\sqrt{\tq}$ since
geometrically it corresponds to a strip of width $2L$, as shown in
figure \ref{fig:overlaps}.

However, if we consider defects with different values of $\abgd$ we do
not get sensible results. 
The overlap of the boundary state in \SVIR2 corresponding to the identity defect
with the boundary state corresponding to the factorising defect
$\kkettbbraa{I_{\NS}}{I_{\NS}}$ will give the partition function 
on the strip of width $2L$ and boundary conditions $I_{\NS}$ on both
sides, that is 
\be
\Tr_{\sTCIM} \left( q^H \: D_I \: \kkettbbraa{I_{\NS}}{I_{\NS}} \right)
= 
\chthree_{1,1}(\sqrt{\tq})
\;.
\ee
But $\chthree_{1,1}(\sqrt q) = q^{-7/480}( 1+q^{3/4}+q+q^{5/4}+q^{3/2}+\ldots )$ cannot be expressed
as a sum of characters of the $c=7/5$ algebra, and so it is not
possible for the two defects $D_I$ and $ \kkettbbraa{I_{\NS}}{I_{\NS}} $
to be represented as boundary states for $\sVir_{10}$ at the same
time.
If we look at table \ref{tab:knownd}, we see that $D_I$ corresponds to
$\kkett{(2,6)_\NS}$ defined with embedding $\iota_{++++}$ but
$\kkettbbraa{I_{\NS}}{I_{\NS}}$ corresponds to $\kkett{(1,6)_\NS}$ with
embedding $\iota_{+++-}$, and so their overlap being calculated as
\be
\bbraa{(2,6)_\NS} q^H \kkett{(1,6)_\NS}
=\sqrt 2\, \chthree_{1,4}(\sqrt{\tq})
\ee
has nothing to do with the required quantity.

\subsection{Identifying new defects}

Now that we have identified all the known defects, we can see that
they all correspond to the nodes $1$ and $6$ on the \ES\ diagram. If we
instead use the nodes $2$ and $3$ on the \ES\ diagram, we find new
defects which are neither topological nor factorising. 

The transmission coefficient was
defined in \cite{QRW} in such a way that a topological defect has $T=1$ and a
factorising boundary condition has $T=0$, and thus it lets us quickly
identify these amongst the boundary states. We can calculate the $T$
(transmission) coefficient and the 
defect entropy for these new defects using the expressions in appendix
\ref{app:states}. If the defect obtained from a boundary state
$\kkett\Psi$ of the following form,
\be
 \kkett\Psi = A \kett{(1,1)\eps} + B \kett{(1,5)\eps} + \ldots 
\;, 
\ee
and the sign is $\eta_{1,5}=-1$, then the transmission coefficient is
\be
 T = \tfrac 12 \Big[ 1 - \eps B/A \Big]
\;,
\ee
whether the map is $\rho_{+++-}$ or $\rho'_{++++}$.

The $g$ values are independent of the embedding and choice of signs
$\eta$ given by
\bea
 g \big( \kkett{(a,1)_{\NS / \wtNS}} \big) &=& 
 \begin{cases}
 \frac{S^{(8)}_{a1}}{\sqrt{S^{(8)}_{1,1}}} = \sqrt{1+\frac{1}{\sqrt{5}}} \: \sin \left( \frac{a\pi}{10} \right) & \text{for } a =1,2,3,4 \\
 \frac{S^{(8)}_{5,1}}{2\sqrt{S^{(8)}_{1,1}}} = \frac{1}{2}\sqrt{1+\frac{1}{\sqrt{5}}} & \text{for } a =5,6
 \end{cases}
 \\
 g \big( \kkett{(a,2)_{\NS / \wtNS}} \big) &=& \sqrt{2+\sqrt{3}} \quad g \big( \kkett{(a,1)_{\NS / \wtNS}} \big)
 \;,
 \\
 g \big( \kkett{(a,3)_{\NS / \wtNS}} \big) &=& (1+\sqrt{3}) \quad g \big( \kkett{(a,1)_{\NS / \wtNS}} \big)
 \;,
 \\
 g \big( \kkett{(a,6)_{\NS / \wtNS}} \big) &=& \sqrt{2} \quad g \big( \kkett{(a,1)_{\NS / \wtNS}} \big)
 \;,
\eea
where the last two relations follow from \eqref{eq:nsdef}.

Note that $\kkett\Psi$ and $(-1)^F\kkett\Psi$ have the same value of
$g$ and $T$.
We find that the boundary
states $\kkett{(a,b)_{\NS / \wtNS}}$ only take four different values for $T$ as in
table \ref{tab:gTSV2}, but a large range of $g$ values.  We
also list the $g$ values for the known topological and factorising defects
in \sTCIM\ in the same table.

If the $g$ value of a boundary state cannot be expressed as a sum of
the $g$ values of known topological and factorising defects, then this
boundary state must correspond to a ``new'' defect. 

\begin{table}
\[
\begin{array}{|c|cc|c|}
\hline
T & \multicolumn{2}{c|}{g} & \hbox{boundary states defined with
  $\iota_{+++-}$ and map $\rho$} \\\hline
&&& \\[\dimexpr-\normalbaselineskip+3pt]
1 & \sqrt 2 & 1.414.. 
  & 2\:\kkett{(2,1)_\NS}, \; 2\:\kkett{(2,1)_\wtNS}\\
 & \tfrac{1 + \sqrt 5}{\sqrt 2} & 2.288...
 & 2\:\kkett{(4,1)_\NS}, \; 2\:\kkett{(4,1)_\wtNS}\\\hline
0 & \left(\tfrac{5 - \sqrt 5}{10}\right)^{1/2} & 0.5257...
  & \kkett{(1,6)_\NS}, \; \kkett{(1,6)_\wtNS}\\ 
 & \left(\tfrac{5 + \sqrt 5}{10}\right)^{1/2} & 0.8506...
 & \kkett{(5,6)_\NS}, \; \kkett{(6,6)_\NS}, \; \kkett{(5,6)_\wtNS}, \; \kkett{(6,6)_\wtNS}\\ 
 & \left(\tfrac{5 + 2\sqrt 5}{5}\right)^{1/2} & 1.3763...
  & \kkett{(3,6)_\NS}, \; \kkett{(3,6)_\wtNS}\\ 
\hline
 \tfrac{\sqrt 3 - 1}{2}
  &\sqrt{2 + \sqrt 3}  & 1.9318... 
    &\kkett{(2,3)_\NS}, \; \kkett{(2,3)_\wtNS} \\
    && 3.1258...  &\kkett{(4,3)_\NS}, \; \kkett{(4,3)_\wtNS} \\
\hline
 \tfrac{3 - \sqrt 3}{2}
  &  & 1.4363...  &2\:\kkett{(1,2)_\NS}, \; 2\:\kkett{(1,2)_\wtNS} \\
  &  & 3.7603...  &2\:\kkett{(3,2)_\NS}, \; 2\:\kkett{(3,2)_\wtNS} \\
  &  & 2.3240...  &2\:\kkett{(5,2)_\NS}, \; 2\:\kkett{(6,2)_\NS}, \; 2\:\kkett{(5,2)_\wtNS}, \; 2\:\kkett{(6,2)_\wtNS} \\
\hline
\multicolumn{4}{c}{}\\
\hline
T & \multicolumn{2}{c|}{g} & \hbox{boundary states defined with 
$\iota_{++++}$  and map $\rho'$} \\\hline
&&& \\[\dimexpr-\normalbaselineskip+3pt]
1 & 1 & 1 
  & \kkett{(2,6)_\NS}, \; \kkett{(2,6)_\wtNS} \\
 & \tfrac{1 + \sqrt 5}{2} & 1.618... 
  & \kkett{(4,6)_\NS}, \; \kkett{(4,6)_\wtNS}\\\hline
0 & \left(\tfrac{5 - \sqrt 5}{5}\right)^{1/2} & 0.7434...
  & 2\:\kkett{(1,1)_\NS}, \; 2\:\kkett{(1,1)_\wtNS} \\ 
 & \left(\tfrac{5 + \sqrt 5}{5}\right)^{1/2} & 1.2030...
  & 2\:\kkett{(5,1)_\NS}, \; 2\:\kkett{(6,1)_\NS}, \;
    2\:\kkett{(5,1)_\wtNS}, \; 2\:\kett{(6,1)_\wtNS}\\ 
 & \left(\tfrac{10 - 2\sqrt 5}{5}\right)^{1/2} & 1.9465...
  & 2\:\kkett{(3,1)_\NS}, \; 2\:\kkett{(3,1)_\wtNS}\\ 
\hline
 \tfrac{\sqrt 3 - 1}{2}
  &1 + \sqrt 3  & 2.732.. 
&2\:\kkett{(2,2)_\NS}, \; 2\:\kkett{(2,2)_\wtNS} \\
    && 4.4205... 
    &2\:\kkett{(4,2)_\NS}, \; 2\:\kkett{(4,2)_\wtNS} \\
\hline
 \tfrac{3 - \sqrt 3}{2}
  &  & 1.0156... &\kkett{(1,3)_\NS}, \; \kkett{(1,3)_\wtNS} \\
  &  & 2.6589...  &\kkett{(3,3)_\NS}, \; \kkett{(3,3)_\wtNS} \\
  &  & 1.6433...  &\kkett{(5,3)_\NS}, \; \kkett{(6,3)_\wtNS} \\
\hline
\end{array}
\]
\caption{$T$ and $g$ values for the \SVIR2 boundary states}
\label{tab:gTSV2}
\end{table}

Again, these defects fall into two sets - those defined from the
boundary state using embedding
$\iota_{++++}$ and map $\rho'$, and those defined with embedding
$\iota_{+++-}$ and map $\rho$.
With each set, the boundary states satisfy Cardy's condition, that is, 
the overlaps of any two boundary states corresponding to the same embedding
$\rho$, or $\rho'$, are non-negative
integer combinations of characters of  $\sVir_{10}$. The overlaps
of states corresponding to different maps do not satisfy Cardy's condition.

Further, the overlaps involving the known (topological and
factorising) defects can be expressed in terms of the
characters of $\sVir_{3}$, but those involving the new defects can
not.

As an example, we consider the overlaps of the boundary states
$\sqrt 2 \: \kkett{(2,1)_\NS}=\kkett{(2,6)_\NS}$ (representing the identity
defect) with 
$\sqrt 2\:\kkett{(1,2)_\NS}=\kkett{(1,3)_\NS}$ and $\sqrt 2\:\kkett{(1,1)_\NS} =
\kkett{(1,6)_\NS}$ (representing $\kkettbbraa{I_{\NS}}{I_{\NS}}$) with $\sqrt
2\:\kkett{(2,2)_\NS} 
= \kkett{(2,3)_\NS}$.

We have
\bea
\bbraa{(2,6)_\NS} q^H \kkett{(2,6)_\NS}
&{=}& 
\big(\chten_{1,1} + 
\chten_{1,5} + 
\chten_{1,7} + 
\chten_{1,11} \nn \\
&& \hspace{3cm}{} + 
\chten_{3,1} + 
\chten_{3,5} + 
\chten_{3,7} + 
\chten_{3,11}
\big)(\tq)\;,
\nn\\
&{=}& 
\big( \chthree_{1,1}(\tq) \big)^2 
+ 
\big( \chthree_{1,3}(\tq) \big)^2 
\\
\bbraa{(2,6)_\NS} q^H \kkett{(1,3)_\NS}
&{=}& 
\big(
\chten_{2,2} + 
\chten_{2,4} + 
2\,\chten_{2,6} + 
\chten_{2,8} + 
\chten_{2,10}
\big)(\tq)\;,
\\
\bbraa{(1,6)_\NS} q^H \kkett{(1,6)_\NS}
&{=}& 
\big(\chten_{1,1} + 
\chten_{1,5} + 
\chten_{1,7} + 
\chten_{1,11} 
\big)(\tq)
\nn\\
&{=}& 
\big( \chthree_{1,1}(\tq) \big)^2 \;,
\\
\bbraa{(1,6)_\NS} q^H \kkett{(2,3)_\NS}
&{=}& 
\big(
\chten_{2,2} + 
\chten_{2,4} + 
2\,\chten_{2,6} + 
\chten_{2,8} + 
\chten_{2,10}
\big)(\tq)\;,
\eea
where $H = (L_0 + \bar L_0 - 7/60)$, $q=\exp(-4\pi L)$ and $\tq =
\exp(-\pi/L)$. 

Since $\hten_{2,2} =\tfrac{1}{80} \neq \hthree_{r,s} + \hthree_{r',s'}$ for any $(r,s), \: (r',s')$ in $\sVir_3$, 
the overlap $\bbraa{(2,6)_\NS} q^H \kkett{(1,3)_\NS}$ 
cannot be expressed as a sum of products of
characters 
$\chthree_{r,s}(\tq) \: \chthree_{r',s'}(\tq)$. 
In addition, since 
$\hten_{2,2} - \tfrac{7}{120} = - \tfrac{11}{240} \neq \frac 12
(\hthree_{r,s} - \tfrac{7}{240})$ for any $(r,s)$ in $\sVir_3$, it cannot be expressed as a sum of
characters $\chthree_{r,s}(\sqrt{\tq})$.

Note that $\bbraa{(2,6)_\NS} q^H \kkett{(1,3)_\NS} = 
\bbraa{(1,6)_\NS} q^H \kkett{(2,3)_\NS}$, which suggest that these overlaps 
are related by the insertion of a topological defect in the doubled
model labelled by the Dynkin nodes $(2,1)$. 

Just for reference, we give the overlaps of the new boundary states with
themselves to show that they satisfy Cardy's condition, but also cannot be
expressed in terms of characters of $\sVir_3$:
\bea
\bbraa{(1,3)_\NS} q^H \kkett{(1,3)_\NS}
&{=}& 
\big(\chten_{1,1} + 
2 \chten_{1,3} + 
3 \chten_{1,5} + 
3 \chten_{1,7} + 
2 \chten_{1,9} + 
\chten_{1,11}
\big)(\tq)\;,
\nn\\
\bbraa{(2,3)_\NS} q^H \kkett{(2,3)_\NS}
&{=}& 
\big(\chten_{1,1} + 
2 \chten_{1,3} + 
3 \chten_{1,5} + 
3 \chten_{1,7} + 
2 \chten_{1,9} + 
\chten_{1,11} 
\nn\\
&{+}& 
\big(\chten_{3,1} + 
2 \chten_{3,3} + 
3 \chten_{3,5} + 
3 \chten_{3,7} + 
2 \chten_{3,9} + 
\chten_{3,11} 
\big)(\tq)\;,
\nn\\
\eea

\subsubsection{New factorising defects in \sTCIM}

While the boundary state
$\kkett{(1,6)_\wtNS}$ can be identified as 
the defect $(-1)^F\kkettbbraa{I_{\NS}}{I_{\NS}}(-1)^F$, this is not
actually the product of two boundary states in $\sTCIM$.
The state $(-1)^F\kkett{I_{\NS}}$ does not satisfy Cardy's constraint -
for example,
its overlap with $\kkett{I_{\NS}}$ is not an integer combination of
characters in the crossed channel:
\be
 \bbraa{I_{\NS}} q^H (-1)^F \kkett{I_{\NS}}
= \sqrt 2\, \chthree_{1,4}(\tq)
\;.
\ee
The defect 
 $(-1)^F\kkettbbraa{I_{\NS}}{I_{\NS}}(-1)^F$
does however satisfy the constraint - for example
\be
  \bbraa{(1,6)_\NS} q^H \kkett{(1,6)_\wtNS}
= \bbraa{I_{\NS}} q^H (-1)^F \kkettbbraa{I_{\NS}}{I_{\NS}} (-1)^F q^H \kkett{I_{\NS}}
= 2\, \chthree_{1,4}(\tq)^2
\;.
\ee

Conversely, the factorising defect $\kkettbbraa{I_\R}{I_\R}$ does not
arise in the tables \ref{tab:knownd}, 
The resolution seems to be that 
these factorising defects are not fundamental and instead we have
\be
  \kkett{(1,6)_\wtNS} \simeq (-1)^F \kkett{I_{\NS}}\bbraa{I_{\NS}}(-1)^F
\;,\;\;
  \kkett{I_{\R}}\bbraa{I_{\R}} = 2 (-1)^F \kkett{I_{\NS}}\bbraa{I_{\NS}}(-1)^F
 \simeq 2 \kkett{(1,6)_\wtNS}
\;.
\ee
This illustrates the possibility that each known factorising and
topological defect in \sTCIM\ gives rise to a superconformal boundary
state in \SVIR2, but the converse need not to be true.

\section{Boundary states in \SVIR2 and extended algebras}
\label{sec:newboundaries}

The boundary states we have discussed can be understood from the point
of view of extended superconformal algebras.
There are two relevant algebras, $\WM(3/2)$ and $\WM(10)$ which is a
subalgebra of $\WM(3/2)$.

\subsection{The algebra $\WM(3/2)$}

The first case to consider is boundary states which preserve the whole
algebra $\WM(3/2)$. Since this is the same as $\sVir_3\otimes\sVir_3$,
we expect to recover the known topological and factorised
defects.
This algebra contains not only the superconformal generator, $G(z)$,
but a fermionic primary field of weight 3/2, $\cW^{(3/2)}(z)$ and its
superpartner of weight 2, $\cT(z)$.
Details are given in appendix \ref{app:chiralalgebra}.

As with the field $G(z)$, we have a choice for the gluing conditions
of the field $\cW^{(3/2)}$, so that we can define gluing conditions
$(\eps,\eps')$ where the Ishibashi state 
$\kett{(h,\tilde h)\eps,\eps'}$ satisfies
\be
( G_m + i \eps \bar G_{-m}) \kett{(h,\tilde h)\eps,\eps'} = 0
\;,\;\;
( \cW^{(3/2)}_m + i \eps' \bar \cW^{(3/2)}_{-m}) \kett{(h,\tilde h)\eps,\eps'} = 0
\;.
\ee

The \SVIR2\ model can be thought of as a model of $\WM(3/2)$ in two
ways.
With the embedding $\abgd=1$, the partition function is diagonal in the
four characters of this algebra,
\be
Z = |\chi_1|^2 + |\chi_3|^2 + |\chi_5|^2 + |\chi_{5'}|^2
\;.
\ee
With the embedding $\abgd=-1$, the partition function is not diagonal, but is instead
\be
Z = |\chi_1|^2 + |\chi_3|^2 + \chi_5 \bar\chi_{5'} + \chi_{5'}\bar\chi_5
\;.
\ee
Either way, we would expect to have four Ishibashi states for each set
of gluing conditions, and hence four boundary states per gluing
condition, but this is not quite the case.
The $\WM(3/2)$ algebra relations include
\be
 \{ G_m,\cW^{(3/2)}_n\}
 = 2 \cT_{m+n}
\ee
and so an Ishibashi state $\kett{(h,\tilde h)\eps,\eps'}$ satisfies
\be
( \cT_0 - \eps \eps' \bar \cT_0 ) \kett{(h,\tilde h)\eps,\eps'} = 0
\;.
\ee
This means that for either embedding, there are only two choices of
gluing conditions for the representations with $h=\tfrac 1{10}$,
so that rather than having 16 Ishibashi states, in fact we only have 12
different Ishibashi states. 

This means we will have 12 independent combinations of these Ishibashi
states into boundary states, which is exactly what we find. There are
12 mutually consistent boundary states which preserve this algebra with
$\abgd=1$ and 12 (different) consistent states with $\abgd=-1$ which
correspond to the known topological and factorising boundary
defects of \sTCIM.

Since we can choose $\cW^{(3/2)}$ and $\bar \cW^{(3/2)}$ so that 
\be
\ket{1,5} = i \eta_{1,5} \cW^{(3/2)}_{-3/2} \: \bar \cW^{(3/2)}_{-3/2} \ket 0
\;,
\ee
then we have
\be
 \cW^{(3/2)}_{3/2} \ket{1,5} = i \eta_{1,5} \bar \cW^{(3/2)}_{-3/2} \ket{0}
\;,
\ee
and so the coefficients of $\kett{1,1)}$ and $\kett{1,5}$ are
related by this gluing condition.
As we can read off table \ref{tab:nums}, 
$\Psi^{(a,b)}_{(1,5)} =
\Psi^{(a,b)}_{(1,1)}$ for $(a,b)$ equal to $(2,6)$ and $(4,6)$, and 
$\Psi^{(a,b)}_{(1,5)} = - \Psi^{(a,b)}_{(1,1)}$ 
for $(a,b)$ equal to $(1,1)$, $(3,1)$, $(5,1)$ and $(6,1)$.

\subsection{The algebra $\WM(10)$}

When we turn to the algebra $\WM(10)$ which has a superprimary field
$\cW$ of weight 10. Since this algebra is invariant under 
$\cW^{(10)}\to - \cW^{(10)}$, we can again choose the gluing condition
\be
 \cW^{(10)}_m \kett h = \pm \bar\cW^{(10)}_{-m} \kett h
\;.
\ee
Since we can choose $\cW^{(10)}$ and $\bar \cW^{(10)}$ so that 
\be
\ket{(1,11)} = \eta_{1,11} \cW^{(10)}_{-10} \bar \cW^{(10)}_{-10} \ket 0
\;,
\ee
then we have
\be
 \cW^{(10)}_{10} \ket{(1,11)} = \eta_{1,11} \bar \cW^{(10)}_{-10} \ket{0}
\;,
\ee
and so the coefficients of $\kett{(1,1)}$ and $\kett{(1,11)}$ are
related by this gluing condition.
As we can read off table \ref{tab:nums}, 
$\Psi^{(a,b)}_{(1,11)} =
\Psi^{(a,b)}_{(1,1)}$ for boundaries $(a,b)$ in $\cI_e$ with $b=2$ and $b=6$, 
but $\Psi^{(a,b)}_{(1,11)} = - \Psi^{(a,b)}_{(1,1)}$ 
for $b=1$ and $b=3$.
Likewise, the coefficients
$\Psi^{(a,b)}_{(1,5)}=\pm\Psi^{(a,b)}_{(1,7)}$,
$\Psi^{(a,b)}_{(3,1)}=\pm\Psi^{(a,b)}_{(3,11)}$
and
$\Psi^{(a,b)}_{(3,5)}=\pm\Psi^{(a,b)}_{(3,7)}$.

There are now 8 different representations of $\WM(10)$ appearing in
\SVIR2, as each representation of $\WM(3/2)$ splits into exactly two
representations of $\WM(10)$. By the same reasoning as for the algebra
$\WM(3/2)$, 4 of these representations will allow all four choices of
gluing condition but 4 will only allow 2 choices, so that there are 24
different Ishibashi states, leading to two sets of 24 mutually
consistent boundary states. These are exactly the full set of boundary
states we have found. We can say that the boundary states we have
discussed in this paper are precisely those which preserve the algebra
$\WM(10)$.

\section{Defects in TCIM from defects in \SVIR2}
\label{sec:tcim}

We have found a set of non-topological, non-factorising defects in
\sTCIM\ from the boundary states $\kett{(a,2)\NS/\wtNS}$ and
$\kett{(a,3)\NS/\wtNS}$ in \SVIR2. We can now use these to construct
defects in TCIM by using the interface operators constructed in section
\ref{sec:tciminterface}: if $\hat D_{\sTCIM}$ is a defect in \sTCIM,
then 
$D_{TCIM}$ defined by
\be
  \hat D_{TCIM} = I\cdot \hat D_{\sTCIM}\cdot I^\dagger
\;,
\ee
is a defect in TCIM.
The $T$ and $g$ values of $\hat D_{TCIM}$ are easy to find, they are
just
\be
  T(\hat D_{TCIM}) = T(\hat D_{\sTCIM})
\;,\;\;
  g(\hat D_{TCIM}) = 2 g(\hat D_{\sTCIM})
\;.
\ee

It is very unlikely that this defect in TCIM is fundamental - it is
instead very likely that it is the superposition of two (possibly identical)
defects. This is exactly what happened in the free fermion case, where
the map from free fermion defects to Ising model defects always
resulted in the superposition of two [or more] defects of the same $g$ value, as
below,
and it is 
also true for the identifiable topological and factorising
defects in \sTCIM, eg 
\be
\begin{array}{l}
I \cdot D_1 \cdot I^\dagger = D_0 + D_{3/2} \;,\\
I \cdot D_{\varphi} \cdot I^\dagger = D_{1/10} + D_{3/5} \;,\\
I \cdot \sqrt{2} (-1)^F \cdot I^\dagger = 2 D_{7/16}  \;,\\
I \cdot \sqrt 2(-1)^F D_{\varphi} \cdot I^\dagger = 2 D_{3/80} \;,\\
I \cdot \kett{I_{NS}}\braa{I_{NS}} \cdot I^\dagger = 
\left( \kett{B_0} + \kett{B_{3/2}}\right)
\left( \braa{B_0} + \braa{B_{3/2}}\right)
\;,\\
\end{array}
\ee

When we come to the new defects in \sTCIM, we cannot say for certain
whether they are fundamental or not, but given the results above it is
very likely that they are not.
Looking at table \ref{tab:gTSV2}, the simplest conformal non-topological
defects we can construct come 
from the boundary states $\kkett{(1,3)\NS/\wtNS}$ with $g=2.03..$.
Let us denote these by $\cD^{\pm}$,
\be
 \cD^+ = I \cdot \rho(\kkett{(1,3)\NS}) \cdot I^\dagger
\;,\;\;
 \cD^- = I \cdot \rho(\kkett{(1,3)\wtNS}) \cdot I^\dagger
\;.
\label{eq:cdcddef}
\ee
From the fact that $\kkett{(1,3)\wtNS} = (-1)^F \kkett{(1,3)\NS}$, it
follows that $ \rho(\kkett{(1,3)\wtNS}) = (-1)^F
\rho(\kkett{(1,3)\NS}) (-1)^{\bar F}$ and so
\be
 \cD^- =  (-1)^F \cD^+ (-1)^F = \tfrac 12 D_{2,1} \cdot \cD^+ \cdot D_{2,1}
\;.
\label{eq:cdcd}
\ee
Using this relation, 
we have
\be
  \bra 0 \cD^+ \ket 0 
= \bra 0 \cD^- \ket 0 
\;,\;\;
    \bra 0 \cD^+ \ket{\tfrac 1{10}} 
= - \bra 0 \cD^- \ket{\tfrac 1{10}} 
\;,
\ee
and from the explicit expressions for the boundary state coefficients in
table \ref{tab:nums}, we thus see  that $\cD^\pm$ are distinct,
different operators.
\begin{align}
  \bra 0 \cD^+ \ket 0 
&= \bra 0 \cD^- \ket 0 
= 2 \Psi_{(1,1)}^{(1,3)} 
= 2 (15)^{-1/4} \left(\frac{(3+\sqrt 3)(\sqrt 5-1)}{2(\sqrt 3-1)}\right)^{1/2}
= 2.031..
\;,\;\;
\nn\\
    \bra 0 \cD^+ \ket{\tfrac 1{10}} 
&= - \bra 0 \cD^- \ket{\tfrac 1{10}} 
= - 2 (15)^{-1/4}\sqrt{ 2 \sqrt 3 - 3}
= - 0.692..
\;.
\end{align}

\subsection{Comparisons with the results of Gang and Yamaguchi}

We can now attempt to compare our results for non-topological,
non-factorising defects with those of Gang and
Yamaguchi. 
The simplest such defects we have found are $\cD^\pm$ defined in 
\eqref{eq:cdcddef} with $g=2.031..$ and $T=(3-\sqrt 3)/2$.
Looking at the list of proposed defects in section 3.2 of \cite{GY} the
only candidates to which we can hope to relate $\cD^\pm$ are those
from the boundary state $\ket{(1,3)}_{A_\pm}$ which have the same
value of $T$ and half the $g$-value.

From the definitions in equation (3.9) of \cite{GY}, the states
$\ket{(1,3)}_{A_\pm}$ have equal and opposite components in the Ramond
sector. Since our defects have no components in the Ramond sector, we
must consider the sum $\ket{(1,3)}_{A_+} + \ket{(1,3)}_{A_-}$ which
has the same $T$ and $g$ values as each of $\cD^\pm$. 

There are no precise definitions given in \cite{GY} on how to obtain a
defect from a boundary state, but we can see that
$\ket{(1,3)}_{A_+} + \ket{(1,3)}_{A_-}$ has zero overlap with the
states $\ket{(5,3,5)_{10}}$ and $\ket{(5',3,5)_{10}}$ 
(in the notation of \cite{GY}) which are equivalent to 
(in our notation) $\ket{(5,5)}$ and $\ket{(5',5)}$.
This means that whatever map $\tilde \rho$ is required to obtain a
defect from a boundary state in the formalism of \cite{GY}, the 
corresponding defect has zero matrix elements between $\bra 0$ and
$\ket{1/10}$ 
\be
\bra 0 
\tilde\rho\Big(\ket{(1,3)}_{A_+} + \ket{(1,3)}_{A_-}\Big) 
\ket{\tfrac 1{10}}
= 0
\;,
\ee
and so cannot be equal to either $\cD^+$ or $\cD^-$.  

Gang and Yamaguchi do not give details on the the precise map
$\tilde\rho$ required to obtain a defect from a boundary state in
their formalism.
We can be sure that the method we use
cannot work, as this will result in defects which are not GSO
projected, that is defects which are not maps from the TCIM to the TCIM. To
illustrate this, we consider the states used in \cite{GY} in
the representation 
\be
\Big[\cH_{1,3}^{3}\otimes\cH_{1,3}^3 \Big]^{\otimes 2}
= 
\Big[
\cH_{3,1}^{10}
\oplus
\cH_{3,5}^{10}
\oplus
\cH_{3,7}^{10}
\oplus
\cH_{3,11}^{10}
\Big]^{\otimes 2}
\ee
The paper \cite{GY} uses coset representations, and each highest
weight representation $\cH_{r,s}^{10}\equiv\cH_{10-r,12-s}^{10}$ of $\sVir_{10}$ splits into two coset
representations, 
\be
\begin{array}{ll}
\cH_{3,1}^{10} = \cH_{(3,1,1)_{10}} \oplus \cH_{(3,3,1)_{10}}
\;,\;\;
&\cH_{3,5}^{10} = \cH_{(3,1,5)_{10}} \oplus \cH_{(3,3,5)_{10}}
\;,\;\;
\\
\cH_{3,7}^{10} = \cH_{(7,1,5)_{10}} \oplus \cH_{(7,3,5)_{10}}
\;,\;\;
&\cH_{3,11}^{10} = \cH_{(7,1,1)_{10}} \oplus \cH_{(7,3,1)_{10}}
\;.\;\;
\end{array}
\ee
Only four of these coset representations appear in the boundary states
of \cite{GY}, with conformal weights as follows
\be
\begin{array}{c|c}
\text{Representation} & \text{Weight} \\ \hline
(3,3,5)_{10} & 1/5 \\
(3,1,1)_{10} & 6/5 \\
(7,3,5)_{10} & 6/5 \\
(7,1,1)_{10} & 31/5
\end{array}
\ee
In our terms, these can be identified with \SVIR2 descendants of the
$\sVir_{10}$ highest weight states,
\be
\begin{array}{rclrcl}
\ket{(3,3,5)_{10}} &=& \ket{(3,5)}\;, &
\ket{(7,3,5)_{10}} &=& \ket{(3,7)}\;, \\
\ket{(3,1,1)_{10}} &=& \ds\frac{i\eta}{7/5} G_{-1/2}\bar G_{-1/2}\ket{(3,1)} \;, &
\ket{(7,1,1)_{10}} &=& \ds\frac{i\eta'}{57/5} G_{-1/2}\bar G_{-1/2}\ket{(3,11)} \;,\\
\end{array}
\label{eq:4states}
\ee
where $\eta$ and $\eta'$ are undetermined signs.
Further, given an embedding $\iota_\abgd$, the states $\ket{(3,7)}$
and $\ket{(3,1)}$ can be identified from appendix \ref{app:states} as
\be
\begin{array}{rcl}
\ket{(3,1)}
&=& \ds\frac{i\eta_{3,1}}{2/5}
  (\alpha G^1_{-1/2} - \beta G^2_{-1/2})
  (\gamma\bar G^1_{-1/2} - \delta\bar G^2_{-1/2})
  \ket{(3,5)} \;,
\\
\ket{(3,7)}
&=& \ds \frac{\eta_{3,7}}{7/5}
 (L^1_{-1} - L^2_{-1} + \frac{\alpha\beta}{1/5}G^1_{-1/2}G^2_{-1/2})
 (\bar L^1_{-1} - \bar L^2_{-1} 
    + \frac{\gamma\delta}{1/5}\bar  G^1_{-1/2}\bar G^2_{-1/2})
  \ket{(3,5)}
\;.
\end{array}
\ee
This means that the state $\ket{(3,1,1)_{10}}$ is
\be
 \ket{(3,1,1)_{10}} = 
\ds \frac{-\eta\,\eta_{3,1}}{(2/5)(7/5)}
 (L^1_{-1} - L^2_{-1} -2{\alpha\beta}G^1_{-1/2}G^2_{-1/2})
 (\bar L^1_{-1} - \bar L^2_{-1} 
    -2{\gamma\delta}\bar  G^1_{-1/2}\bar G^2_{-1/2})
  \ket{(3,5)}
\;.
\ee
Putting these together with the results in appendix \ref{app:states},
and the fact that the boundary state 
\be
  \kett{(3,3,5)_{10}}
= \ket{(3,5)} + \frac{1}{2/5}L_{-1}\bar L_{-1}\ket{(3,5)} + \ldots
\;,
\ee
we can find the expansion up to level one of defect given by a
combination of boundary states constructed from the 
four states \eqref{eq:4states}:
\newcommand{\kb}{ \ketbra {\tfrac 1{10}}{\tfrac 1{10}} }
\begin{align}
  \kett\Psi 
&= 
  A \kett{(3,3,5)_{10}}
+ B \kett{(3,1,1)_{10}}
+ C \kett{(7,3,5)_{10}}
+ D \kett{(7,1,1)_{10}}
\;,
\\
\rho_\abgd\Big({\kett\Psi}\Big)
&=
   A \kb
\nn\\
&+ \Big( 
\frac{A}{2/5}
- \frac{B \eta \eta_{3,1}}{(2/5)(7/5)}
+ \frac{C \eta_{3,7}}{7/5}
\Big)
\Big[ 
L_{-1}\bar L_{-1} \kb
+  \kb \bar L_{1} L_{1}
\Big]
\nn\\
&+ \Big( 
\frac{A}{2/5}
+ \frac{B \eta \eta_{3,1}}{(2/5)(7/5)}
- \frac{C \eta_{3,7}}{7/5}
\Big)
\Big[ 
L_{-1} \kb L_1
+  \bar L_{-1} \kb \bar L_{1}
\Big]
\nn\\
&
+ i \alpha\beta
\frac{( B \eta \eta_{3,1} + C \eta_{3,7})}{7/25}
\Big[
  G_{-1/2}\kb \bar G_{1/2} L_1 
- \bar L_{-1} G_{-1/2}\kb \bar G_{1/2} \Big]
\Big]
\nn\\
&
+ i \gamma\delta 
\frac{( B \eta \eta_{3,1} + C \eta_{3,7})}{7/25}
\Big[
   L_{-1} \bar G_{-1/2}\kb  G_{1/2} 
-  \bar G_{-1/2}\kb \bar G_{1/2} \bar L_1 
\Big]
\nn\\
&
+ \abgd
\Big( 
\frac{2B \eta \eta_{3,1}}{7/25}
- \frac{C \eta_{3,7}}{7/125}
\Big)
\Big[
G_{-1/2}\bar G_{-1/2} \kb \bar G_{1/2} G_{1/2}
\Big]
+\ldots
\label{eq:lots}
\end{align}
The expression \eqref{eq:lots} is only GSO projected if 
$B\eta\eta_{3,1}+C\eta_{3,7} = 0$, otherwise it is not.
We can fix $\eta\eta_{3,1}$ and $\eta_{3.7}$ by comparing 
\eqref{eq:lots} with equation (3.20) of \cite{GY}.
Equation (3.20) says that the expression \eqref{eq:lots} should be
purely transmitting for $A=B=1,C=-1$ and purely reflecting for
$B=-1,A=C=1$, from which we 
deduce that $\eta\eta_{3,1} = \eta_{3.7}=1$. 
We can now decide if the defects arising from the boundary states of
\cite{GY} are GSO projected or not by looking at the ratio of the
coefficients $B$ and $C$ of the states $\kett{(3,1,1)_{10}}$ and
$\kett{(7,3,5)_{10}}$.
If this ratio is $-1$, the resulting defect can be GSO projected, if it
is not $-1$ then it is not GSO projected:
\be
\begin{array}{c|c}
B = - C,\; \text{GSO projected} &
B \neq - C,\; \text{not GSO projected} \\ \hline
\ket{(2,6)}_{A_\pm}, \ket{(4,6)}_{A_\pm}
& 
\ket{(1,3)}_{A_\pm}, \ket{(3,3)}_{A_\pm},
\ket{(5,3)}_{A_\pm}, \ket{(6,3)}_{A_\pm}
\\
\ket{(1,1)}_B,
\ket{(3,1)}_B,
\ket{(5,1)}_B,
\ket{(6,1)}_B
&
\ket{(2,2)}_B,
\ket{(4,2)}_B
\end{array}
\ee
Those which are GSO projected correspond to topological or factorising
defects; none of the ``new'' defects proposed in \cite{GY} lead to GSO
projected defects in our formalism, and so it is difficult for us to
make a stronger comparison with the proposals of \cite{GY}.

\section{Conclusions}
\label{sec:conclusions}

We have constructed GSO-projected defects in the tri-critical Ising model from
defects in the Neveu-Schwarz sector of the supersymmetric tri-critical
Ising model using interface operators. Our construction uses many
elements from the paper of Gang and Yamaguchi \cite{GY} but in the end
the defects we propose are not the same as theirs. 
There is some doubt over the complete validity of their approach as it
leads to factorised defects outside the normal classification, and
using our methods would result in the new defects proposed in
\cite{GY} not being properly GSO projects, but we
must stress that we have not shown that their non-topological defects
are incorrect, simply that they are not the same as ours. 

As part of our construction, we found evidence for two
non-commensurate sets of boundary states in \SVIR2 corresponding to
two inequivalent embeddings of $c=7/5$ algebra into two copies of
$c=7/10$. We have identified half of these boundary states as known
objects, the remaining half are new and lead to non-topological and
non-factorising defects in the tri-critical Ising model
We hope that these new defects will include the conjectured `C'
defect in \cite{KRW}.

We think it should be possible to derive the boundary states we
have proposed for \SVIR2 using topological field theory methods, in
the spirit of  
as well as compare our method with the construction of fermionic
models of Novak and Runkel \cite{NR} using topological field theory
methods which 
incorporate  spin structure.

The next steps would be to extend our approach of explicit
construction of boundary states and consideration of extended algebras
to include the Ramond  sector of the supersymmetric tri-critical Ising
model and obtain defects in the tri-critical Ising model using GSO
projection, rather than interface operators and compare these directly
with the results of \cite{GY}.

{\bf\large Acknowledgements}

We would like to thank Ingo Runkel for very helpful discussions and
for critical comments on the manuscript.

\appendix
\section{The chiral algebra of \SVIR2}
\label{app:chiralalgebra}

The chiral algebra of \SVIR2 is, of course, generated by two copies of
the superconformal algebra, with superconformal fields $G^1$ and
$G^2$. It can also be viewed as generated by a superconformal
generator $G(z)$  and three super-primary fields
$\cW^{(3/2)}$, $\cW^{(7/2)}$ and $\cW^{(10)}$, of weights $3/2$, $7/2$
and $10$ respectively. 

The choice of $G$ is fixed by the two signs $\alpha$ and $\beta$, 
\be
 G(z) = \alpha G^1(z) + \beta G^2(z)
\;.
\ee
With this choice, the super-primary fields $\cW^{(3/2)}$ and
$\cW^{(7/2)}$ can be defined by the states 
\bea
 \ket{\cW^{3/2}} &=& (\alpha G^1_{-3/2} - \beta G^2_{-3/2})\vac
\\
 \ket{\cW^{7/2}} &=&  \Big[
\alpha (L^1_{-2}G^1_{-3/2} - \frac 34 G^1_{-7/2}) 
 + \beta (L^2_{-2}G^2_{-3/2} - \frac 34 G^2_{-7/2}) 
\nn\\
&& - \frac {17}2( \beta L^1_{-2}G^2_{-3/2} + \alpha L^2_{-2}G^1_{-3/2})
\Big] \vac
\eea
-- the expression for $\ket{\cW^{10}}$ is too lengthy to give here,
and is not unique due to null states in the vacuum representation at $c=7/10$.
The states $\ket{G}$ and $\ket{\cW^{(7/2)}}$ are even under interchanging $G^1 \leftrightarrow G^2$, and 
$\ket{\cW^{(3/2)}}$ and $\ket{\cW^{(10)}}$ are odd.

Since $G$ and $\cW^{3/2}$ generate the whole chiral algebra on their
own, the fields $\cW^{7/2}$ and $\cW^{(10)}$ can be expressed in terms
of $G$ and $\cW^{3/2}$, but they can also be considered as fields in
their own right.

The super-partner to $\cW^{(3/2)}$ is $\cT(z)$ defined by 
\be
\ket{\cT} = \tfrac 12 G_{-1/2} \ket{ \cW^{(3/2)}}
 = (L^1_{-2} - L^2_{-2}) \ket 0
\;,\;\;
 \cT(z) =  T^1(z) -  T^2(z)
\;.
\ee
The representations of $\cW^{(3/2)}$ are thus labelled by the
eigenvalues $h$ of $L_0 = L^1_0 + L^2_0$ and 
$\tilde h$ of $\cT_0 =  L^1_0 - L^2_0$; in terms of the
eigenvalues $h^i$ of $L^i_0$ we clearly have 
\be
h = h^1 + h^2
\;,\;\;
\tilde h = h^1 - h^2
\;.
\ee
Since there are two NS representations of $\sVir_3$, there are four NS
representations of $\WM(3/2)$ at $c=7/5$ which we label $\{1,3,5,5'\}$
with highest weight  eigenvalues and characters as in table
\ref{tab:wm32}

\begin{table}
\[
\renewcommand{\arraystretch}{1.5}
\begin{array}{c|cl}
\hbox{label} & (h,\tilde h) & \multicolumn{1}{c}{\hbox{character}} \\
\hline
1 & (0,0) & \chi_1 = \chten_{1,1} + \chten_{1,5} + \chten_{1,7}
+ \chten_{1,11} = ( \chthree_{1,1} )^2
\\
3 & (\tfrac 15,0) & \chi_3 = \chten_{3,1} + \chten_{3,5} + \chten_{3,7}
+ \chten_{3,11} = ( \chthree_{1,3} )^2
\\
5 & (\tfrac 1{10},\tfrac 1{10}) & \chi_5 = \chten_{5,1} + \chten_{5,5}= \chthree_{1,1} \chthree_{1,3} 
\\
5' & (\tfrac 1{10},-\tfrac 1{10}) & \chi_{5'} = \chten_{5,1} +
\chten_{5,5}= \chthree_{1,1} \chthree_{1,3} 
\end{array}
\]
\caption{The NS representations of $\WM(3/2)$ at $c=7/5$.}
\label{tab:wm32}
\end{table}

There are two interesting subalgebras of this chiral algebra -
the super W-algebra SW(7/2), where the superconformal algebra is extended by the  single field $\cW^{7/2}$ of weight $7/2$, 
and the super W-algebra SW(10), where the superconformal algebra is extended by the  single field $\cW^{7/2}$ of weight $7/2$.

These can be proven to be closed algebras without calculating the commutation relations explicitly.

In the first case, $\WM(7/2)$ consists of all fields in \SVIR2 which are invariant under interchanging the fields $G^1$ and $G^2$. 
SW(7/2) was considered as an abstract super W-algebra in 
\cite{FS,H1,BEHH}
where it was shown to be consistent for $c=7/5$. 
 
In the second case, $\WM(10)$ is closed as the fusion rules of the
superconformal algebra at $c=7/2$ are $[1,11] * [1,11] = [1,1]$, that
is the field $\cW^{(10)}$ is a simple current for the superconformal
algebra. 
There are eight representations of $\WM(10)$ as each representation of
$\WM(3/2)$ is reducible into two representations of $\WM(10)$ with
labels and characters as in table \ref{tab:wm10}.
We give the value of $h$, the eigenvalue of $L_0$, only, the
eigenvalue of $\cW^{(10)}$ being too hard to calculate.

\begin{table}[htb]
\[
\renewcommand{\arraystretch}{1.5}
\begin{array}{c|cl}
\hbox{label} & h & \multicolumn{1}{c}{\hbox{character}} \\
\hline
1 & 0 & \chi_1 = \chten_{1,1}  + \chten_{1,1}
\\
\tilde 1 & \tfrac 32 & \chi_{\tilde 1} = \chten_{1,5}  + \chten_{1,7}
\\
3 & \tfrac 7{10} & \chi_3 = \chten_{3,1} + \chten_{3,11}
\\
\tilde 3 & \tfrac 15 & \chi_{\tilde 3} = \chten_{3,5} + \chten_{3,7} 
\\
5 & \tfrac {13}5 & \chi_5 = \chten_{5,1}
\\
\tilde 5 & \tfrac 1{10} & \chi_{\tilde 5} =  \chten_{5,5}
\\
5' & \tfrac {13}5 & \chi_{5'} = \chten_{5,1}
\\
\tilde 5' & \tfrac 1{10} & \chi_{\tilde 5'} =  \chten_{5,5}
\end{array}
\]
\caption{The NS representations of $\WM(10)$ at $c=7/5$}
\label{tab:wm10}
\end{table}

\section{Conventions for the free-fermion and the Ising model}
\label{app:ff}

If a free fermion is single-valued on a path around the origin,
then it has an expansion over modes $\psi_m$ with $m \in \mathbb Z +
1/2$,
\be 
\psi(z) = \sum_{m \in \mathbb Z + 1/2} \psi_m z^{-m-1/2}
\;.
\ee
These modes form the Neveu-Schwarz free-fermion algebra,
\be
\{ \psi_m , \psi_n \} = \delta_{m+n,0}
\ee 
which has a single unitary
irreducible representation, 
$\cH_{\NS}$ with character
\be
  \chi_{\NS}(q) 
= \Tr_{\cH_{\NS}}(q^{L_0 - c/24})
= q^{-1/48}\prod_{m=0}^\infty(1 + q^{m+1/2})
\;.
\ee
We also define $\chi_{\wtNS}$ as the trace with the insertion of
$(-1)^F$,
\be
  \chii_{\wtNS}(q) 
= \Tr_{\cH_{\NS}}(q^{L_0 - c/24}(-1)^F)
= q^{-1/48}\prod_{m=0}^\infty(1 - q^{m+1/2})
\;.
\ee

If a free fermion instead changes sign on a path around the origin, it
has an expansion in modes $\psi_m$, $m\in \mathbb Z$, 
\be 
\psi(z) = \sum_{m \in \mathbb Z} \psi_m z^{-m-1/2}
\;.
\ee
These satisfy the  Ramond free-fermion algebra
$\{\psi_m,\psi_n\}=\delta_{m+n,0}$.
This algebra has two unitary irreducible highest weight
representations, $\cH_{\R_\pm}$ with highest weights
$\ket{1/16}_{\pm}$ satisfying  
\be
 \psi_0 \ket{1/16}_{\pm} = \pm \frac{1}{\sqrt 2} \ket{1/16}_{\pm}
\;.
\ee
These have the same character
\be
  \chii_{\R}(q) 
= \Tr_{\cH_{\R_{\pm}}}(q^{L_0 - c/24})
= q^{1/24}\prod_{m=1}^\infty(1 + q^{m})
\;.
\ee
Note that there is an alternative and widely used convention which
includes a factor of $\sqrt 2$ in the definition,
$\chii_\R{}^\text{alternative} = \sqrt 2 \chii_\R$. This alternative
definition has the advantage of making the modular $S$ matrix in
equation \eqref{eq:s2} symmetric, but the disadvantage that it is not
the trace of $q^{L_0 - c/24}$ over a representation.

The three unitary irreducible highest weight representations of the
Virasoro algebra with $c=1/2$ have weights $h \in\{0, \: 1/2, \: 1/16\}$ and
their characters are
\be
\chiim_0 = \frac 12(\chii_{\NS} + \chii_{\wtNS})
\;,\;\;
\chiim_{1/2} = \frac 12(\chii_{\NS} - \chii_{\wtNS})
\;,\;\;
\chiim_{1/16} = \chii_{\R}
\;.
\ee

These characters are related under modular transformation 
$\tau \to -1/\tau$, that is
$q = \exp(2\pi i \tau) \to \tq = \exp(- 2 \pi i / \tau)$,
by
\be
\begin{pmatrix}
\chiim_{0} \\ \chiim_{1/2} \\ \chiim_{1/16}
\end{pmatrix}(q)
= 
\begin{pmatrix}
1/2 & 1/2 & 1/\sqrt 2 \\
1/2 & 1/2 & -1/\sqrt 2 \\
1/\sqrt 2 & -1/\sqrt 2 & 0
\end{pmatrix}
\begin{pmatrix}
\chiim_{0} \\ \chiim_{1/2} \\ \chiim_{1/16}
\end{pmatrix}(\tq)
\label{eq:s1}
\ee
\be
\begin{pmatrix}
\chii_{\NS} \\ \chii_{\wtNS} \\ \chii_\R
\end{pmatrix}(q)
= 
\begin{pmatrix}
1   & 0 & 0 \\
0   & 0 & \sqrt 2 \\
0   & 1/\sqrt 2  & 0
\end{pmatrix}
\begin{pmatrix}
\chii_{\NS} \\ \chii_{\wtNS} \\ \chii_\R
\end{pmatrix}(\tq)
\label{eq:s2}
\ee

\section{Conventions for the super Virasoro minimal models}
\label{app:tcim}

\subsection{Characters}

For the super Virasoro algebra with the central charge $0 \leq c <
3/2$, irreducible modules are unitary at discrete points, and
corresponding highest weight modules are labelled by $(c,h)$, both of
which are parametrised by the integers $m, \: r, \: s$ as 
\be 
 c = \frac{3}{2} \left( 1 - \frac{8}{m(m+2)} \right) 
\label{eq:SVirC} 
\ee
with $m=2,3,4,\dots$, and 
\be 
 h^{(m)}_{r,s} = \frac{((m+2)r-ms)^2 -  4}{8m(m+2)} 
 + \frac{1}{32} \left( 1 - (-1)^{r-s} \right) 
\label{eq:SVirH} 
\ee 
where $1 \leq r \leq m-1$ and $1 \leq s \leq m+1$. 
As usual, due to the Kac table symmetry $h^{(m)}_{r,s} =
h^{(m)}_{m-r,m+2-s}$, we need identification of the Kac labels $(r,s)
\sim (m-r,m+2-s)$. When we denote the super Virasoro algebra by
$\sVir_m$, it is understood to take the irreducible modules with
$(c,h)$ specified by \eqref{eq:SVirC} and \eqref{eq:SVirH}. When $r-s
\in 2\mathbb{Z}$, a representation is in the Neveu-Schwarz sector,
which corresponds to the modes $G_n$ with $n \in \mathbb{Z} + 1/2$,
and when $r-s \in 2\mathbb{Z}+1$, a representation is in the Ramond
sector, which corresponds to $G_n$ with $n \in \mathbb{Z}$.

Since \sVir is $Z_2$ graded by fermion parity of the generators ($L_n$
are bosonic and $G_m$ are fermionic), it is natural to consider $Z_2$
graded modules. We may introduce an operator $(-1)^F$ on a module and
take a basis, in which the highest weight state $\vec{h^{(m)}_{r,s}}$
is an eigenvector of $(-1)^F$ with the eigenvalue $\veps(r,s) = \pm1$
and $\{(-1)^F,G_m\} = 0$. 

For a highest weight module $\cH_\NS$ in the Neveu-Schwarz sector,
which is generated from $\vec{h^{(m)}_{r,s}}$, we define its
character by 
\be
	\ch^{m}_{r,s} (q) \equiv \Tr_{\cH_\NS} \:
        q^{L_0-\frac{c}{24}} = q^{-\frac{c}{24}}
        \sum_{n=-\infty}^\infty \left( q^{h(2mn+r,s)} - q^{h(2mn-r,s)}
        \right) \prod_{l=1}^\infty \frac{1+q^{l-\frac{1}{2}}}{1-q^l},
        \nn 
\ee
where $h(r,s) = h^{(m)}_{r,s}$ and the explicit formula on the right
hand side is given in \cite{GKO}. We also define the following
quantity associated to this module 
\begin{multline*}
	\chTi^{m}_{r,s} (q) \equiv \Tr_{\cH_\NS} \: (-1)^F q^{L_0-\frac{c}{24}} \\
 = \veps(r,s) \: q^{-\frac{c}{24}} \sum_{n=-\infty}^\infty (-1)^{mn} 
  \left( q^{h(2mn+r,s)} - (-1)^{rs} q^{h(2mn-r,s)} \right) 
  \prod_{l=1}^\infty \frac{1-q^{l-\frac{1}{2}}}{1-q^l},
\end{multline*}
where $\veps(r,s)=\pm$ is the eigenvalue of $(-1)^F$ on the highest
weight vector. Note that when $\veps(r,s)=1$, the series expansion of
$\chTi^{(m)}_{r,s}$ always starts from $q^{h-\frac{c}{24}}(1-\dots)$. 

Due to the zero modes, some care is needed when defining a character
for a highest module $\cH_\R$ in the Ramond sector, which is generated
from $\vec{h^{(m)}_{r,s}}$. When $h^{(m)}_{r,s} \neq c/24$, $L_0$
eigensubspaces of $\cH_\R$ are two-dimensional in which we can take
two basis vectors to carry opposite fermion parity. On the other hand,
$L_0$ eigensubspaces are one-dimensional when $h^{(m)}_{r,s} = c/24$,
which happens for $m \in 2\mathbb{Z}$ and $(r,s) = (m/2,m/2+1)$ -- we
call this representation the fixed point of a Kac table. We simply
define the following function 
\begin{gather}
	\ch^{m}_{r,s} (q) 
= q^{-\frac{c}{24}} \sum_{n=-\infty}^\infty 
  \left( q^{h(2mn+r,s)} - q^{h(2mn-r,s)} \right) 
  \prod_{l=1}^\infty \frac{1+q^l}{1-q^l}, \nn 
\end{gather}
for $r-s \in 2\mathbb{Z}+1$. The expansion of this function is of the
form $q^{h-\frac{c}{24}}(1+\dots)$. We view this as a ``character'' of
$\cH_\R$ in the sense that 
\begin{align*}
  \Tr_{\cH_\R} \: q^{L_0-\frac{c}{24}} 
&= 2\:\ch^{m}_{r,s} (q) \qquad \text{when } h \neq c /24 \\
  \Tr_{\cH_\R} \: q^{L_0-\frac{c}{24}} 
&= \ch^{m}_{r,s} (q) \qquad \text{when } h = c /24.
\end{align*}
We choose this normalisation so that it is
easy to see if Cardy's condition is satisfied or not.

\subsection{Modular transformations}

When we denote a representation of \sVir by its Kac label, we take the
``bottom half'' of the Kac table. That is, we take the following sets
of Kac labels 
\begin{gather}
	(r,s) \in \cI_\NS \text{ when } r+s \in 2\mathbb{Z} \text{ and } \bc 1 \leq s \leq m+1 &\text{for } r < \frac{m}{2} \\ 1 \leq s \leq \frac{m}{2} &\text{for } r=\frac{m}{2} \ec \nn \\
	(r,s) \in \cI_\R \text{ when } r+s \in 2\mathbb{Z}+1 \text{ and } \bc 1 \leq s \leq m+1 &\text{for } r < \frac{m}{2} \\ 1 \leq s \leq \frac{m}{2}+1 &\text{for } r=\frac{m}{2} \ec \nn
\end{gather}
Note that we take this convention only to make it clear that two
distinct Kac labels correspond to different representations, and there
is no ``physical'' reason to do so. For example, in the \DS--\ES\
theory, it may be more natural to take $r \in \{1,3,5,5',7,9 \}$ and
$s \in \{1,7\}$. 

We define the modular S-matrix elements as follows
\begin{align*}
	\ch^{\NS}_{r,s} (\tilde{q}) 
&= \sum_{(r',s')\in \cI_\NS} S^{[\NS,\NS]}_{(r,s)(r',s')} \: \ch^\NS_{r',s'} (q) \;, \\
	\chTi^\TiNS_{r,s} (\tilde{q}) 
&= \sum_{(r',s')\in \cI_\R} S^{[\TiNS,\R]}_{(r,s)(r',s')} \: \ch^\R_{r',s'} (q) \;, \\
	\ch^\R_{r,s} (\tilde{q}) 
&= \sum_{(r',s')\in \cI_\NS} S^{[\R,\TiNS]}_{(r,s)(r',s')} \: \chTi^\TiNS_{r',s'} (q) \;,
\end{align*}
which can be written explicitly as
\begin{align*}
	S^{[\NS,\NS]}_{(r_1,s_1)(r_2,s_2)} 
&= \frac{4}{\sqrt{m(m+2)}} 
  \sin\left(\frac{\pi r_1 r_2}{m}\right) 
  \sin\left(\frac{\pi s_1 s_2}{m+2}\right) \;, \\
	S^{[\TiNS,\R]}_{(r_1,s_1)(r_2,s_2)} 
&= \veps(r_1,s_1) \: (-1)^{\frac{r_1-s_1}{2}} 
  \frac{4 \sqrt 2\,G(r_2,s_2)}{\sqrt{m(m+2)}} 
  \sin\left(\frac{\pi r_1 r_2}{m}\right) 
  \sin\left(\frac{\pi s_1 s_2}{m+2}\right) \;, \\
	S^{[\R,\TiNS]}_{(r_1,s_1)(r_2,s_2)} 
&= \veps(r_2,s_2) \: (-1)^{\frac{r_2-s_2}{2}} 
  \frac{2\sqrt 2}{\sqrt{m(m+2)}} 
  \sin\left(\frac{\pi r_1 r_2}{m}\right) 
  \sin\left(\frac{\pi s_1 s_2}{m+2}\right) \;,
\end{align*}
where
\be
	G(r,s) = \bc \tfrac 12
  & \text{if } r=\frac{m}{2} \text{ and } s=\frac{m}{2}+1 \\ 
1 & \text{otherwise} \ec 
\;. \nn
\ee
Note that $\ch_i^\R (q)$ is not quite a modular function but $\sqrt{2}
\ch_i^\R (q)$ is. Therefore, the above modular S matrix is
non-symmetric but squares to $\One$. It is possible to make the S
matrix symmetric by introducing the modified Ramond character
$\sqrt{2} \ch_i^\R (q)$ when $i$ is not the fixed  point, but we do
not do that in this paper. 

In terms of $\hat{su}(2)_k$ modular S matrix elements
\be
	S^{(k)}_{ij} 
= \sqrt{\frac{2}{k+2}}\sin\left( \frac{ij\pi}{k+2} \right) \nn
\ee
where $i,j = 1,2,\dots,k+1$, \sVir S matrix elements can be written as
\begin{align*}
  S^{[\NS,\NS]}_{(r_1,s_1)(r_2,s_2)} 
 &= 2 S^{(m-2)}_{r_1 r_2} S^{(m)}_{s_1 s_2} \;, \\
  S^{[\TiNS,\R]}_{(r_1,s_1)(r_2,s_2)} 
 &= \veps(r_1,s_1) \:
       (-1)^{\frac{r_1-s_1}{2}} \: \sqrt 2 G(r_2,s_2) \: 
      2 S^{(m-2)}_{r_1 r_2} S^{(m)}_{s_1 s_2} \;, \\
	S^{[\R,\TiNS]}_{(r_1,s_1)(r_2,s_2)} &= \veps(r_2,s_2) \:
        (-1)^{\frac{r_2-s_2}{2}} \: \sqrt 2 S^{(m-2)}_{r_1 r_2} S^{(m)}_{s_1 s_2} \;.
\end{align*}

\subsection{Fermion parity assignment of NS highest weight vectors}
In most cases, a choice of $\veps(r,s)$ for NS highest weight
vectors is irrelevant. Usually, NS highest weight vectors
$\vec{r,s}$ are taken to be bosonic (i.e. $G_{-1/2} \vec{r,s}$
and $G_{-3/2} \vac$ are fermionic). However, we take the
following convention: 

\begin{itemize}
\item For $m$ odd,
\begin{align*}
  r+s \in 4\mathbb{Z} +2 
 &\rightarrow \vec{r,s} \text{ bosonic i.e. } \veps(r,s) = 1 \\
	r+s \in 4\mathbb{Z} 
  &\rightarrow \vec{r,s} \text{ fermionic i.e. } \veps(r,s) = -1
\end{align*}
(In particular, $\vec{1,3}=\vec{2,2}$ with $h=\frac{1}{10}$ is
fermionic in $m=3$.) 
\item For $m=10$ with the {\DS-\ES} bulk partition function,
\begin{align*}
	(r,s) = (1,5),(1,7),(3,1),(3,11),(5,5),(5,7),(7,1),(7,11),(9,5),(9,7) &\rightarrow \text{fermionic} \\
	\text{others} &\rightarrow \text{bosonic}
\end{align*}
\end{itemize}
The first choice for $m$ odd cases makes all the fusion coefficients
${\left( {N_{\TiNS\:\TiNS}}^\TiNS \right)_{ij}}^k$  non-negative. 
However, there is no obvious procedure to make all these
coefficients non-negative for $m$ even cases. The second choice for
$m=10$ comes from two observations: modular transformations of the
bulk partition function and character identities between $m=3$ and
$m=10$. 
\begin{itemize}
\item Consider the {\DS-\ES} bulk partition function,
\begin{align*}
Z &= 
  \frac{1}{2} \left( Z_\NS + Z_\TiNS \right) + Z_\R \\
Z_\NS &= 
  \left| \chten_{1,1} + \chten_{1,5} + \chten_{1,7} + \chten_{1,11} \right|^2 
+ \left| \chten_{3,1} + \chten_{3,5} + \chten_{3,7} + \chten_{3,11} \right|^2 
+ 2 \left| \chten_{5,1} + \chten_{5,5} \right|^2 \\
Z_\R &= 
  2\left| \chten_{1,4} + \chten_{1,8} \right|^2 
+ 2\left| \chten_{3,4} + \chten_{3,8} \right|^2 
+ 4 \left| \chten_{5,4} \right|^2
\end{align*}
If we demand $Z_\TiNS$ to have the same form as $Z_\NS$, we need
\begin{align*}
	\veps(1,1) = \veps(1,11) &= -\veps(1,5) = -\veps(1,7) \\
	\veps(3,1) = \veps(3,11) &= -\veps(3,5) = -\veps(3,7) \\
	\veps(5,1) &= -\veps(5,5)
\end{align*}
to ensure modular S transformation $\frac{1}{2} Z_\TiNS \leftrightarrow Z_\R$.
\item From the $\NS$ character identities between $m=3$ and $m=10$, if
  we want something similar for $\TiNS$ characters, that is  (again
  with $q$ real)
\begin{align*}
	 \left(\chTi^{3}_{1,1}\right)^2 
&= \chTi^{10}_{1,1} + \chTi^{10}_{1,5}  + \chTi^{10}_{1,7} + \chTi^{10}_{1,11} \\
	\left( \chTi^{3}_{1,3} \right)^2 
&= \chTi^{10}_{3,1} + \chTi^{10}_{3,5} + \chTi^{10}_{3,7} + \chTi^{10}_{3,11}
\end{align*}
then they fix $\veps(1,1)=1$, $\veps(3,1)=-1$, etc. Furthermore, if we
take $\veps(1,3)=-1$ for $m=3$,  
\be
  \chTi^{3}_{1,1} \cdot \chTi^{3}_{1,3} 
= \chTi^{10}_{5,1} + \chTi^{10}_{5,5} \nn
\ee
fixes $\veps(5,1)=1$ and $\veps(5,5)=-1$.
\end{itemize}
The above arguments fix $\veps(r,s)$ of the $\NS$ representations with
$(r,s)$ appearing in the {\DS-\ES} bulk partition function. For the
other $\NS$ representations, we simply pick $\veps(r,s)=1$.

\section{The folding map relating boundaries and defects}
\label{app:glue}

We want to 
relate two copies of the superconformal algebra
defined on the exterior of the unit circle with one copy outside and one
inside. We shall do this by considering a family of M\"obius maps $w
\mapsto z(w)$, such that the image of the real axis changes smoothly
from the real axis to the unit circle. We can take such a map to be
defined by
\be
 w = 2 i R \left( \frac{ z - i/R }{z - i/R + 2 i R} \right)
\;.
\label{eq:w}
\ee
For $R=\infty$, this is the identity map; for $R=1$ this maps the real
axis to the unit circle. This map further has the property that the
derivative at the origin is 1,
\be
 \left.\frac{\partial z}{\partial w}\right|_{w=0} = 1
\;.
\ee
The map relating generators of the folded model, $G^2$, and the
unfolded model, $\bar G$, is
\be
 G^2(w)|_{w=a} = \bar G(\bar w)|_{w = \bar a}
\;.
\ee
We would like to relate the modes $G^2_m$ and $\bar G_m$ in the
expansions of the fields
\be
 z^{3/2} G^2(z) = \sum_m G^2_m z^{-m}
\;,\;\;
 \bar z^{3/2} \bar G(\bar z) = \sum_m \bar G_m \bar z^{-m}
\;,
\ee
when $R=1$. Under the map \eqref{eq:w}, the relation becomes
\be
 z^{3/2} G^2(z)  \Big|_{w=a}
= + \left( \frac{(2 i R - a)(2 a R^2 + 2 i R - a)}
                {(2 i R + a)(2 a R^2 - 2 i R - a)}
     \right)^{3/2}
    \bar z^{3/2} \bar G(\bar z)\Big |_{w=\bar a}
\;,
\label{eq:z}
\ee
where the `$+$' sign is chosen so that the map is correct at
$R=\infty$.
At $R=1$, we have 
\be
 \bar z|_{w=a} = \frac{1}{a}
\;,
\ee
and so, 
\be
a^{3/2} G^2(a) = - i a^{-3/2} \bar G(1/a)
\;,\;\;
G^2_m = - i \bar G_{-m}
\;,
\ee
where the factor of $-i$ comes from requiring the relation
\eqref{eq:z} continue smoothly to $R=1$.
Likewise, we find
\be
\bar G^2_m = + i G_{-m}
\;.
\ee

\section{Explicit expansions of boundary states}
\label{app:states}

We give here the explicit expansions of some boundary states for
\SVIR2 and their images under the maps $\iota$ and $\rho$. 
These expressions are needed to fix the constants $\eta_{1,5}$,
$\eta_{3,1}$ etc as well as to calculate the transmission
coefficient $T$.
Throughout this section, we shall use $c' = 2c$.

\newcommand{\db}{\displaybreak[1]}
\begin{align}
  & \!\!\!\!\!\! \ket{(1,1)} 
= \ket 0 \\
 & \!\!\!\! \kett{(1,1)\eps}
= 
  \ket 0 
- \frac{i \eps}{2c'/3} G_{-3/2}\bar G_{-3/2} \ket 0 
+ \frac{1}{c'/2}L_{-2}\bar L_{-2}\ket 0
- \frac{i \eps}{2c'} G_{-5/2}\bar G_{-5/2} \ket 0
\nn\\
& 
+ \frac{1}{2c'} L_{-3}\bar L_{-3} \ket 0
- \frac{3 i \eps}{c'(c'+12)} L_{-2}G_{-3/2}\bar L_{-2}\bar
G_{-3/2}\ket 0
\nn\\
&
- \frac{81 i \eps}{c'(c'{+}12)(21 {+} 4c')}
\left[ L_{-2}G_{-3/2} - \tfrac{c'+12}9 G_{-7/2} \right]
\!\!\left[ \bar L_{-2}\bar G_{-3/2} - \tfrac{c'+12}9 \bar G_{-7/2} \right]
\ket 0
+ \ldots
\db\\
\iota&(\kett{(1,1)\eps}) 
= 
  \ket 0 
- \frac{i \eps}{4c/3} (\alpha G^1_{-3/2} + \beta G^2_{-3/2})
                       (\gamma\bar G^1_{-3/2} + \delta\bar G^2_{-3/2}) \ket 0 
\nn\\
&
+ \frac{1}{c}(L^1_{-2} + L^2_{-2})(\bar L^1_{-2} + \bar L^2_{-2})\ket 0
- \frac{i \eps}{4c} \gex{-5/2}\gbex{-5/2}\ket 0
\nn\\
& 
+ \frac{1}{4c} \lex{-3}\lbex{-3} \ket 0
\nn\\
&
- \frac{3 i \eps}{4c(c+6)} \lex{-2}\gex{-3/2}\lbex{-2}\gbex{-3/2}\ket 0
\nn\\
&
- \frac{81 i \eps}{4c(c{+}6)(21 {+} 8c)}
\left[ \lex{-2}\gex{-3/2} - \tfrac{2(c+6)}9 \gex{-7/2} \right]
\nn\\
&\;\;\;\;\;\;\;
\left[ \lbex{-2}\gbex{-3/2} - \tfrac{2(c+6)}9 \gbex{-7/2} \right]
\ket 0
+ \ldots
\db \\
\rho  (&\kett{(1,1)\eps})  = 
\ketbra{0}{0}
\nn\\
&- \frac{i\eps}{4c/3}
  \Big[
  \alpha\gamma G_{-3/2}\bar G_{-3/2}\ketbra 00
+i\alpha\delta G_{-3/2}\ketbra 00 G_{3/2}
+i\beta\gamma \bar G_{-3/2}\ketbra 00 \bar G_{3/2}
+ \beta\delta \ketbra 00 \bar G_{3/2} G_{3/2} 
  \Big]
\nn\\
& + \frac{1}{c} 
  \Big[ 
  L_{-2}\bar L_{-2} \ketbra 00
+ L_{-2}\ketbra 00 L_2
+ \bar L_{-2} \ketbra 00 \bar L_2
+ \ketbra 00 \bar L_2 L_2
  \Big]
\nn\\
&- \frac{i\eps}{4c}
  \Big[
  \alpha\gamma G_{-5/2}\bar G_{-5/2}\ketbra 00
+i\alpha\delta G_{-5/2}\ketbra 00 G_{3/2}
+i\beta\gamma \bar G_{-5/2}\ketbra 00 \bar G_{3/2}
+ \beta\delta \ketbra 00 \bar G_{3/2} G_{3/2} 
  \Big]
\nn\\
& + \frac{1}{4c} 
  \Big[ 
  L_{-3}\bar L_{-3} \ketbra 00
+ L_{-3}\ketbra 00 L_3
+ \bar L_{-3} \ketbra 00 \bar L_3
+ \ketbra 00 \bar L_3 L_3
  \Big]
\nn\\
& + \ldots 
\end{align}
\begin{align}
\ket{(1,5)} & = \ket{\tfrac 32}\\
\kett{(1,5)\eps} & = \ket{\tfrac 32}
-\frac{i\eps}{3}G_{-1/2}\bar G_{-1/2} \ket{\tfrac 32}
+ \frac{1}{3}L_{-1}\bar L_{-1} \ket{\tfrac 32}
+\ldots
\\
\iota(\ket{(1,5)}) & = 
\frac{i\,\eta_{1,5}}{4c/3}
  (\alpha G^1_{-3/2}-\beta G^2_{-3/2})(\gamma\bar G^1_{-3/2} - \delta\bar G^2_{-3/2})\ket 0  \\
\iota(\kett{(1,5)\eps}) & = 
\frac{i\,\eta_{1,5}}{4c/3}
  (\alpha G^1_{-3/2}-\beta G^2_{-3/2})(\gamma\bar G^1_{-3/2} - \delta\bar G^2_{-3/2})\ket 0  
-\frac{\eps\eta_{1,5}}{c}(L^1_{-1}-L^2_{-1})(\bar L^1_{-1}\bar L^2_{-1})\ket 0
+ \ldots
\end{align}
\begin{align}
\rho&(\kett{(1,5)\eps}) = 
\nn\\
&
\;\;\;\frac{i\,\eta_{1,5}}{4c/3}
\Big[
  \alpha\gamma G_{-1/2}\bar G_{-3/2}\ketbra 00
+ \beta\delta \ketbra 00 \bar G_{3/2} G_{3/2} 
-i\alpha\delta G_{-3/2}\ketbra 00 G_{3/2}
-i\beta\gamma \bar G_{-3/2}\ketbra 00 \bar G_{3/2}
  \Big]
\nn\\
& - \frac{\eps\eta_{1,5}}{c} 
  \Big[ 
  L_{-2}\bar L_{-2} \ketbra 00
- L_{-2}\ketbra 00 L_2
- \bar L_{-2} \ketbra 00 \bar L_2
+ \ketbra 00 \bar L_2 L_2
  \Big] + \ldots 
\end{align}
We now consider the sector corresponding to $\cH_{1/10}\otimes\cH_{1/10}$.
We give the results in terms of a state of weight $2h$, but of course in this
particular case $h=1/10$,  the states are identified as
\be
\ket{2h}=\ket{\tfrac 15} = \ket{(3,5)}
\;,\;\;
\ket{2h{+}\tfrac 12}=\ket{\tfrac 7{10}}=\ket{(3,1)}
\;,\;\;
\ket{2h{+}1} = \ket{\tfrac 65}= \ket{(3,7)}
\;,
\ee
and the constants are $\eta\equiv\eta_{3,1}$ and $\eta'\equiv\eta_{3,7}$.
\begin{align}
\kett{(2h)\eps} & =
\ket{2h} - \frac{i\eps}{4h}G_{-1/2}\bar G_{-1/2}\ket{2h}
+ \frac{1}{4h}L_{-1}\bar L_{-1}\ket{2h}+\ldots \\
\iota(\kett{(2h)\eps}) & = 
  \ket {2h}
- \frac{i \eps}{4h} (\alpha G^1_{-1/2} + \beta G^2_{-1/2})
                       (\gamma\bar G^1_{-1/2} + \delta\bar G^2_{-1/2}) \ket {2h}
\nn\\
&
+ \frac{1}{4h}(L^1_{-1} + L^2_{-1})(\bar L^1_{-1} + \bar L^2_{-1})\ket {2h}
+\ldots
\\
\rho  (\kett{(2h)\eps})  &= 
\ketbra hh
\nn\\
&- \frac{i\eps}{4h}
  \Big[
  \alpha\gamma G_{-1/2}\bar G_{-1/2}\ketbra hh
+i\alpha\delta G_{-1/2}\ketbra hh G_{1/2}
+i\beta\gamma \bar G_{-1/2}\ketbra hh \bar G_{1/2}
+ \beta\delta \ketbra hh \bar G_{1/2} G_{1/2} 
  \Big]
\nn\\
& + \frac{1}{4h} 
  \Big[ 
  L_{-1}\bar L_{-1} \ketbra hh
+ L_{-1}\ketbra hh L_1
+ \bar L_{-1} \ketbra hh \bar L_1
+ \ketbra hh \bar L_1 L_1
  \Big] + \ldots 
\nn\\
\kett{(2h{+}\tfrac 12)\eps} & =
  \ket{2h{+}\tfrac 12} 
- \frac{i\eps}{4h{+}1}G_{-1/2}\bar G_{-1/2}\ket{2h{+}\tfrac 12} 
+ \ldots 
\db\\
\iota(\ket{2h{+}\tfrac 12} &=
\frac{i\eta}{4h}
  (\alpha G^1_{-1/2}-\beta G^2_{-1/2})
  (\gamma \bar G^1_{-1/2} -\delta \bar G^2_{-1/2}) \ket{2h}
\db\\
\iota(\kett{(2h{+}\tfrac 12)\eps}) & = 
\frac{i\eta}{4h}
  (\alpha G^1_{-1/2}-\beta G^2_{-1/2})
  (\gamma \bar G^1_{-1/2} -\delta \bar G^2_{-1/2}) \ket{2h}
\nn
\db\\
&- \frac{\eps\eta}{4h(4h+1)} 
    (L^1_{-1}-L^2_{-1} - 2 \alpha\beta G^1_{-1/2}G^2_{-1/2})
    (\bar L^1_{-1}-\bar L^2_{-1} -2\gamma\delta \bar G^1_{-1/2}\bar G^2_{-1/2})
  \ket{2h}
+\ldots
\db\\
\rho  (\kett{(2h+\tfrac 12)\eps})  &= 
\nn\\
&\frac{\eta}{4h}
  \Big[
 i \alpha\gamma G_{-1/2}\bar G_{-1/2}\ketbra hh
+i \beta\delta \ketbra hh \bar G_{1/2} G_{1/2} 
+ \alpha\delta G_{-1/2}\ketbra hh G_{1/2}
+ \beta\gamma \bar G_{-1/2}\ketbra hh \bar G_{1/2}
  \Big]
\nn\\
& - \frac{\eps\eta}{4h(4h{+}1)} 
  \Big[ 
  L_{-1}\bar L_{-1} \ketbra hh
+ L_{-1}\ketbra hh L_1
+ \bar L_{-1} \ketbra hh \bar L_1
+ \ketbra hh \bar L_1 L_1
  \Big] 
\nn\\
& + \frac{2i\eps\eta\alpha\beta}{4h(4h{+}1)} 
  \Big[ 
  G_{-1/2}\ketbra hh \bar G_{1/2} L_1 
- \bar L_{-1} G_{-1/2}\ketbra hh \bar G_{1/2} \Big]
\nn\\
& + \frac{2i\eps\eta\gamma\delta}{4h(4h{+}1)} 
  \Big[ 
   L_{-1} \bar G_{-1/2}\ketbra hh G_{1/2} 
-  \bar G_{-1/2}\ketbra hh \bar G_{1/2} \bar L_1 
\Big]
\nn\\
& + \frac{4\eps\eta\alpha\beta\gamma\delta}{4h(4h{+}1)} 
  \Big[ 
   G_{-1/2} \bar G_{-1/2}\ketbra hh \bar G_{1/2} G_{1/2}
\Big] + \ldots
\nn\\
\kett{(2h{+}1)\eps} & = \ket{2h{+}1} + \ldots \\
\iota(\kett{(2h{+}1)\eps}) & = 
\frac{\eta'}{4h{+}1}
  (L^1_{-1} - L^2_{-1} + \frac{\alpha\beta}{2h}G^1_{-1/2}G^2_{-1/2})
 (\bar L^1_{-1}-\bar L^2_{-1}+\frac{\gamma\delta}{2h}\bar G^1_{-1/2}\bar G^2_{-1/2}) 
\ket{2h}+\ldots
\\
\rho(\kett{(2h{+}1)\eps}) & = 
\frac{\eta'}{4h{+}1}
 \Big[ L_{-1}\bar L_{-1}\ketbra hh
 - L_{-1}\ketbra hh L_1
 - \bar L_{-1}\ketbra hh \bar L_1
 + \ketbra hh \bar L_1 L_1 \Big] + \ldots
\\
& + \frac{i\eta'\alpha\beta}{2h(4h{+}1)} 
  \Big[ 
  G_{-1/2}\ketbra hh \bar G_{1/2} L_1 
- \bar L_{-1} G_{-1/2}\ketbra hh \bar G_{1/2} \Big]
\nn\\
& + \frac{i\eta'\gamma\delta}{2h(4h{+}1)} 
  \Big[ 
   L_{-1} \bar G_{-1/2}\ketbra hh G_{1/2} 
-  \bar G_{-1/2}\ketbra hh \bar G_{1/2} \bar L_1 
\Big]
\nn\\
& - \frac{\eta'\alpha\beta\gamma\delta}{4h^2(4h{+}1)} 
  \Big[ 
   G_{-1/2} \bar G_{-1/2}\ketbra hh \bar G_{1/2} G_{1/2}
\Big] + \ldots
\end{align}

\section{The matrices $\Psi^{(a,b)}_{(r,s)}$}
\label{app:bcs}

The matrices $\Psi^{(a,b)}_{(r,s)}$ are given in terms of the
eigenvectors of adjacency matrices of the Dynkin diagrams of \DS\ and
\ES\ in equation \eqref{eq:Psis}:
\be
  \Psi^{(a,b)}_{(r,s)} 
= \frac{ \psi^r_a(D_6)\,\psi^s_b(E_6) }
       {\sqrt{S_{1r}^{(8)} S_{1s}^{(10)} }}
\;,
\ee
We repeat here for convenience the vectors $\psi^r_a(G)$ given in
\cite{GY}:


The eigenvectors of the {\DS} adjacency matrix 
$\psi_a^r(\DS)$ are given by  
\begin{align*}
 \psi_a^r(\DS) &= \sqrt{2} S^{(8)}_{ar} \quad \text{for } a,r \neq 5 
& \psi_a^{5^\pm}(\DS) &= S^{(8)}_{a5} \quad \text{for } a \neq 5 \\
 \psi_{5^\pm}^r(\DS) &= \frac{1}{\sqrt{2}} S^{(8)}_{5r} \quad \text{for } r \neq 5 
& \psi_{5^\eps}^{5^{\eps'}}(\DS)&= \frac{1}{2} \left( S^{(8)}_{55} -\eps\eps' \right)
\end{align*}
where $a = 1,2,3,4,5^+,5^-$ ($a=5^\pm$ correspond to $5$ and $6$ nodes
on the {\DS} Dynkin diagram), $r \in \cE(\DS) = \{1,3,5,5',7,9\}$
($r=5^\pm$ above correspond to $5$ and $5'$), and $S^{(8)}_{ij}$ is
the $\hat{\text{su}}(2)_8$ modular S matrix elements, 
\be
	S^{(k)}_{ij} = \sqrt{\frac{2}{k+2}} \sin\left( \frac{\pi ij}{k+2} \right) \nn
\ee
Explicitly, the entries in $\psi_a^r(\DS)$ are
\be
{\arraycolsep=0.3em\def\arraystretch{1.8}
\begin{array}{c|cccccc}
	a \text{ {\textbackslash} } r & 1 & 3 & 5^+ \:(=5) & 5^- \:(=5') & 7 & 9 \\ \hline
	1 & \frac{-1+\sqrt{5}}{2 \sqrt{10}} & \frac{1}{2} \sqrt{\frac{3}{5}+\frac{1}{\sqrt{5}}} & \frac{1}{\sqrt{5}} & \frac{1}{\sqrt{5}} & \frac{1}{2} \sqrt{\frac{3}{5}+\frac{1}{\sqrt{5}}} & \frac{-1+\sqrt{5}}{2 \sqrt{10}} \\
	2 & \frac{1}{2} \sqrt{1-\frac{1}{\sqrt{5}}} & \frac{1}{2} \sqrt{1+\frac{1}{\sqrt{5}}} & 0 & 0 & -\frac{1}{2} \sqrt{1+\frac{1}{\sqrt{5}}} & -\frac{1}{2} \sqrt{1-\frac{1}{\sqrt{5}}} \\
	3 & \frac{1}{2} \sqrt{\frac{3}{5}+\frac{1}{\sqrt{5}}} & \frac{-1+\sqrt{5}}{2 \sqrt{10}} & -\frac{1}{\sqrt{5}} & -\frac{1}{\sqrt{5}} & \frac{-1+\sqrt{5}}{2 \sqrt{10}} & \frac{1}{2} \sqrt{\frac{3}{5}+\frac{1}{\sqrt{5}}} \\
	4 & \frac{1}{2} \sqrt{1+\frac{1}{\sqrt{5}}} & -\frac{1}{2} \sqrt{1-\frac{1}{\sqrt{5}}} & 0 & 0 & \frac{1}{2} \sqrt{1-\frac{1}{\sqrt{5}}} & -\frac{1}{2} \sqrt{1+\frac{1}{\sqrt{5}}} \\
	5^+ \:(=5) & \frac{1}{\sqrt{10}} & -\frac{1}{\sqrt{10}} & \frac{1}{10} \left(-5+\sqrt{5}\right) & \frac{1}{10} \left(5+\sqrt{5}\right) & -\frac{1}{\sqrt{10}} & \frac{1}{\sqrt{10}} \\
	5^- \:(=6) & \frac{1}{\sqrt{10}} & -\frac{1}{\sqrt{10}} & \frac{1}{10} \left(5+\sqrt{5}\right) & \frac{1}{10} \left(-5+\sqrt{5}\right) & -\frac{1}{\sqrt{10}} & \frac{1}{\sqrt{10}} \\
\end{array}
}
\nn\ee

The eigenvectors of the {\ES} adjacency matrix 
$\psi_b^s(\ES)$ are given by 
\be
{\arraycolsep=0.3em\def\arraystretch{1.8}
\begin{array}{c|cccccc}
	b \text{ {\textbackslash} } s & 1 & 4 & 5 & 7 & 8 & 11 \\ \hline
	1 & a & \frac{1}{2} & b & b & \frac{1}{2} & a \\
	2 & b & \frac{1}{2} & a & -a & -\frac{1}{2} & -b \\
	3 & c & 0 & -d & -d & 0 & c \\
	4 & b & -\frac{1}{2} & a & -a & \frac{1}{2} & -b \\
	5 & a & -\frac{1}{2} & b & b & -\frac{1}{2} & a \\
	6 & d & 0 & -c & c & 0 & -d \\
\end{array}
\quad \text{where} \quad
\begin{array}{cccc}
	a &= \frac{1}{2} \sqrt{\frac{3-\sqrt{3}}{6}} & b &= \frac{1}{2} \sqrt{\frac{3+\sqrt{3}}{6}} \\
	c &= \frac{1}{2} \sqrt{\frac{3+\sqrt{3}}{3}} & d &= \frac{1}{2} \sqrt{\frac{3-\sqrt{3}}{3}}
\end{array}
}
\nn\ee

Putting these together, we can calculate the entries of $\Psi$. Since
it is helpful to have an overview of the properties of $\Psi$ when
discussing the boundary states from the extended algebra point of view, we
include a table of the approximate numerical values in table \ref{tab:nums}.

\begin{landscape}
\begin{table}
\[
\begin{array}{c|rrrrrrrrrrrr}
& \multicolumn{12}{c}{(r,s)} \\
(a,b) & (1,1) & (1,5) & (1,7) & (1,11) & (3,1) & (3,5) & (3,7) & (3,11) &
 (5,1) & (5,5) & (5',1) & (5',5) \\ \hline
 (1,1) & 0.3717 & 0.3717 & 0.3717 & 0.3717 & 0.6015 & 0.6015 & 0.6015 & 0.6015 & 0.4729 & 0.4729 & 0.4729 & 0.4729 \\
 (1,2) & 0.7182 & -0.1924 & -0.1924 & 0.7182 & 1.162 & -0.3114 & -0.3114 & 1.162 & 0.9135 & -0.2448 & 0.9135 & -0.2448 \\
 (1,3) & 1.016 & -0.2721 & -0.2721 & 1.016 & 1.643 & -0.4403 & -0.4403 & 1.643 & 1.292 & -0.3462 & 1.292 & -0.3462 \\
 (1,6) & 0.5257 & 0.5257 & 0.5257 & 0.5257 & 0.8507 & 0.8507 & 0.8507 & 0.8507 & 0.6687 & 0.6687 & 0.6687 & 0.6687 \\
 (2,1) & 0.7071 & -0.7071 & 0.7071 & -0.7071 & 0.7071 & -0.7071 & 0.7071 & -0.7071 & 0 & 0 & 0 & 0 \\
 (2,2) & 1.366 & 0.3660 & -0.3660 & -1.366 & 1.366 & 0.3660 & -0.3660 & -1.366 & 0 & 0 & 0 & 0 \\
 (2,3) & 1.932 & 0.5176 & -0.5176 & -1.932 & 1.932 & 0.5176 & -0.5176 & -1.932 & 0 & 0 & 0 & 0 \\
 (2,6) & 1.000 & -1.000 & 1.000 & -1.000 & 1.000 & -1.000 & 1.000 & -1.000 & 0 & 0 & 0 & 0 \\
 (3,1) & 0.9732 & 0.9732 & 0.9732 & 0.9732 & 0.2298 & 0.2298 & 0.2298 & 0.2298 & -0.4729 & -0.4729 & -0.4729 & -0.4729 \\
 (3,2) & 1.880 & -0.5038 & -0.5038 & 1.880 & 0.4438 & -0.1189 & -0.1189 & 0.4438 & -0.9135 & 0.2448 & -0.9135 & 0.2448 \\
 (3,3) & 2.659 & -0.7125 & -0.7125 & 2.659 & 0.6277 & -0.1682 & -0.1682 & 0.6277 & -1.292 & 0.3462 & -1.292 & 0.3462 \\
 (3,6) & 1.376 & 1.376 & 1.376 & 1.376 & 0.3249 & 0.3249 & 0.3249 & 0.3249 & -0.6687 & -0.6687 & -0.6687 & -0.6687 \\
 (4,1) & 1.144 & -1.144 & 1.144 & -1.144 & -0.4370 & 0.4370 & -0.4370 & 0.4370 & 0 & 0 & 0 & 0 \\
 (4,2) & 2.210 & 0.5922 & -0.5922 & -2.210 & -0.8443 & -0.2262 & 0.2262 & 0.8443 & 0 & 0 & 0 & 0 \\
 (4,3) & 3.126 & 0.8376 & -0.8376 & -3.126 & -1.194 & -0.3199 & 0.3199 & 1.194 & 0 & 0 & 0 & 0 \\
 (4,6) & 1.618 & -1.618 & 1.618 & -1.618 & -0.6180 & 0.6180 & -0.6180 & 0.6180 & 0 & 0 & 0 & 0 \\
 (5,1) & 0.6015 & 0.6015 & 0.6015 & 0.6015 & -0.3717 & -0.3717 & -0.3717 & -0.3717 & -0.2923 & -0.2923 & 0.7651 & 0.7651 \\
 (5,2) & 1.162 & -0.3114 & -0.3114 & 1.162 & -0.7182 & 0.1924 & 0.1924 & -0.7182 & -0.5646 & 0.1513 & 1.478 & -0.3961 \\
 (5,3) & 1.643 & -0.4403 & -0.4403 & 1.643 & -1.016 & 0.2721 & 0.2721 & -1.016 & -0.7984 & 0.2139 & 2.090 & -0.5601 \\
 (5,6) & 0.8507 & 0.8507 & 0.8507 & 0.8507 & -0.5257 & -0.5257 & -0.5257 & -0.5257 & -0.4133 & -0.4133 & 1.082 & 1.082 \\
 (6,1) & 0.6015 & 0.6015 & 0.6015 & 0.6015 & -0.3717 & -0.3717 & -0.3717 & -0.3717 & 0.7651 & 0.7651 & -0.2923 & -0.2923 \\
 (6,2) & 1.162 & -0.3114 & -0.3114 & 1.162 & -0.7182 & 0.1924 & 0.1924 & -0.7182 & 1.478 & -0.3961 & -0.5646 & 0.1513 \\
 (6,3) & 1.643 & -0.4403 & -0.4403 & 1.643 & -1.016 & 0.2721 & 0.2721 & -1.016 & 2.090 & -0.5601 & -0.7984 & 0.2139 \\
 (6,6) & 0.8507 & 0.8507 & 0.8507 & 0.8507 & -0.5257 & -0.5257 & -0.5257 & -0.5257 & 1.082 & 1.082 & -0.4133 & -0.4133
\end{array}
\]
\caption{Numerical values of the boundary state coefficients $ \Psi^{(a,b)}_{(r,s)} $}
\label{tab:nums}
\end{table}
\end{landscape}

\section{Character identities}
\label{app:charids}

Relations expressing products of
NS characters of $\sVir_3$ as sums of NS characters of $\sVir_{10}$:
\bea
\chten_{1,1} + 
\chten_{1,5} + 
\chten_{1,7} + 
\chten_{1,11} 
&=&
(\chthree_{1,1})^2
\\ 
\chten_{3,1} + 
\chten_{3,5} + 
\chten_{3,7} + 
\chten_{3,11} 
&=&
(\chthree_{1,3})^2
\\
\chten_{5,1} + \chten_{5,5} &=&
\chthree_{1,1} \cdot \chthree_{1,3}
\eea

Relations expressing products of
Ramond characters for $\sVir_3$ as sums of Ramond characters for $\sVir_{10}$
:
\bea
\chten_{3,4} + 
\chten_{3,8} 
&=&
(\chthree_{1,2})^2
\\ 
\chten_{1,4} + 
\chten_{1,8} 
&=&
(\chthree_{1,4})^2
\\
\chten_{5,4}
&=&
\chthree_{1,2} \cdot \chthree_{1,4}
\eea

Relations expressing Ramond characters for
$\sVir_3$ at $\sqrt q$ as sums of characters for $\sVir_{10}$. Note
that the same 
characters for $\sVir_3$ can be expressed as sums of characters in both the
Ramond and NS sectors of $\sVir_{10}$.
\bea
\chten_{2,4} + \chten_{2,8} &=& \chthree_{1,4}(\sqrt q)
\\
\chten_{2,1} + \chten_{2,5} 
+ \chten_{2,7} + \chten_{2,11} 
&=& \chthree_{1,4}(\sqrt q)
\\
\chten_{4,4} + \chten_{4,8} &=& \chthree_{1,2}(\sqrt q)
\\
\chten_{4,1} + \chten_{4,5} 
+ \chten_{4,7} + \chten_{4,11} 
&=& \chthree_{1,2}(\sqrt q)
\eea

\newpage
\bibliographystyle{JHEP}
\begingroup\raggedright\endgroup

\end{document}